\documentclass{aa}  

\usepackage{graphicx}
\usepackage{txfonts}
\begin{document}

   \title{
   HD\,101584: Circumstellar characteristics and evolutionary status
   }

   \titlerunning{Circumstellar environment and nature of HD\,101584}

   \author{H.~Olofsson  \inst{1}
          \and 
          T.~Khouri    \inst{1}
           \and
          M.~Maercker \inst{1}
          \and
          P.~Bergman   \inst{1}
          \and
          L.~Doan  \inst{2}
          \and
          D.~Tafoya  \inst{3}
          \and
          W.H.T.~Vlemmings \inst{1}
          \and      
          E.M.L.~Humphreys \inst{4}
          \and
          M.~Lindqvist \inst{1}
          \and
          L.~Nyman \inst{5}
          \and
          S.~Ramstedt \inst{2}          
}

   \institute{Dept of Space, Earth and Environment, Chalmers Univ. of Technology,
              Onsala Space Observatory, SE-43992 Onsala, Sweden\\
              \email{hans.olofsson@chalmers.se}
         \and
         Dept of Physics and Astronomy, Uppsala University, Box 516, SE-75120 Uppsala, Sweden
         \and
         National Astronomical Observatory of Japan, 2-21-1 Osawa, Mitaka, Tokyo 181-8588, Japan
         \and
         ESO, Karl-Schwarzschild-Str. 2, D-85748 Garching bei M{\"u}nchen, Germany
         \and
         ESO, Alonso de Cordova 3107, Vitacura, Santiago, Chile               
}

   \date{Received 17 December 2018; accepted 30 January 2019}

 \abstract{}{}{}{}{} 
 
  \abstract
   {There is growing evidence that red giant evolution is often affected by an interplay with a nearby companion, in some cases taking the form of a common-envelope evolution.}
   {We have performed a study of the characteristics of the circumstellar environment of the binary object HD\,101584, that provides information on a likely evolutionary scenario. }
   {We have obtained and analyzed ALMA observations, complemented with observations using APEX, of a large number of molecular lines. An analysis of the spectral energy distribution has also been performed.}
  {Emissions from 12 molecular species (not counting isotopologues) have been observed, and most of them  mapped with angular resolutions in the range 0\farcs1 to 0\farcs6. Four circumstellar components are identified: {\it i)} a central compact source of size $\approx$\,0\farcs15, {\it ii)} an expanding equatorial density enhancement (a flattened density distribution in the plane of the orbit) of size $\approx$\,3\arcsec , {\it iii)} a bipolar high-velocity outflow ($\approx$\,150\,km\,s$^{-1}$), and {\it iv)} an hourglass structure. The outflow is directed almost along the line of sight. There is evidence of a second bipolar outflow. The mass of the circumstellar gas is $\approx$\,0.5\,[$D$/1\,kpc]$^2$\,$M_\odot$, about half of it lies in the equatorial density enhancement. The dust mass is $\approx$\,0.01\,[$D$/1\,kpc]$^2$\,$M_\odot$, and a substantial fraction of this is in the form of large-sized, up to 1\,mm, grains. The estimated kinetic age of the outflow is $\approx$\,770\,[$D$/1\,kpc]\,yr. The kinetic energy and the scalar momentum of the accelerated gas are estimated to be 7$\times$10$^{45}$\,[$D$/1\,kpc]$^2$\,erg and  10$^{39}$\,[$D$/1\,kpc]$^2$\,g\,cm\,s$^{-1}$, respectively. }
   {We provide good evidence that the binary system HD\,101584 is in a post-common-envelope-evolution phase, that ended before a stellar merger. Isotope ratios combined with stellar mass estimates suggest that the primary star's evolution was terminated already on the first red giant branch (RGB). Most of the energy required to drive the outflowing gas was probably released when material fell towards the companion.  }


   \maketitle
%
%
%
\section{Introduction}

It has become increasingly apparent that the presence of a nearby companion has a large impact on stellar evolution during various phases. A particularly interesting case is that of red giant evolution when the primary star becomes so large that the companion is (or at least is close to being) engulfed by the tenuous stellar hydrogen envelope of the giant star. This leads to a common-envelope (CE) evolution process which is difficult to understand in detail and very difficult to simulate numerically \citep{ivanetal13}. There is good evidence of this phenomenon occurring. Companions around objects beyond the asymptotic giant branch (post-AGB) are ubiquitous in for example the Magellanic Clouds (MCs) \citep{vanw03, kamaetal14, kamaetal15}, and they show an unexpected period distribution covering the range 100 to 2000 days, meaning that the companion lies closer than typical stellar radii on the AGB \citep{vanwetal09}. In addition, they have dusty circumbinary disks. Recently, characteristics of the same kind have been found also for stars earlier in their evolution, for example, beyond the first red giant branch (post-RGB) \citep{kamaetal16}. These are examples where the red giant evolution was terminated and the companion survived, but there are likely also cases that ended with a complete stellar merger.

The described scenario may be closely related to well-known phenomena, for example, planetary nebulae (PNe). That they are the descendants of AGB stars is well established \citep{shkl78}, but it remains to be shown what the necessary conditions for this to actually happen are, since not all AGB stars become PNe. The increasingly energetic radiation from the star (essentially its core at this stage) as it evolves off the AGB is expected to produce a PN, with its highly excited and ionised circumstellar gas produced by an effective mass loss during the AGB, provided the conditions are the right ones \citep{kwoketal78, vanw03}. In addition to this, there is, in many cases, an associated spectacular transformation of the circumstellar morphology and kinematics that poses an explanatory problem, where the effects of a nearby companion is often introduced. 

There are other objects with spectacular circumstellar characteristics, whose explanation call for a companion and most likely a CE-evolution scenario. An example is the Boomerang nebula, potentially a stellar merger following RGB evolution of its primary, with its massive, high-velocity outflow of extremely cold gas \citep{sahaetal17b}.  Another type of possibly related objects are the red novae, explained to be the results of stellar mergers following CE evolution. Such objects also show circumstellar characteristics, for example, very-high-velocity molecular outflows, that are difficult to explain by single stellar evolution \citep{kamietal18a, kamietal18b}. 

AGB stars and PNe are the most well-studied objects of the type discussed here. We will describe them in some more detail, and note that many of the discussed phenomena may have a more general application. AGB stars are largely spherical although inhomogeneities in their extended atmospheres appear to be a common phenomenon \citep{ohnaetal16b, khouetal16a, vlemetal17}, most likely an effect of convection and pulsation \citep{freyetal17, liljetal18}. Also their circumstellar envelopes (CSEs) of gas and microscopic dust grains have an overall spherical symmetry \citep{cacaetal10}. Notable exceptions to this general behaviour exist though, for example, the carbon star V~Hya \citep{sahaetal16}, the S-star $\pi^1$~Gru \citep{doanetal17} and the M-star L$_2$~Pup \citep{kervetal15}, all three being semi-regular variables. The two former have bipolar high-velocity outflows and equatorial density enhancements, while the latter has a 10\,au-sized circumbinary disk. Further, detections of spiral patterns or arcs in the circumstellar gas and dust distributions are becoming increasingly common \citep{maurhugg06, maeretal12, kimetal17, ramsetal17, gueletal18}.

This is in stark contrast to the PNe where complex geometrical patterns and directed flows of high-velocity gas are the rule rather than the exception \citep{sahaetal11a}. This applies in particular to the objects in transition from the AGB to the PN stage, the proto-PNe, where no examples of overall sphericity have been found \citep{sahaetal07}. However, even though the PNe morphology is often complex, the presence of an equatorial density enhancement, and often an axial symmetry in the orthogonal direction in the form of a high-velocity bipolar outflow, are key features. Both these components are also commonly found already in the proto-PNe. Notable examples are AFGL618 \citep{coxetal03, leeetal13}, OH231.8+4.2 \citep{alcoetal01, bujaetal02, sacoetal18}, M1-6 \citep{huggetal00}, M1-92 \citep{bujaetal97, alcoetal07}, M2-9 \citep{cacaetal17}, and M2-56 \citep{cacaetal02}, and the young PN NGC6302 \citep{sagaetal17}. Among the lower-mass post-AGB objects the presence of disks appear ubiquitous \citep{hilletal14, hilletal17, gorletal15, bujaetal18}, and also jets are found \citep{bolletal17}. Molecular bipolar outflows are more scarce, and if present of low velocity \citep{bujaetal13}. The disks are often stable and in Keplerian rotation, for example, the Red Rectangle \citep{bujaetal16} and IW~Car \citep{bujaetal17}. Less well-characterised sources like IRAS\,08005--2356 \citep{sahapate15}, IRAS\,16342--3814 \citep{sahaetal17a}, and IRAS\,22026+5306 \citep{sahaetal06} have young bipolar outflows of very high velocities.

There is a general belief that the axial symmetry has its root in the AGB star not being alone \citep{vanw03, doucetal15, joneboff17}. A companion in orbit provides the energy and the angular momentum required to produce equatorial density enhancements and high-velocity bipolar outflows \citep{soke97, soke98a, dema09}, either through a CE evolution \citep{sokelivi94, nordblac06, soke15} or through increasing the rotation and/or magnetic field of the primary \citep{blacetal01}. The outflows are thought to be driven by jets, originating from an accretion disk \citep{reyelope99, chametal18}, that sculpt the CSE, and hence produce the apparently complex geometrical patterns seen in the circumstellar gas of PNe \citep{sahatrau98, leesaha03}. Most likely, this phenomenon starts at the very end of the AGB evolution when the mass of the AGB star reaches its minimum, the radius its maximum, and the mass-loss rate its maximum. The alternative explanation where the AGB star itself produces the energy and the angular momentum of the outflow appears less likely \citep{nordblac06}. Unfortunately, the identification of a (close) binary companion around an AGB star is a very difficult observational task. Indirect evidence in the form of UV and x-ray emission exists \citep{sahaetal08, sahaetal15}, although there could be other causes for the emission \citep{montetal17}.

The objects with spectacular circumstellar characteristics discussed here are not only physically complex, they are also chemically complex. It is well-known that AGB CSEs are rich in different molecular species at the end of the AGB evolution; more than 100 are now detected \citep[][and references therein]{cernetal00a, velietal17, debeolof18}. This is the effect of a number of different processes, such as stellar atmosphere equilibrium chemistry, extended atmosphere non-equilibrium chemistry, and photo-induced circumstellar chemistry \citep[for example,][]{mill16}. Additional processes become active during the post-AGB phase, for example, increased UV radiation and high-velocity outflows that lead to shocks. The result is that they often show molecular species that are not detected (or tend to be much weaker) in AGB CSEs \citep{pardetal07, edwaziur13, sacoetal15, velietal15}. The red novae have a circumstellar medium that is most likely formed in the process leading up to the stellar merger. These envelopes are also relatively rich in different molecular species \citep{kamietal18b}, and the first detection ever of a radioactive molecule, $^{26}$AlF, was done in such an environment \citep{kamietal18a}.

The object of this paper, HD\,101584, show many of the characteristics traditionally associated with a post-AGB object, but now also found in connection with post-RGB objects. The star is warm and there is a substantial far-IR excess due to a dusty circumstellar environment, and in addition there is  good evidence of a companion. Its characteristics are summarised in Sect.~\ref{s:hd_characteristics}. In this paper we present observations performed with the Atacama Large Millimeter/submillimeter Array (ALMA) in observing bands centred on the $^{12}$CO and $^{13}$CO $J$\,=\,\mbox{2--1} lines (from now on the more common isotope is meant unless rarer isotopes are specifically mentioned, that is, $^{12}$C$^{16}$O\,=\,CO). In total about 13\,GHz, in high spectral resolution mode, have been covered using ALMA.  Continuum emission at 1.3\,mm is recovered from line-free regions. We further report complementary observations of line emission and continuum emission at 350\,$\mu$m using the Atacama Pathfinder Experiment telescope (APEX).

%
%
%
\section{HD101584}

\subsection{The characteristics}
\label{s:hd_characteristics}

\object{HD\,101584} (V885~Cen, IRAS\,11385--5517) is bright at optical wavelengths ($V$\,$\approx$\,7 mag) and with an estimated effective temperature of $\approx$\,8500\,K it has a spectral type A6Ia classification \citep{sivaetal99,kipp05}. Its large far-IR excess led \citet{partpott86} to infer an evolutionary status at, or shortly after, the end of the AGB, a conclusion corroborated by \citet{bakketal96a} who presented optical and infrared data, and proposed that HD\,101584 most likely has evolved from the asymptotic giant branch (AGB) at most a few hundred years ago. They estimated a (present-day) mass of $\approx$\,0.55\,$M_\odot$, and a luminosity of $\approx$\,5000\,$L_\odot$. This identification is consistent with its location well above the galactic plane (galactic latitude of 6$^\circ$) and its high space velocity. \citet{olofetal17} provided evidence that HD\,101584 has gone through CNO-processing on the red giant branch and is of low initial mass ($\approx$\,1\,$M_\odot$), based on observations of circumstellar CO isotopologues. \citet{kipp05} found the abundances of C, N, O, Na, and Mg to be close to solar, while the Si abundance is sub-solar by a factor of 20, possibly an effect of accretion of depleted gas. 

In this paper we will argue that a post-RGB star is an alternative evolutionary status of HD\,101584. The reason for this is that we envision a geometrically different circumstellar environment than that assumed by \citet{bakketal96a}. The consequence is less circumstellar extinction, and the effect is that for a given luminosity the star must be located further away. This will impose some issues with a post-AGB interpretation as will be discussed below. 

Photometric and radial-velocity variations show that HD\,101584 has a companion. \citet{bakketal96b} used the former and estimated a period of 218$^{\rm d}$, while \citet{diazetal07} found a period of 144$^{\rm d}$ using the latter. The radial velocity estimate is presumably more accurate, but the data themselves have never been published and it is therefore not possible to estimate their uncertainty. Using these results, and assuming an 0.6\,$M_\odot$ primary star and an almost face-on orientation of the orbit plane (based on the circumstellar morphology, see below), \citet{olofetal15} estimated a binary separation of $\approx$\,0.7\,au and a companion mass of $\approx$\,0.6\,$M_\odot$, suggesting that it is a low-mass main-sequence star or possibly a low-luminosity white dwarf (WD), consistent with the absence of spectroscopic emission from the companion. 

\citet{bakketal96b} inferred an essentially edge-on circumbinary disk to explain the presence of optical absorption lines, but this orientation of the orbit plane appears less likely for a number of reasons. Images from the Hubble Space Telescope (HST) show a diffuse circumstellar environment with evidence of an essentially circular ring of radius $\approx$\,1\farcs5 roughly centred on the star \citep{sahaetal07, siodetal08}, suggesting more of a face-on orientation. Further, the central star is bright despite significant amount of circumstellar material \citep{olofetal15}, presumably because the polar axis of the system is oriented towards us and the region around it has been (at least) partially evacuated. In this paper we will provide additional arguments for the face-on orientation.

The molecular line emissions reveal much more morphological information than the visual images, and in addition provide kinematical information.  \citet{olofnyma99} obtained high-quality $^{12}$CO and $^{13}$CO $J$\,=\,\mbox{1--0} and \mbox{2--1} single-dish map data, and inferred the presence of relatively compact emission covering a velocity range of $\approx$\,100\,km\,s$^{-1}$, including a prominent central narrow-line feature, and a high-velocity ($\approx$\,150\,km\,s$^{-1}$) bipolar outflow having an east-west orientation and a Hubble-like velocity gradient. The most blue- and red-shifted emissions lie $\approx$\,5$\arcsec$ to the W and E, respectively. The full complexity of the circumstellar material was revealed through ALMA observations in frequency regions centred on the $^{12}$CO and $^{13}$CO $J$\,=\,\mbox{2--1} lines \citep{olofetal15}. A double-peaked OH 1667\,MHz maser line, with a total velocity coverage of $\approx$\,80\,km\,s$^{-1}$, was imaged by \citet{telietal92}. The integrated OH emission is centred on the star (within 0\farcs3), and the velocities of the maser spots increase systematically along a position angle (PA) $\approx$\,$-60^\circ$ with the most blue- and red-shifted emission at $\approx$\,2$\arcsec$ to the SE and the NW, respectively, that is, essentially in the opposite direction to the CO outflow. Therefore, \citet{zijletal01} proposed that the OH maser emission comes from a second bipolar outflow.

So far, there are 12 circumstellar molecular species (not counting isotopologues) detected towards HD\,101584, Table~\ref{t:species} \citep[][and this paper]{olofetal17}. This is based on data covering only a 13\,GHz spectral range in ALMA band 6, complemented with targeted detections of HCN and HCO$^+$ using APEX. The different molecular line emissions sample different regions of the circumstellar medium depending on chemistry and excitation. \citet{olofetal17} found the extreme-velocity spots of the high-velocity flow to be particularly rich in various species, and presented the first detection of methanol in an AGB-related object. In terms of detected species and their relative abundances they resemble the chemistry found in the so-called ``bullet-regions'' of bipolar outflows associated with young stellar objects (YSOs) \citep{tafabach11}. The chemistry of the circumstellar environment of HD\,101584 will be discussed in a forthcoming paper.

In summary, the circumstellar environment of HD\,101584 consists of a central component and an orthogonal, bipolar, molecular outflow. OH emission, abundant oxygen-bearing molecules, and a 10$\,\mu$m feature indicate an O-rich (C/O$<$1) circumstellar medium, that is, consistent with the chemistry of the primary star \citep{sivaetal99}. 

\begin{table}
\caption{Molecular species detected in the circumstellar environment of HD\,101584 (not counting isotopologues), divided into the different components introduced in Sect.~\ref{s:decomp}.}
\centering
\begin{tabular}{l l}
\hline
CCS: CO, CS, SiO, SiS, SO, SO$_2$, OCS, H$_2$S \\
EDE: CO, CS, SiO, SO, SO$_2$, H$_2$S \\
HGS: CO \\
HVO: CO, CS, SiO, SO, OCS, HCN, HCO$^+$, H$_2$CO, CH$_3$OH \\
\hline
\end{tabular}
\label{t:species}
\end{table}

%
%
%
%
\subsection{The scenario}
\label{s:hd_scenario}

Based on the spectacular circumstellar characteristics of HD\,101584, the following scenario for the evolution of the object has emerged \citep{olofetal15}. The companion (of low mass and in a relatively close orbit) was eventually captured a few hundred years ago, for example, when the red giant star reached a critical size. It spiralled in towards the red giant, but stopped before it fell into the core of the primary. In this process, the outer parts of the red giant was ejected and most of the material formed an equatorial density enhancement in the plane of the binary system. The cease of the inward motion of the companion was likely connected to this. A smaller fraction of the circumstellar mass is now seen in the form of a high-velocity, bipolar outflow. During this CE evolution, the red giant evolution of HD\,101584 was terminated and its core is becoming gradually revealed. HD\,101584 may serve as an example where one version of the CE scenario can be studied observationally in some detail.

%
%
%
%
\subsection{The distance}
\label{s:hd_distance}

Based on the identification of HD\,101584 as a young post-AGB object and assuming a spherical dust envelope providing significant extinction, \citet{bakketal96a} estimated the distance of HD\,101584 to be about 0.7\,kpc. However, our ALMA data rather suggest that the dust is located in a thick disk seen almost face-on, hence providing much less extinction along the line of sight, see Sect.~\ref{s:sed_results}. As a consequence, for a given luminosity the star must be placed at a larger distance. This will lead to some problems with the post-AGB interpretation as discussed below, but opens up the possibility that HD\,101584 is instead a post-RGB object of lower luminosity. We will therefore investigate two cases, a 500\,$L_\odot$ (the post-RGB case) and a 5000\,$L_\odot$ (the post-AGB case) star. The corresponding distances, taking into account the circumstellar extinction we estimate in Sect.~\ref{s:sed}, are 0.56\,kpc and 1.8\,kpc, respectively.

The recent Gaia release-2 data suggest a distance of 2.0\,(+0.19,--0.16)\,kpc \citep{gaiaetal18}. However, there are reasons why the Gaia result may not be correct in this particular case. First, the star is bright and estimated Gaia parallaxes for 7$^{\rm m}$ stars are expected to be less reliable. Second, the estimated size of the orbit is of the same magnitude as the parallax. Third, even though the uncertainty of the result is formally small (parallax equals 0.48\,$\pm$\,0.04), the goodness of fit (13.6) and the chi-square (741) values are very large. 

Therefore, we regard the Gaia estimate sufficiently uncertain to warrant giving all the distance-dependent results as their values at 1\,kpc and the scaling of these values with distance in this paper. We will discuss the consequences of the uncertain distance for the evolutionary status of HD\,101584 in Sect.~\ref{s:evol_status}. Note that some quantities are constant, irrespective of the distance, for example, $L_\ast/D^2$ and $R_\ast/D$, where $D$ is the distance, and $L_\ast$ and $R_\ast$ the stellar luminosity and radius, respectively.
%
%
%
%
\section{Observations}
\label{s:obs_desc}
%
\subsection{ALMA}

The ALMA data were obtained during cycles~1 (May 2014, TA1) and 3 (October 2015, TA2; September 2016, TA3) with 35 to 39 antennas of the 12\,m main array in two frequency settings in band 6, one for the $^{12}$CO($J$\,=\,\mbox{2--1}) line (both cycles) and one for the $^{13}$CO($J$\,=\,\mbox{2--1}) line (only cycle~1). In both settings, the data set contains four 1.875\,GHz spectral windows with 3840 channels each. The baselines range from 13 to 16196\,m. This means a highest angular resolution of 0\farcs025, and a maximum recoverable scale of $\approx$\,8\arcsec . Bandpass calibration was performed on J1107-4449, and gain calibration on J1131-5818 (TA1) and J1132-5606 (TA2 and TA3). Flux calibration was done using Ceres and Titan (TA1), J1131-5818 (TA2), and J1150-5416 (TA3). Based on the calibrator fluxes, we estimate the absolute flux calibration to be accurate to within 5\%. However, the uncertainties in the reported flux densities are significantly larger than this. This is due to a combination of uncertainties introduced in the cleaning process and the difficulty in discriminating emission from the different components identified in the circumstellar medium of HD\,101584. For this reason we do not report any formal error estimates since they would not reflect the real uncertainties that we estimate are at least of the order 20\,\%.

The data were reduced using various versions of CASA over the years, the last one being 4.7.3. After corrections for the time and frequency dependence of the system temperatures, and rapid atmospheric variations at each antenna using water vapour radiometer data, bandpass and gain calibration were done. For the $^{12}$CO($J$\,=\,\mbox{2--1}) setting, data obtained in three different configurations were combined. Subsequently, for each individual tuning, self-calibration was performed on the strong continuum. Imaging was done using the CASA clean algorithm after a continuum subtraction was performed on the emission line data. The final line images were created using Briggs robust weighting. This resulted in close to circular beam sizes of about 0\farcs65$\times$0\farcs55 (8$^\circ$) and 0\farcs09$\times$0\farcs08 (12$^\circ$) for the cycle~1 and combined cycles~1 and 3 data, respectively. A beamsize of 0\farcs15$\times$0\farcs14 (10$^\circ$) is used for the presented continuum data. Typical channel rms noises are $\approx$\,2\,mJy\,beam$^{-1}$ and $\approx$\,0.7\,mJy\,beam$^{-1}$ for the cycle~1 and combined cycles~1 and 3 data at 1.5\,km\,s$^{-1}$ resolution, respectively.
%
%
%
%
%
\subsection{APEX}

Complementary molecular line and continuum data on HD\,101584 were obtained using APEX \citep{gustetal06}. The Swedish heterodyne facility instruments SHeFI \citep[A1,A2,A3;][]{vassetal08} and SEPIA \citep[B5,B9;][]{belietal18} were used together with the facility FFT spectrometer covering about 4\,GHz. The observations were made from August 2015 to August 2017 in dual-beamswitch mode with a beam throw of 2$\arcmin$. In May 2018 observations with the PI230 receiver and a 16\,GHz FFT spectrometer were used. Regular pointing checks were made on strong CO line emitters and continuum sources. Typically, the pointing was found to be consistent with the pointing model within 3$\arcsec$. The antenna temperature, $T_{\mathrm A}^{\star}$, is corrected for atmospheric attenuation. The uncertainty in the absolute intensity scale is estimated to be about $\pm 20$\%. APEX telescope characteristics [beam width ($\theta_{\rm b}$), main beam efficiency ($\eta_{\rm mb}$), and Jy to K conversion] at representative observing frequencies are given in Table~\ref{t:apex}. Low-order polynomial baselines were subtracted from the spectra.

Finally, we have used the ArTeMiS bolometer camera to measure the 350\,$\mu$m flux of HD\,101584. ArTeMiS is an ESO PI sub-mm camera arranged in 16$\times$18 sub-arrays operating at 200, 350, and 450\,$\mu$m \citep{reveetal14}. We observed for 3.5 hours on 26 Nov. 2016 at 350\,$\mu$m in spiral-raster-mapping mode under good weather conditions with precipitable water vapour in the range 0.4--0.6\,mm. The resulting image has an angular resolution of 8\arcsec , thus covering well the dust continuum emission region of HD\,101584. The data were reduced using the ArTeMiS data reduction package provided by the ArTeMis team. G305.80--0.24 (aka B13134) was observed as a flux calibrator, and the uncertainty of the flux calibration is estimated to be 30\%.

\begin{table}
\caption{APEX characteristics at representative observational frequencies.}
\centering
\begin{tabular}{l c l l}
\hline \hline
Frequency       & $\theta_{\rm b}$         & $\eta_{\rm mb}$  &   Jy/K  \\
$[$GHz]         & [\arcsec ]  \\
\hline 
170             & 37           &  0.75            & 38    \\
185             & 34           &  0.75            & 38    \\
200             & 31           &  0.75            & 38    \\
220             & 28           &  0.75            & 39    \\
260             & 24           &  0.75            & 39    \\
300             & 21           &  0.74            & 40    \\
345             & 18           &  0.73            & 41    \\
460             & 14           &  0.60            & 48    \\
690             & \phantom{0}9 &  0.46            & 63    \\
\hline
\end{tabular}
\label{t:apex}
\end{table}

\begin{table*}
\caption{Molecular lines observed towards HD\,101584.}
\centering
\begin{tabular}{l l c c l}
\hline \hline
Molecule           & Line                                         & Freq.     & $E_{\rm u}$\,$^1$  & Telescope  \\
                   &                                              & [GHz]     & [K]          &  \\
\hline 
CO                 & $J\,=\,2-1$                                  & 230.538   & \phantom{0}17           & ALMA\\
                   & $J\,=\,3-2$                                  & 345.796   & \phantom{0}33           & APEX\\
                   & $J\,=\,4-3$                                  & 461.041   & \phantom{0}55           & APEX  \\
                   & $J\,=\,6-5$                                  & 691.473   & 116                     & APEX  \\
$^{13}$CO          & $J\,=\,2-1$                                  & 220.399   & \phantom{0}16           & ALMA, APEX \\
                   & $J\,=\,3-2$                                  & 330.588   & \phantom{0}32           & APEX  \\
C$^{17}$O          & $J\,=\,2-1$                                  & 224.714   & \phantom{0}16           & APEX  \\
C$^{18}$O          & $J\,=\,2-1$                                  & 219.560   & \phantom{0}16           & ALMA, APEX    \\
$^{13}$C$^{17}$O   & $J\,=\,2-1$                                  & 214.574   & \phantom{0}15           & ALMA   \\
SiO                & $J\,=\,5-4$                                  & 217.105   & \phantom{0}31           & ALMA \\
$^{29}$SiO         & $J\,=\,5-4$                                  & 214.386   & \phantom{0}31           & ALMA \\
SiS                & $J\,=\,12-11$                                & 217.818   & \phantom{0}68           & ALMA \\
                   & $J\,=\,13-12$                                & 235.961   & \phantom{0}79           & ALMA  \\
CS                 & $J\,=\,4-3$                                  & 195.954   & \phantom{0}24           & APEX   \\
$^{13}$CS          & $J\,=\,5-4$                                  & 231.221   & \phantom{0}33           & ALMA   \\
SO                 & $N_J\,=\,5_5-4_4$                            & 215.221   & \phantom{0}44           & ALMA  \\
                   & $N_J\,=\,5_6-4_5$                            & 219.949   & \phantom{0}35           & ALMA  \\
                   & $N_J\,=\,6_5-5_4$                            & 251.826   & \phantom{0}51           & APEX \\
                   & $N_J\,=\,8_7-7_7$                            & 214.357   & \phantom{0}81           & ALMA \\
$^{33}$SO          & $N_J\,=\,5_6-4_5$                            & 217.831   & \phantom{0}35           & ALMA \\                   
$^{34}$SO          & $N_J\,=\,5_6-4_5$                            & 215.840   & \phantom{0}34           & ALMA \\
SO$_2$             & $J_{K_{\rm a},K_{\rm c}}\,=\,4_{22}-3_{13}$        & 235.152   & \phantom{00}9           & ALMA  \\
                   & $J_{K_{\rm a},K_{\rm c}}\,=\,16_{1,15}-15_{2,14}$  & 236.217   & 131           & ALMA \\
                   & $J_{K_{\rm a},K_{\rm c}}\,=\,16_{3,13}-16_{2,14}$  & 214.689   & 148          & ALMA  \\
                   & $J_{K_{\rm a},K_{\rm c}}\,=\,22_{2,20}-22_{1,21}$  & 216.643   & 249          & ALMA \\
                   & $J_{K_{\rm a},K_{\rm c}}\,=\,28_{3,25}-28_{2,26}$  & 234.187   & 403          & ALMA   \\
OCS                & $J\,=\,18-17$                                & 218.903   & 100          & ALMA \\                                         
			       & $J\,=\,19-18$                                & 231.061   & 111          & ALMA \\
HCN                & $J\,=\,3-2$                                  & 265.886   & \phantom{0}26           & APEX  \\ 
HCO$^+$            & $J\,=\,3-2$                                  & 267.558   & \phantom{0}26           & APEX  \\
$p$-H$_2$O\,$^2$   & $J_{K_{\rm a},K_{\rm c}}\,=\,3_{13}-2_{20}$        & 183.313   & 200          & APEX  \\
$p$-H$_2$S         & $J_{K_{\rm a},K_{\rm c}}\,=\,2_{20}-2_{11}$        & 216.710   & \phantom{0}84           & ALMA, APEX  \\
                   & $J_{K_{\rm a},K_{\rm c}}\,=\,2_{02}-1_{11}$        & 687.303   & \phantom{0}55           & APEX   \\
$o$-H$_2$S         & $J_{K_{\rm a},K_{\rm c}}\,=\,1_{10}-1_{01}$        & 168.763   & \phantom{0}28           & APEX  \\
                   & $J_{K_{\rm a},K_{\rm c}}\,=\,3_{30}-3_{21}$        & 300.506   & 169           & APEX   \\
$p$-H$_2^{33}$S    & $J_{K_{\rm a},K_{\rm c}}\,=\,2_{20}-2_{11}$        & 215.503   & \phantom{0}84          & ALMA   \\
                   & $J_{K_{\rm a},K_{\rm c}}\,=\,2_{02}-1_{11}$        & 687.164   & \phantom{0}55           & APEX   \\
$o$-H$_2^{33}$S    & $J_{K_{\rm a},K_{\rm c}}\,=\,1_{10}-1_{01}$        & 168.319   & \phantom{0}28           & APEX  \\
$p$-H$_2^{34}$S    & $J_{K_{\rm a},K_{\rm c}}\,=\,2_{20}-2_{11}$        & 214.377   & \phantom{0}84           & ALMA   \\
                   & $J_{K_{\rm a},K_{\rm c}}\,=\,2_{02}-1_{11}$        & 687.025   & \phantom{0}55           & APEX   \\
$o$-H$_2^{34}$S    & $J_{K_{\rm a},K_{\rm c}}\,=\,1_{10}-1_{01}$        & 167.911   & \phantom{0}28           & APEX   \\
$p$-H$_2$CO        & $J_{K_{\rm a},K_{\rm c}}\,=\,3_{03}-2_{02}$        & 218.222   & \phantom{0}21           & ALMA   \\
                   & $J_{K_{\rm a},K_{\rm c}}\,=\,3_{22}-2_{21}$        & 218.476   & \phantom{0}68           & ALMA           \\
                   & $J_{K_{\rm a},K_{\rm c}}\,=\,3_{21}-2_{20}$        & 218.760   & \phantom{0}68           & ALMA    \\
$o$-H$_2^{13}$CO   & $J_{K_{\rm a},K_{\rm c}}\,=\,3_{12}-2_{11}$        & 218.909   & \phantom{0}33           & ALMA      \\
$E$-CH$_3$OH       & $J_K\,=\,4_2-3_1$                                  & 218.440   & \phantom{0}45           & ALMA    \\
                   & $J_K\,=\,8_{-1}-7_0$                               & 229.759   & \phantom{0}89           & ALMA  \\
\hline
\end{tabular}
\label{t:obs_lines}
\tablefoot{(1) Energy of the upper energy level. (2) Only an upper limit is obtained.}
\end{table*}
%
%
\subsection{Observed lines}

The ALMA data cover the following frequency ranges: 214.02--215.90, 216.53--218.81, 219.13--221.01, 229.57--231.43, 231.63--236.70\,GHz. In these ranges we have identified 30 lines. Only one line remains unidentified. The APEX data cover selected lines, 20 of them [only upper limit for the H$_2$O(\mbox{$3_{13}-2_{20}$}) line], chosen to complement the ALMA data. Table~\ref{t:obs_lines} summarises all the identified lines. They are typical for an oxygen-rich circumstellar chemistry with a significant contribution of sulphur species, but also weaker lines from carbon-species, other than CO, are present \citep{mill16}. However, the presence of lines from H$_2$CO and CH$_3$OH point also to a non-standard circumstellar chemistry.
%
%
%
%
\subsection{Other data}

In our analysis, we will also make use of the CO and $^{13}$CO $J$\,=\,\mbox{1--0} and the CO $J$\,=\,\mbox{2--1} data obtained with the Swedish-ESO Submillimetre Telescope (SEST) and published by \citet{olofnyma99}. In addition, we have constructed a spectral energy distribution (SED) using archive data and our ALMA and APEX data.
%
%
%
%
\subsection{Missing flux in ALMA data}
\label{s:alma_flux}

\begin{figure}
\centering
   \includegraphics[width=8cm]{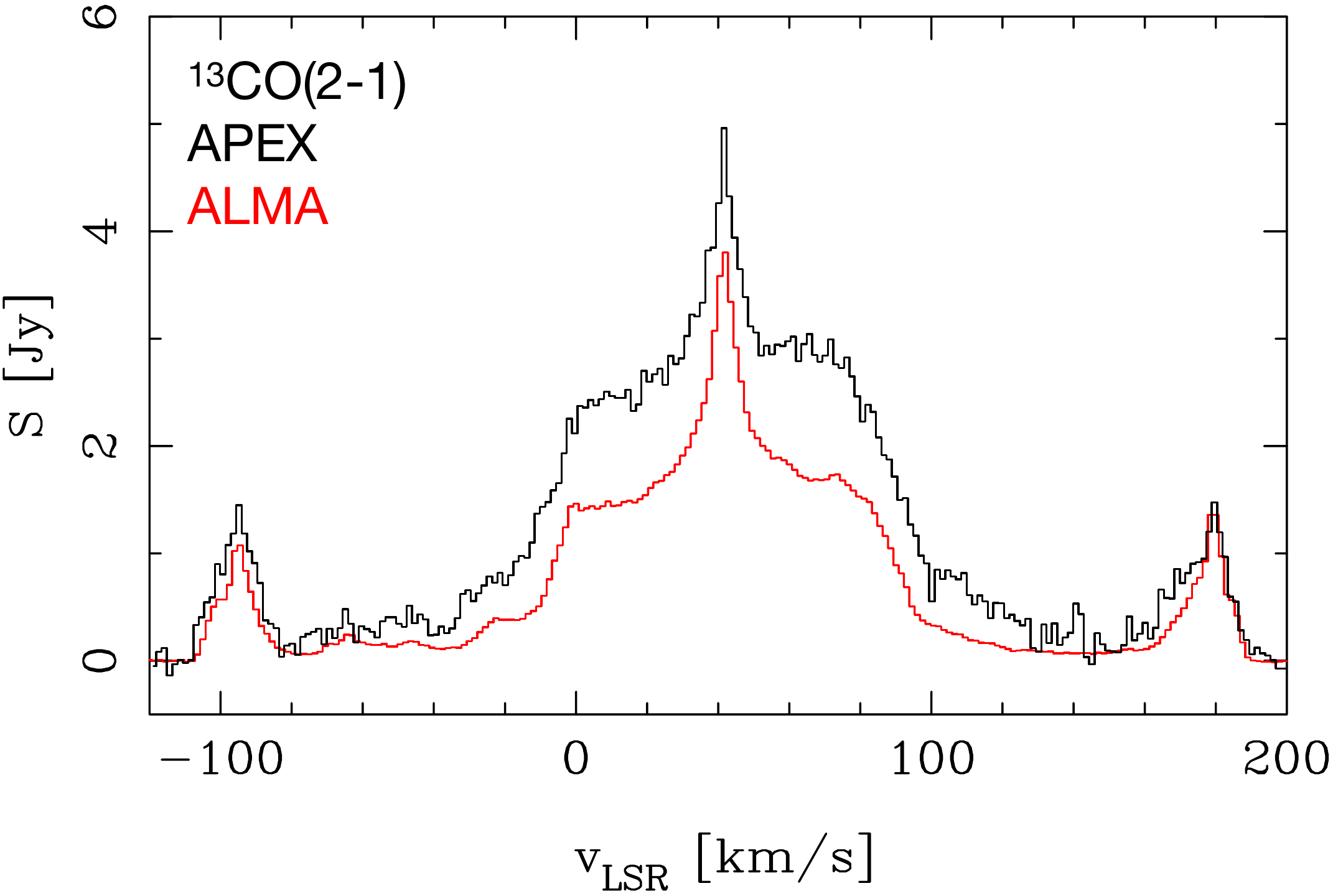}
    \caption{$^{13}$CO(\mbox{2--1}) spectra of HD\,101584 at 1.5\,km\,s$^{-1}$ resolution obtained with APEX (black histogram) and ALMA (red histogram). The ALMA spectrum is integrated over the whole source. 
    }
   \label{f:co_apex_global}
\end{figure}

The amount of missing flux in the ALMA data varies over the line profile as illustrated in Fig.~\ref{f:co_apex_global}, where we show the $^{13}$CO(\mbox{2--1}) lines obtained with APEX and ALMA (the latter is integrated over the source). We have chosen this line because it is not as optically thick as that of the main isotopologue, and both the ALMA and APEX observations have a high signal-to-noise ratio. 

The narrow central feature that stands on top of a broader plateau of emission has the same integrated intensity in the velocity range $\mid$\,$\upsilon-\upsilon_{\rm sys}$\,$\mid$\,$\le$\,10\,km\,s$^{-1}$ ($\upsilon_{\rm sys}$\,$\approx$\,41.7\,$\pm$\,0.2\,km\,s$^{-1}$; all velocities in this paper are with respect to the local standard of rest) in the APEX and ALMA data (with the plateau emission subtracted, and within the combined uncertainties). The ALMA to APEX line intensity ratio is 1.0$\pm$0.15. This indicates that the relative calibration between the data sets is as good as can be expected. At the extreme velocities, 130\,$\le$\,$\mid$\,$\upsilon-\upsilon_{\rm sys}$\,$\mid$\,$\le$\,150\,km\,s$^{-1}$ on either side of the systemic velocity, the amount of lost flux in the ALMA data is $\approx$\,30\,\%. Most of the flux in the ALMA data is lost at intermediate velocities. About 45\,\% is lost in the velocity ranges 20\,$\le$\,$\mid$\,$\upsilon-\upsilon_{\rm sys}$\,$\mid$\,$\le$\,130\,km\,s$^{-1}$ on either side of the systemic velocity. In terms of total flux in the $^{13}$CO(\mbox{2--1}) line, integrated over the full velocity range, the amount of flux lost in the ALMA data is $\approx$\,40\,\%. [Note, the beams of the APEX and ALMA antennae are the same so there is no need for a primary beam correction in this comparison.]

%
   \begin{figure*}
   \includegraphics[width=\textwidth]{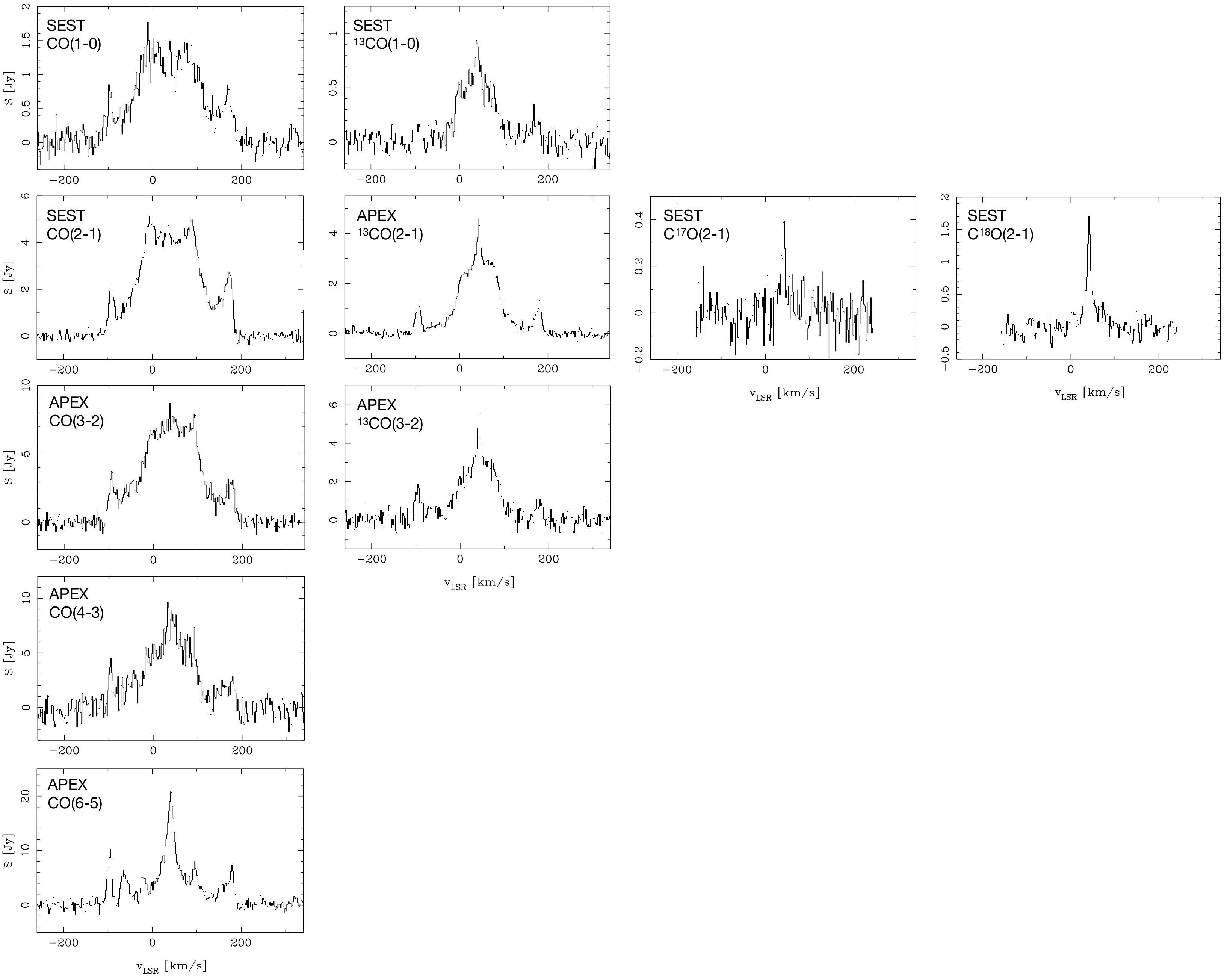}
      \caption{Single-dish CO isotopologue spectra of HD\,101584 at 2\,km\,s$^{-1}$ resolution. {\bf Left:} CO \mbox{1--0}, \mbox{2--1}, \mbox{3--2}, \mbox{4--3}, and \mbox{6-5} spectra from top to bottom. {\bf Middle left:} $^{13}$CO \mbox{1--0}, \mbox{2--1}, and \mbox{3--2} spectra from top to bottom. {\bf Middle right:} C$^{17}$O \mbox{2--1} spectrum. {\bf Right:} C$^{18}$O \mbox{2--1} spectrum. We note that the emission at the velocity extremes in the CO \mbox{4--3} and \mbox{6--5} lines my be suppressed by up to $\approx$\,20\,\% and 50\,\%, respectively, due to the relative sizes of the source and the beam (the beam is always pointed towards the centre of the source).}
   \label{f:co_spec}
   \end{figure*}   

   \begin{figure*}
   \centering
    \includegraphics[width=9cm]{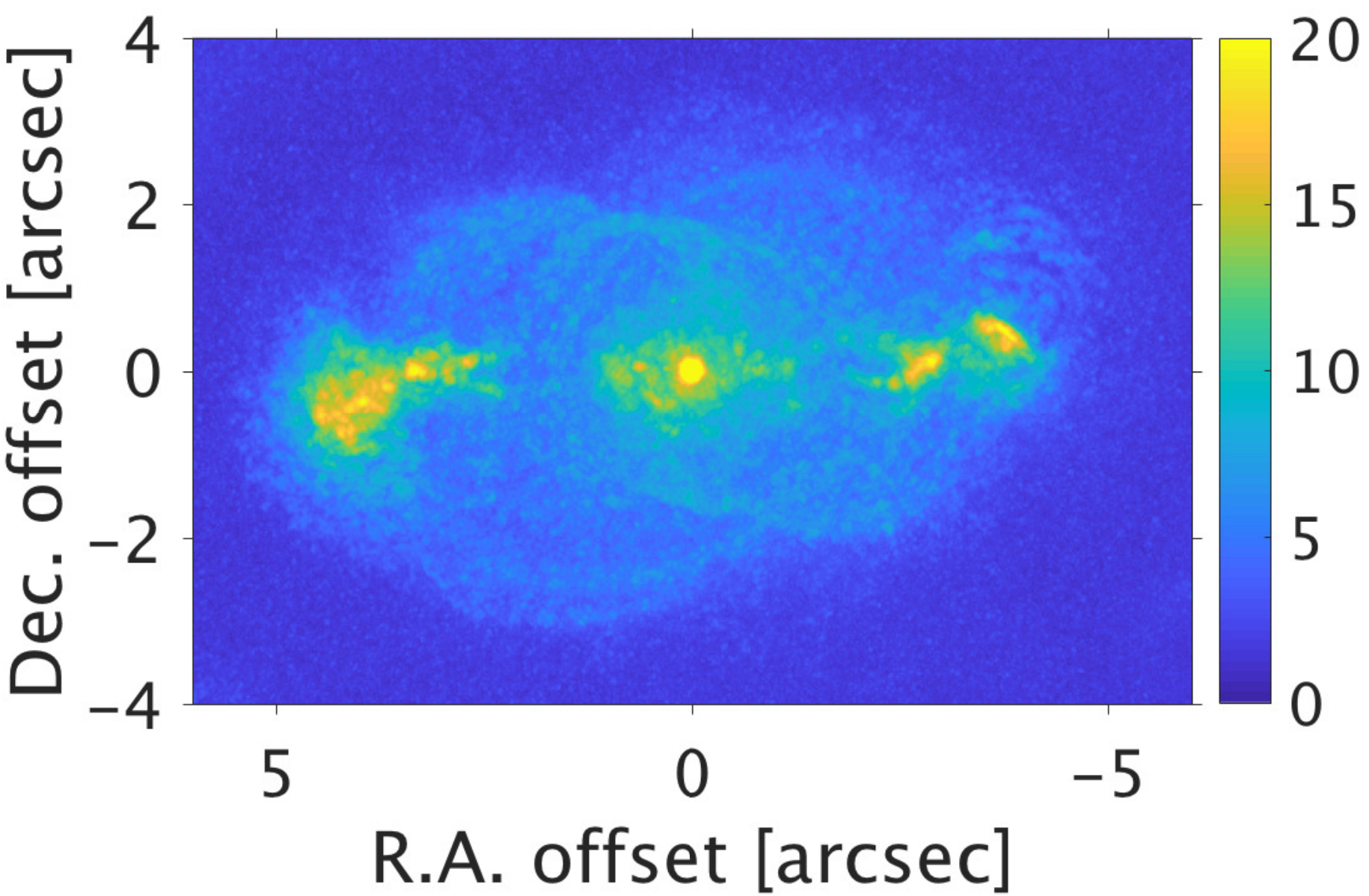} \hspace{5mm} \includegraphics[width=7.9cm]{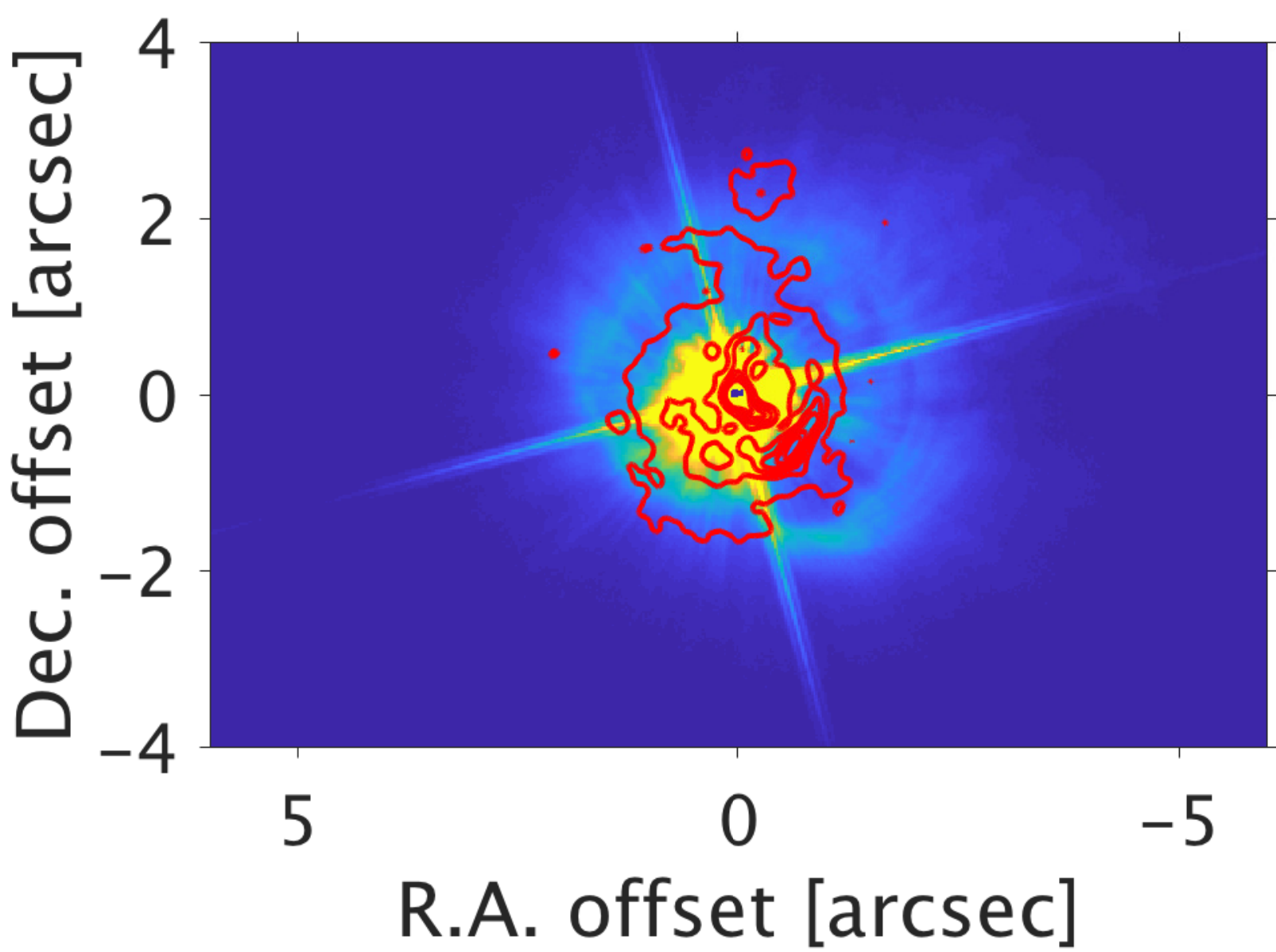}
      \caption{{\bf Left:} CO(\mbox{2--1}) maximum-intensity image at 0\farcs085 resolution. {\bf Right:} 1.3\,mm continuum image (red contours starting at 0.2\,mJy\,beam$^{-1}$ and with a spacing of 0.3\,mJy\,beam$^{-1}$) at 0\farcs15 resolution overlaid an F606W HST image from \citet{sahaetal07}. The latter has been shifted so that the diffraction cross coincides with the continuum peak. The CO line intensity peak at the centre coincides with the continuum peak within an uncertainty of about 0\farcs01.}
   \label{f:morphology}
   \end{figure*}   
%
%
%
%
\section{The circumstellar morphology and kinematics}

The complexity of the circumstellar environment of HD\,101584 is evident already in the single-dish rotational-line data of the CO isotopologues. Figure~\ref{f:co_spec} summarises data obtained with SEST and APEX. The differences in line shape can be attributed to the differences in optical depth of the lines and to some extent to the excitation (for example, the CO(\mbox{6--5}) line clearly samples warmer gas), and to the existence of a number of components with different kinematics and physical conditions. The main CO isotopologue line emission is dominated by an $\approx$\,100\,km\,s$^{-1}$ broad component centred at $\approx$\,40\,km\,s$^{-1}$. This component tapers gradually into two distinct features at the extreme velocities, $\approx$\,140\,km\,s$^{-1}$ offset on either side of the centre (particularly prominent in the CO and $^{13}$CO \mbox{2--1} line data). These features resemble the ``bullet'' emissions seen towards many high-velocity outflows of YSOs \citep{bachetal91a, bachetal91b}. The total emission covers $\approx$\,310\,km\,s$^{-1}$. On the contrary, the line emissions from the rarer CO isotopologues are dominated by a narrow central feature, for example, the C$^{18}$O(\mbox{2--1}) line is well fitted by a Gaussian having a full width at half maximum (FWHM) of 6.5\,km\,s$^{-1}$ centred at 41.5\,km\,s$^{-1}$. As judged by the CO isotopologue line intensity ratios, this feature has the highest optical depth in the CO lines. 
%
%
%
\subsection{The overall morphology}

The ALMA data adds considerable morphological information. An overview is presented in Fig.~\ref{f:morphology} that shows the circumstellar molecular gas, as traced by the CO(\mbox{2--1}) maximum line intensity, and dust, as traced by the 1.3\,mm continuum intensity, distributions around HD\,101584. Although, these are not simply interpreted in the forms of gas and dust density distributions they provide substantially more information than the visual image in scattered light presented in Fig.~\ref{f:morphology}, and serve as a base for our description of the circumstellar medium around HD\,101584. The difference in morphology between the line and continuum maps is primarily due to the much lower optical depths in the extended emission. Adding kinematical information allows a decomposition into separate components as described below.

   \begin{figure*}
   \centering
   \includegraphics[width=17cm]{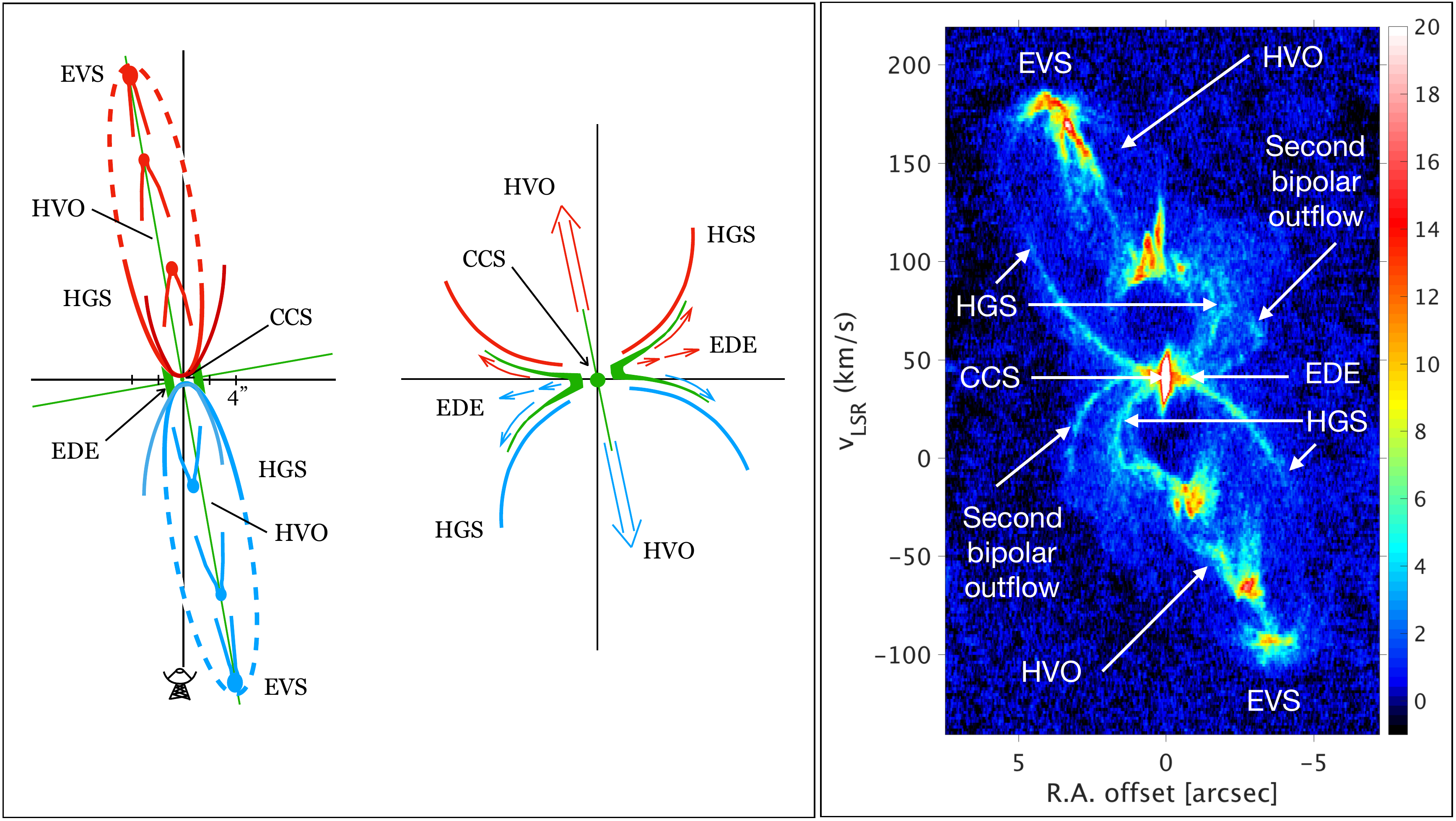}
      \caption{{\bf Left panel:} Sketch of the morphology of the circumstellar medium of HD\,101584 (not drawn to scale). The full extent and the identified morphological components: central compact source (CCS), equatorial density enhancement (EDE), hourglass structure (HGS) which forms the inner part of two diametrically orientated bubbles, and bipolar, high-velocity outflow (HVO) with extreme-velocity spots (EVSs), as well as a blow-up of the central region highlighting the CCS, EDE, and inner part of the HGS, are indicated. An hourglass structure suggesting a tentative second bipolar outflow, with a velocity gradient opposite to that of the HVO, is also shown.  {\bf Right panel:} CO(\mbox{2--1}) PV-diagram along PA\,=\,90$^\circ$ at resolutions of 0\farcs085 and 1.5\,km\,s$^{-1}$ with the different components indicated.  The flux scale is in mJy\,beam$^{-1}$.}
   \label{f:sketch_co21}
   \end{figure*}   

%
%
%
%
\subsection{Decomposition into different circumstellar components}
\label{s:decomp}

The combined morphological and kinematical information in the molecular line images make it possible to identify a number of distinct components in the circumstellar medium. The different components are exemplified below through selected molecular-line channel maps and position-velocity (PV) diagrams, specifically chosen because they highlight the different components that we will discuss and outline the reasons for their interpretations. We have identified the following components: 
\begin{itemize}
\item
CCS: A central compact source within a radius of $\approx$\,0\farcs1 of the centre, Sect.~\ref{s:obs_ccs}.
\item 
EDE: An equatorial density enhancement\footnote{The term `equatorial' is used here to reflect that we believe that this is a flattened density distribution, for example, a disk or a torus, that lies at the waist of the bipolar outflow and in the plane of the binary orbit.} of diameter $\approx$\,3\arcsec\ and centred on the CCS, Sect.~\ref{s:obs_ede}.
\item 
HVO: A bipolar, high-velocity outflow at PA\,$\approx$\,90$^\circ$, that is terminated in two extreme-velocity spots (EVSs) at $\approx$\,4\arcsec\ on each side of the CCS, Sects~\ref{s:obs_hvo} and \ref{s:obs_evs}.
\item 
HGS: An hourglass structure surrounding the initial $\approx$\,2\arcsec\ of the HVO, that develops into bubbles that close at the EVSs, Sect.~\ref{s:obs_hgs}.
\end{itemize}
A sketch of the proposed source structure and the nomenclature used in the remaining part of the paper is shown in Fig.~\ref{f:sketch_co21} (left panel), and the different components are also indicated in the CO(\mbox{2--1}) PV-diagram obtained along the major axis of the outflow (PA\,=\,90$^{\circ}$), Fig.~\ref{f:sketch_co21} (right panel).

\begin{figure*}
   \centering
    \includegraphics[width=15cm]{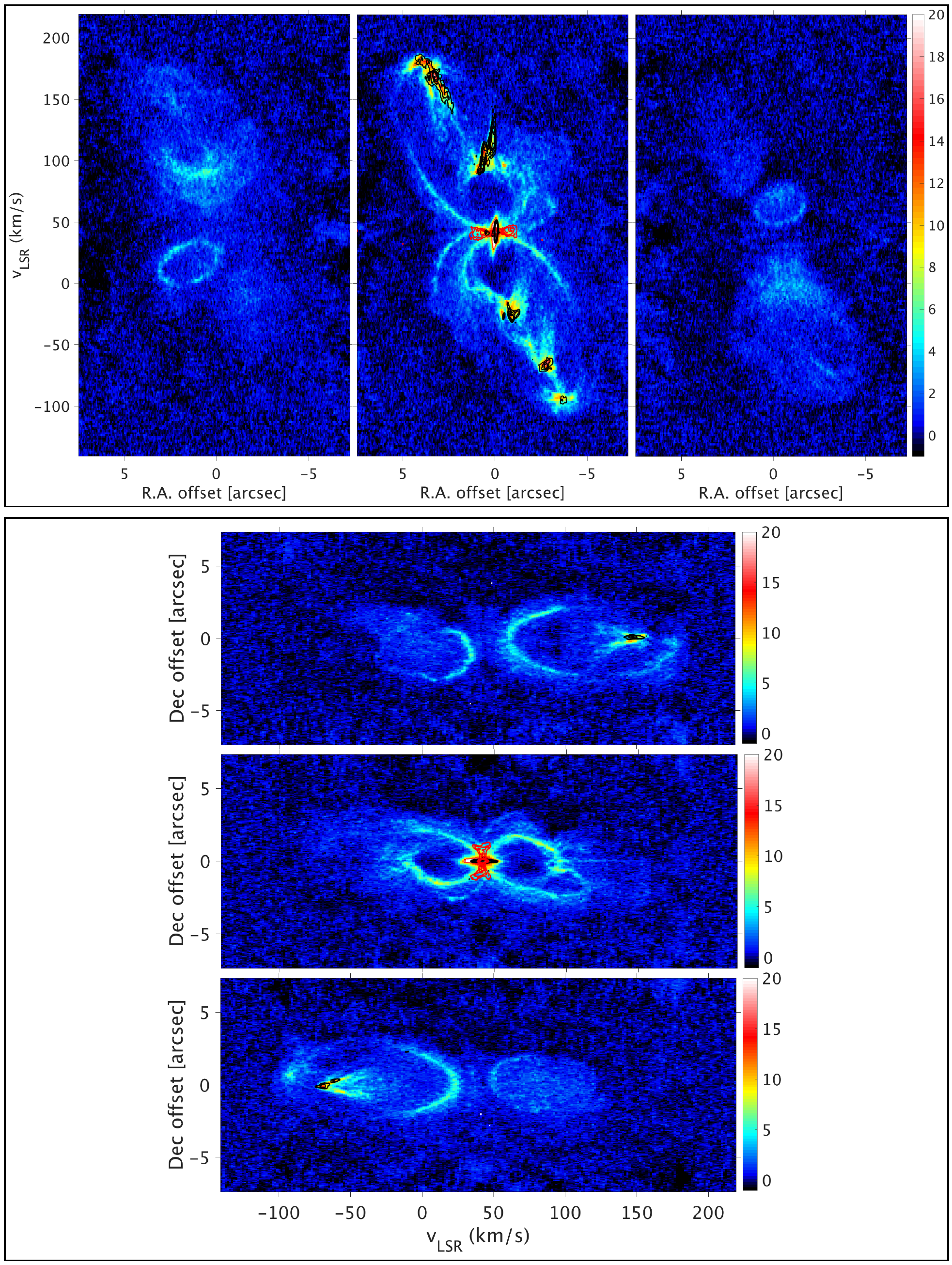}
         \caption{{\bf Upper panel:} PV-diagrams in the right ascension direction as seen in the CO(\mbox{2--1}) (colour), SiO(\mbox{5--4}) (black contours), and $p$-H$_2$S(\mbox{$2_{20}-2_{11}$}) (red contours) lines at declination offsets of 2\farcs5\,S (left), mid plane (middle), and 2\farcs5\,N (right). {\bf Lower panel:} The same in the declination direction at right ascension offsets of 2\farcs5\,E (top), mid plane (middle), and 2\farcs5\,W (bottom). The flux scale is in mJy\,beam$^{-1}$, and the contours start at 2\,mJy\,beam$^{-1}$ with a spacing of 2\,mJy\,beam$^{-1}$.}
   \label{f:pv_ra_de}
   \end{figure*}

The source structure, as well as how different lines probe different components, is further illustrated through six PV-diagrams of the CO(\mbox{2--1}), SiO(\mbox{5--4}), and $p$-H$_2$S(\mbox{$2_{20}-2_{11}$}) line emissions that cut through the circumstellar medium of HD\,101584. Figure~\ref{f:pv_ra_de} (upper panel) shows the morphology in the (R.A.,$\upsilon_{\rm z}$)\,-\,plane at three different declinations (2\farcs5\,S, mid plane, and 2\farcs5\,N). There are several noteworthy features here. In the mid plane the HVO stretches along the line of sight in the CO(\mbox{2--1}) and  SiO(\mbox{5--4}) lines, and it is clearly seen that the bright spots in the SiO line emission coincide with those of the CO line emission on either side of the source centre. The HGS is evident in the CO data, as well as the fact that it is the inner part of a bipolar bubble-like structure that closes at the EVSs. The $p$-H$_2$S line emission is confined to the region where the HGS closes towards the centre, interpreted by us as the EDE component. The CCS is seen in all three lines (it is particularly prominent in higher-excitation SO$_2$ lines as illustrated in Fig.~\ref{f:so2_channels}). The PV-diagrams N and S of the mid plane shows a bipolar bubble-like structure that is inclined with respect to the mid plane and has a velocity gradient opposite to that of the HVO. It will be further discussed in Sect.~\ref{s:multi_polar} in terms of a second bipolar outflow. In the lower panel of Fig.~\ref{f:pv_ra_de} we see the morphology as seen in the (Decl.,$\upsilon_{\rm z}$)\,-\,plane at three different right ascensions [2\farcs5\,E, mid plane (vertical to the mid plane of the upper panel), and 2\farcs5\,W]. This shows once again the HGS component developing into bubbles that close at the EVSs, and the presence of a second bipolar bubble structure with a reversed velocity gradient, in the CO line emission. The CO(\mbox{2--1})and SiO(\mbox{5--4}) channel maps are presented in Figs~\ref{f:co_channels} and \ref{f:sio_channels}.

%
%
%
%
\subsection{Continuum emission}
\label{s:obs_continuum}

The CCS and EDE components are particularly prominent in the 1.3\,mm continuum, but there is also weak emission from parts of the HGS component, and extended diffuse emission that contributes significantly to the total flux, Fig.~\ref{f:morphology}. The HVO component is not present in the continuum image. On the other hand, there is a low-surface-brightness region about 2\farcs5\,N of the centre that has no counterpart in the molecular gas. 

The central peak of the 1.3\,mm continuum has the coordinates $\alpha$(2000) = 11$^{\rm h}$40$^{\rm m}$58\fs7908 and $\delta$(2000) = --55$^\circ$34'25\farcs802 (with an error of 0\farcs004 in both directions). Within the uncertainties of both measurements this agrees with the Gaia position for HD101584 $\alpha$(2000) = 11$^{\rm h}$40$^{\rm m}$58\fs8052 and $\delta$(2000) = --55$^\circ$34'25\farcs813. Thus, we draw the reasonable conclusion that the continuum peaks at the position of HD\,101584. All position offsets in this paper refers to the continuum peak position.

We estimate that the 1.3\,mm continuum fluxes are 12\,mJy in the CCS (0\farcs3 aperture), 120\,mJy in the EDE (3\arcsec\ aperture, but excluding the CCS), and 70\,mJy outside the EDE (but only covering the region where structures in the emission are clearly seen). The total flux density is therefore 202\,mJy. However, there is substantial extended low-brightness emission for which the reliability is uncertain, for example, within a 10\arcsec\ aperture the total flux density is 245\,mJy. In addition, it is possible that flux is missing in our ALMA 1.3\,mm data due to extended emission. The ArTeMiS observations do not resolve the emission, and we can only report a total flux density of 9\,Jy (350\,$\mu$m, 8\arcsec\ beam). The flux estimates are summarised in Table~\ref{t:obs_cont}.

The 1.3\,mm continuum image is overlayed the F606W HST image in Fig.~\ref{f:morphology} assuming that the diffraction cross in the latter is the position of HD\,101584, which coincides with the position of the continuum peak (within the errors of the position estimates). Some similarities are noticeable, and the predominance of scattered light to the west is naturally explained by the fact that this is the side of the HVO facing towards us as shown by the molecular line data. The eastern side is exposed to higher circumstellar extinction.
%
%
%
%
\begin{table}
\caption{Continuum measurements}
\centering
\begin{tabular}{c l c c}
\hline \hline
Wavelength      & Instrument       & Aperture               &  $S$  \\
$[$$\mu$m]        &                & [\arcsec ]             &  [Jy] \\
\hline 
\phantom{1}350  & ArTeMiS          & \phantom{0}8\phantom{.00}              &  \phantom{10}9\phantom{.000}\\
1300            & ALMA             & total\,$^1$                  &  \phantom{10}0.20\phantom{0}\\
1300            & ALMA             & \phantom{0}3\phantom{.00}               &  \phantom{10}0.12\phantom{0} \\
1300            & ALMA             & \phantom{0}0.3\phantom{0}               &  \phantom{10}0.012\\
1300            & ALMA             & \phantom{0}0.05              &  \phantom{10}0.007\\
\hline
\end{tabular}
\label{t:obs_cont}
\tablefoot{(1) Inside the region where structure is seen in the continuum, that is, excluding extended very-low-brightness emission.}
\end{table}
%
%
%
%
\subsection{The central compact source, CCS}
\label{s:obs_ccs}

The CCS component is very compact in the 1.3\,mm continuum. About 60\,\% of the flux density within an aperture of 0\farcs3 comes from a region which is not resolved even when putting larger weight on the longest baselines: 7.0\,mJy resides in a 0\farcs027$\times$0\farcs026 (PA\,=\,100$^\circ$) source (FWHM of a Gaussian 2D fit) when observed with a 0\farcs025 beam. This means a Gaussian source size of $\la$\,0\farcs01 ($\la$\,10\,[$D$/1\,kpc]\,au), meaning that this component is circumbinary in nature (the binary separation being of the order 1\,mas). This is close in size to the mid-IR source studied by \citet{hilletal17}. At 10.7\,$\mu$m they measure a source size of 0\farcs028 (assuming a disk of uniform brightness) with a brightness temperature of $\approx$\,650\,K using VLTI/MIDI.

\begin{figure}
   \centering
   \includegraphics[width=8cm]{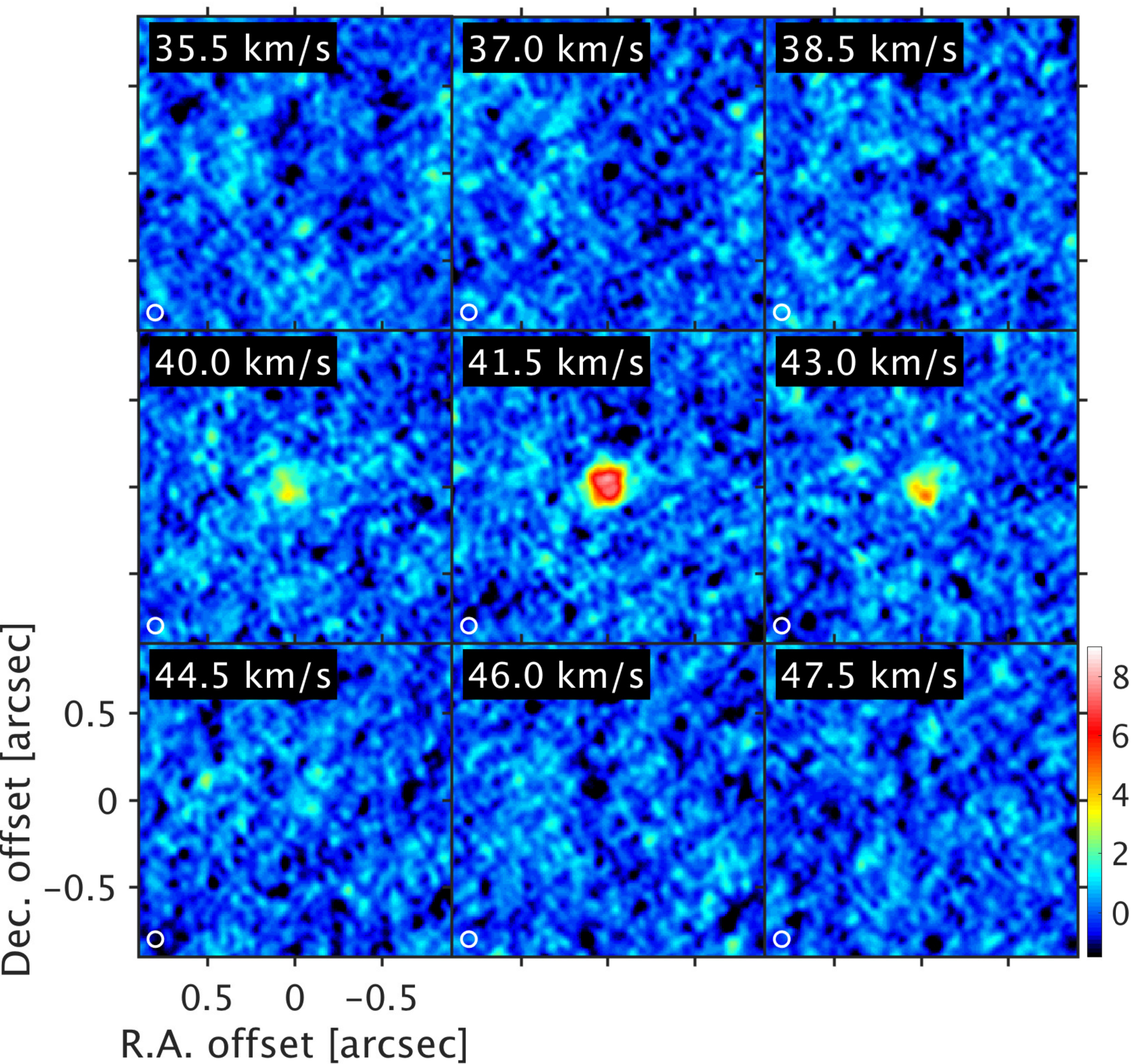}
      \caption{SO$_2$(\mbox{16$_{3,13}$--16$_{2,14}$}) channel maps with a width and spacing of 1.5\,km\,s$^{-1}$ at a resolution of 0\farcs085 (the beam is shown in the lower left corner of each panel). The flux scale is in mJy\,beam$^{-1}$. Narrow line emission from the CCS dominates here.}
   \label{f:so2_channels}
   \end{figure}

The CCS is seen in most of the molecular line emissions with emission peaks that coincide with the continuum peak. The results are summarised in Table~\ref{t:obs_ccs}. It is particularly dominating in the emission of the higher-energy SO$_2$ lines that originate only from this component. As an example the channel maps of the SO$_2$(\mbox{16$_{3,13}$--16$_{2,14}$}) line, that has an upper energy level at 148\,K, is shown in Fig.~\ref{f:so2_channels}. All the detected line emissions are resolved with deconvolved source sizes (FWHMs of 2D Gaussian fits) of $\approx$0\farcs15 ($\approx$\,150\,[$D$/1\,kpc]\,au). This is substantially larger than the compact continuum emission which probably arises in the inner, warmer region of the CCS. Notable exceptions are the SiO and $^{29}$SiO \mbox{5--4} line emissions that are significantly smaller (still larger than the continuum source), perhaps indicating that they probe preferentially a warmer region closer to the centre. 

\begin{figure}
   \centering
   \includegraphics[width=7cm]{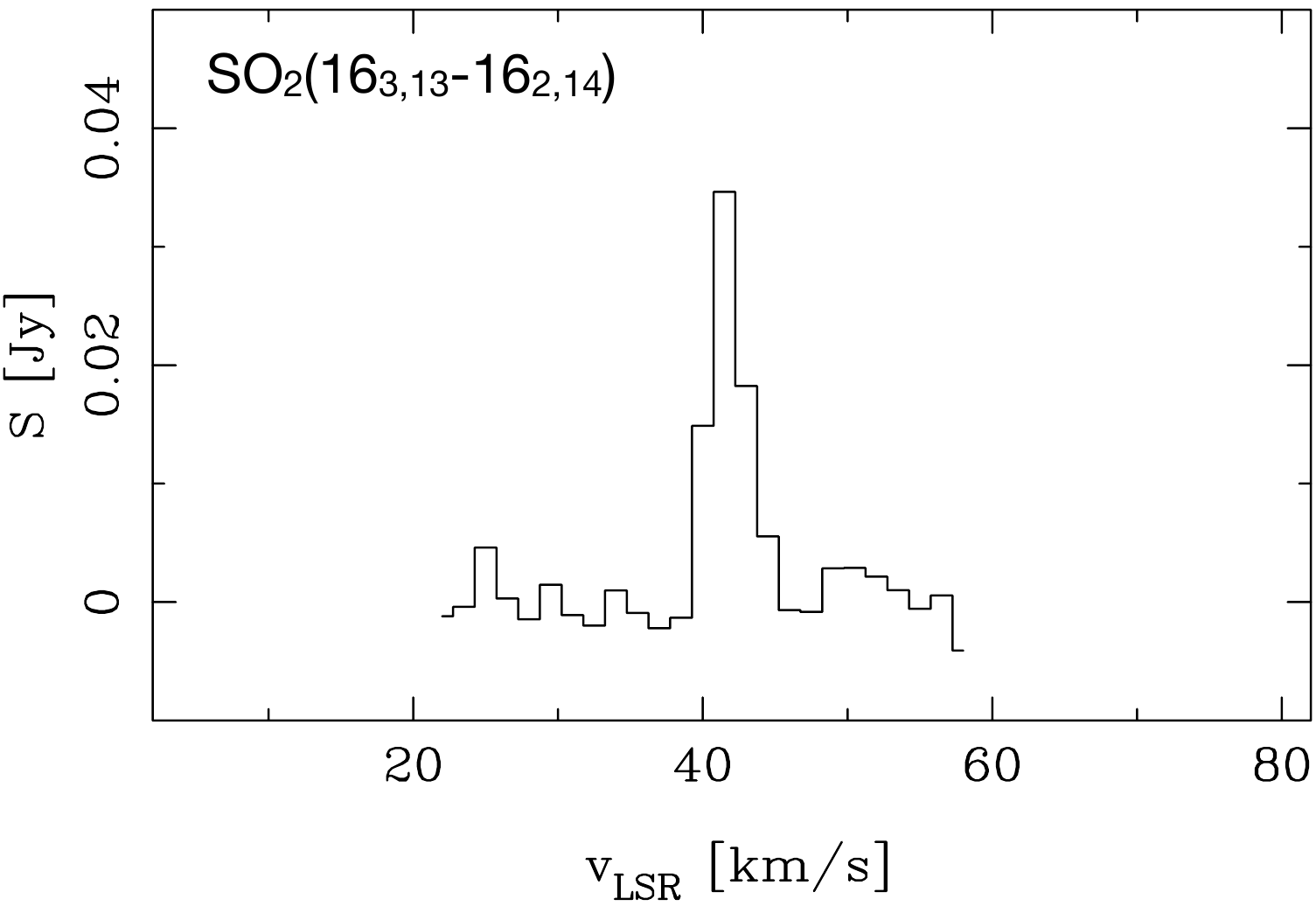}
      \caption{SO$_2$(\mbox{16$_{3,13}$--16$_{2,14}$}) line, obtained with ALMA within an aperture of 0\farcs3 centred on the continuum peak, at a resolution of 1.5\,km\,s$^{-1}$. This line is characteristic of the narrow-line-width emission from the CCS component.}
   \label{f:so2_spec_ccs}
   \end{figure}

Except for the CO, $^{13}$CO, SiO, and $^{29}$SiO lines, all lines are narrow, the average FWHM\,=\,3.0\,km\,s$^{-1}$, corresponds to a deconvolved FWHM\,=\,2.6\,km\,s$^{-1}$ at a spectral resolution of 1.5\,km\,s$^{-1}$. The SO$_2$(\mbox{$16_{3,13}-16_{2,14}$}) line is shown as an example in Fig.~\ref{f:so2_spec_ccs}. It is not clear why the CO and SiO lines are significantly broader, but confusion with emission along the line of sight could be part of an explanation (certainly in the case of CO).

The $^{13}$C$^{17}$O(\mbox{2--1}), $^{13}$CS(\mbox{5--4}), and SiS(\mbox{12--11}, \mbox{13--12}) lines all have the velocity characteristics of the CCS, but they are not peaked at the centre. Instead they show a patchy and extended structure over $\approx$\,0\farcs5. This may be an effect of these lines being among the weaker ones, S/N\,$\approx$\,5 integrated over the area. 

\begin{table*}
\caption{Molecular line emission from the CCS component\,$^1$.}
\centering
\begin{tabular}{l c c c c c }
\hline \hline
Line                                     & Aperture\,$^2$ & $S$              & \phantom{0}$\Delta \upsilon$\,$^3$ & $I$ & Size\,$^4$  \\
                                         & [\arcsec ]               & [Jy]             & [km\,s$^{-1}$]  & [Jy\,km\,s$^{-1}$] &  \\
\hline 
CO(2--1)                                 & 0.2       & 0.15\phantom{0}  & 18.7            &  3.0\phantom{00} & 0\farcs15 \\
$^{13}$CO(2--1)                          & 0.6       & 0.35\phantom{0}  & \phantom{1}6.2  &  2.3\phantom{00} & \ldots \\
C$^{18}$O(2--1)                          & 0.6       & 0.20\phantom{0}  & \phantom{1}2.9  &  0.61\phantom{0} & \ldots \\
$^{13}$C$^{17}$O(2--1)\,$^5$             & 0.6       & 0.015            & \phantom{1}3.7  &  0.059 \\
SiO(5--4)                                & 0.2       & 0.046            & 11.9            &  0.58\phantom{0} &   0\farcs10 \\
$^{29}$SiO(5--4)                         & 0.2       & 0.013            & 10.0            &  0.14\phantom{0} & 0\farcs11\\
SiS(12--11)\,$^5$                        & 0.6       & 0.011            & \phantom{1}3.3  &  0.038 & \ldots \\
SiS(13--12)\,$^5$                        & 0.6       & 0.010            & \phantom{1}4.2  &  0.045 & \ldots \\
$^{13}$CS(5--4)\,$^5$                    & 0.6       & 0.027            & \phantom{1}2.5  &  0.072 & \ldots \\
SO($5_5-4_4$)                            & 0.2       & 0.073            & \phantom{1}4.4  &  0.34\phantom{0} & 0\farcs15\\
SO($5_6-4_5$)                            & 0.6       & 0.10\phantom{0}  & \phantom{1}3.7  &  0.41\phantom{0} & \ldots \\
SO($6_5-5_4$)\,$^6$                      & 25\phantom{000} & 0.22\phantom{0}  & \phantom{1}5.1  &  1.2\phantom{00} & \ldots \\
SO($8_7-7_7$)                            & 0.2       & 0.011            & \phantom{1}2.5  &  0.029 & \ldots \\
$^{33}$SO(5$_6$--4$_5$)\,$^7$            & 0.6       & \ldots              & \ldots             &  0.019 & \ldots \\
$^{34}$SO(5$_6$--4$_5$)                  & 0.2       & 0.025            & \phantom{1}2.7  &  0.072 & 0\farcs19\\
SO$_2$(4$_{22}$--3$_{13}$)               & 0.6       & 0.043            & \phantom{1}2.3  &  0.11\phantom{0} & \ldots \\
SO$_2$(16$_{1,15}$--15$_{2,14}$)         & 0.6       & 0.043            & \phantom{1}2.2  &  0.10\phantom{0} & \ldots \\
SO$_2$(16$_{3,13}$--16$_{2,14}$)         & 0.2       & 0.025            & \phantom{1}2.6  &  0.069 & 0\farcs17 \\
SO$_2$(22$_{2,20}$--22$_{1,21}$)         & 0.2       & 0.015            & \phantom{1}3.2  &  0.051 & 0\farcs15\\
SO$_2$(28$_{3,25}$--28$_{2,26}$)         & 0.6       & 0.019            & \phantom{1}1.8  &  0.036 & \ldots \\
OCS(19--18)                              & 0.2       & 0.023            & \phantom{1}3.2  &  0.078 & 0\farcs14\\
$p$-H$_2$S(2$_{20}$--2$_{11}$)             & 0.2     & 0.084            & \phantom{1}3.6  &  0.32\phantom{0} & 0\farcs15 \\
$p$-H$_2^{33}$S(2$_{20}$--2$_{11}$)\,$^7$  & 0.2     & \ldots              & \ldots            &  0.15\phantom{0} & \ldots \\
$p$-H$_2^{34}$S(2$_{20}$--2$_{11}$)        & 0.2     & 0.060            & \phantom{1}2.9  &  0.18\phantom{0} & 0\farcs26\\
\hline
\end{tabular}
\label{t:obs_ccs}
\tablefoot{(1) See Sect.~\ref{s:obs_desc} for a discussion of the flux uncertainties. (2) The choice of aperture reflects the angular resolution of the ALMA data: the resolutions are $\approx$\,0\farcs085 and $\approx$\,0\farcs55 at 0\farcs2 and 0\farcs6 aperture, respectively. The apertures are centred on the continuum peak. (3) FWHM of Gaussian fit to the line within the given aperture at 1.5\,km\,s$^{-1}$ resolution. (4) Mean of the deconvolved FWHMs of a 2D Gaussian fit to the high-angular-resolution ALMA data. (5) These emissions are patchy and extended over a region of $\approx$\,0\farcs5. (6) This is based on APEX data and the split into emission from the CCS and EDE components obtained using Gaussian decomposition is uncertain. (7) The integrated intensity of the sum of the hyperfine components.}
\end{table*}

%
%
%
%
\subsection{The equatorial density enhancement, EDE}
\label{s:obs_ede}

\begin{figure*}
  \centering
  \includegraphics[width=14cm]{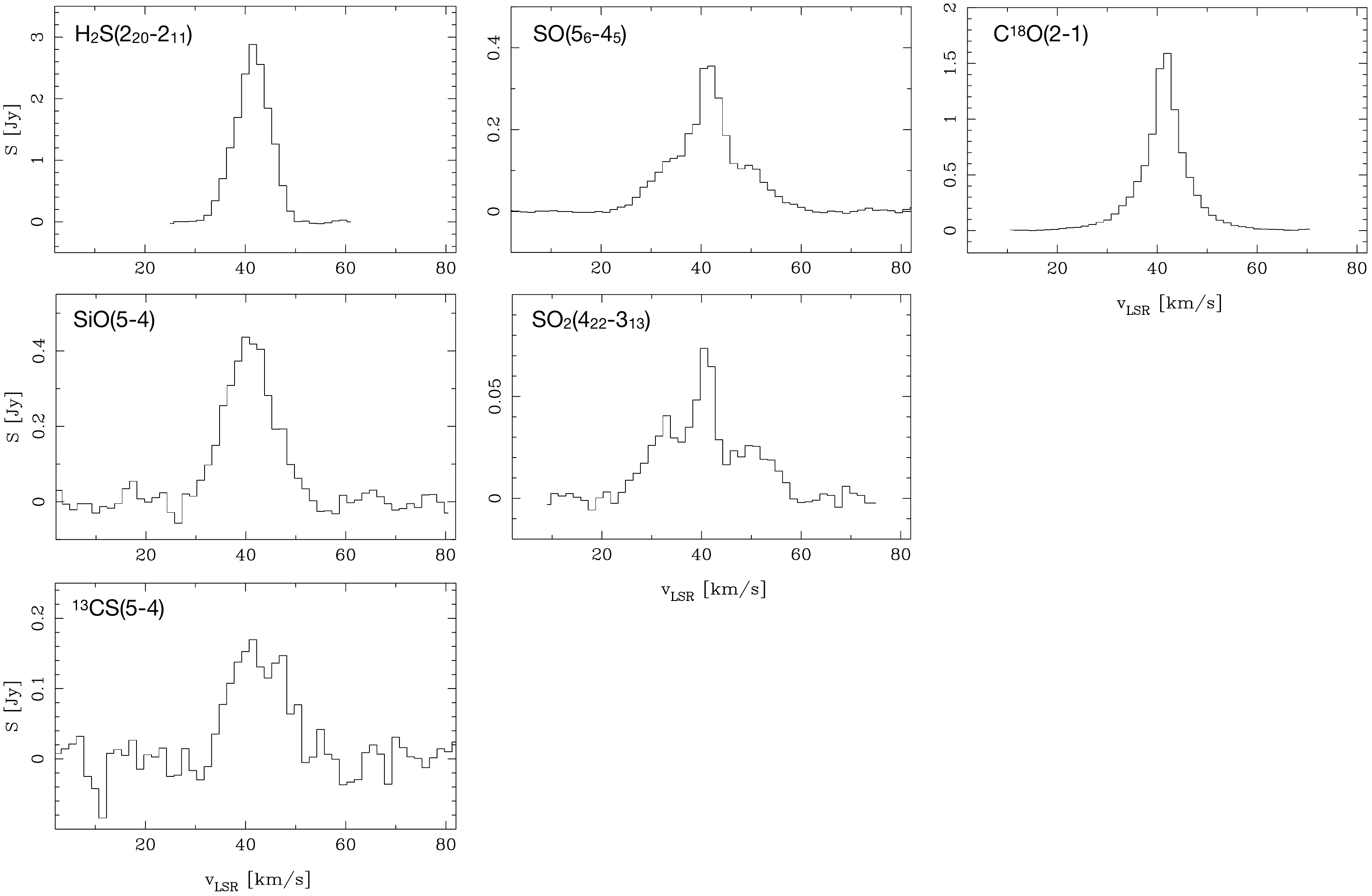}
      \caption{Three types of line profiles seen towards the EDE. All spectra are obtained with ALMA within an aperture of 3\farcs0 and with 1.5\,km\,s$^{-1}$ resolution. {\bf Left:} $p$-H$_2$S(\mbox{2$_{20}$--2$_{11}$}), SiO(\mbox{5--4}), and $^{13}$CS(\mbox{5--4}) lines from top to bottom. {\bf Middle:} SO(\mbox{5$_6$--4$_5$}) and SO$_2$(\mbox{$4_{22}-3_{13}$}) lines from top to middle. {\bf Right:} C$^{18}$O(\mbox{2--1}) line. }
   \label{f:ede_spectra}
   \end{figure*}

The EDE lies at the waist of the HGS and the HVO. It is the only component of the circumstellar medium around HD\,101584 that is not particularly prominent in the CO(\mbox{2--1}) line, nor in any of the other CO isotopologue lines. This is most likely due to contamination by emission from the HGS, a component only seen in the CO lines (see below). On the contrary, the EDE component is particularly prominent in the $p$-H$_2$S(\mbox{2$_{20}$--2$_{11}$}) line emission, and we will infer most of its characteristics using the emission of this line. In fact, the global $p$-H$_2$S(\mbox{2$_{20}$--2$_{11}$}) line is third in peak strength, only the global CO and $^{13}$CO \mbox{2--1} lines are stronger in our ALMA data. The EDE component is not easily seen in the map data of any of the other line emissions. We have therefore identified the line emission from this component using line profiles obtained within a central 3\arcsec\ aperture. Three different types of line profiles can be identified, as exemplified in Fig.~\ref{f:ede_spectra}. The $p$-H$_2$S(\mbox{2$_{20}$--2$_{11}$}), $^{13}$CS(\mbox{5--4}) and SiO(\mbox{5--4}) lines are close to triangular with well-defined full widths at zero power (FWZP) of 15--20\,km\,s$^{-1}$. The SO(\mbox{$5_6-5_4$}) and SO$_2$(\mbox{$4_{22}-3_{13}$}) lines show a narrow feature (largely due to the CCS emission) centred on a plateau with well-defined FWZPs of $\approx$\,40\,km\,s$^{-1}$. Finally, the C$^{18}$O(\mbox{2--1}) line is relatively narrow at the centre but have extended wings with no well-defined FWZP. The results for the different line emissions are summarised in Table~\ref{t:obs_ede}. 

\begin{table*}
\caption{Molecular line emission from the EDE\,$^1$}
\centering
\begin{tabular}{l c c c l}
\hline \hline
Line                                    & Aperture\,$^2$ & $S$               & FWZP\,$^3$       & \phantom{000}$I$   \\
                                        & [\arcsec ]    & [Jy]              & [km\,s$^{-1}$]   & [Jy\,km\,s$^{-1}$] \\
\hline 
C$^{18}$O(2--1)\,$^4$                   & \phantom{0}3        & 1.4\phantom{11}   &  \ldots             &  \phantom{0}12 \\
SiO(5--4)                               & \phantom{0}3        & 0.38\phantom{1}   &            22.3  &  \phantom{00}4.5\\
CS(4--3)                                & 32        & 0.35\phantom{1}   &            12.6  &  \phantom{00}5.0 \\
$^{13}$CS(5--4)                         & \phantom{0}3        & 0.11\phantom{1}   &  18.0  &  \phantom{00}1.9 \\                        
SO($5_5-4_4$)                           & \phantom{0}3        & 0.30\phantom{1}   &  30.2  &  \phantom{00}5.4\\
SO($5_6-4_5$)                           & \phantom{0}3        & 0.18\phantom{1}   &  31.4  &  \phantom{00}4.0\\
SO($6_5-5_4$)\,$^5$                     & 24        & 0.08\phantom{1}   &            20.0  &  \phantom{00}1.7\\
$^{34}$SO($5_6-4_5$)                    & \phantom{0}3        & 0.08:\phantom{:}  &  28:\phantom{0}   & \phantom{00}1.5\\
SO$_2$($4_{22}-3_{13}$)                 & \phantom{0}3        & 0.035             &  35:\phantom{0}   &  \phantom{00}0.81\\
$o$-H$_2$S($1_{10}-1_{01}$)             & 37        & 1.9\phantom{11}   &            15.4  &  \phantom{0}19 \\
$p$-H$_2$S($2_{20}-2_{11}$)             & \phantom{0}3        & 2.8\phantom{11}   &  17.9  &  \phantom{0}23 \\
                                        & 28        & 2.4\phantom{11}   &            15.3  &  \phantom{0}17 \\
$o$-H$_2$S($3_{30}-3_{21}$)             & 21        & 2.8\phantom{11}   &            12.7  &  \phantom{0}17 \\
$p$-H$_2$S($2_{02}-1_{11}$)             & \phantom{0}9        & 5:\phantom{111}   &   6:   &  \phantom{0}25: \\
$o$-H$_2^{33}$S($1_{10}-1_{01}$)\,$^6$  & 37                  & \ldots            &   \ldots               &  \phantom{0}12 \\
$p$-H$_2^{33}$S($2_{20}-2_{11}$)\,$^6$  & \phantom{0}3        & \ldots            &   \ldots              &  \phantom{00}3.4 \\
$p$-H$_2^{33}$S($2_{02}-1_{11}$)\,$^6$  & \phantom{0}9        & \ldots            &   \ldots               &  \phantom{0}24: \\
$o$-H$_2^{34}$S($1_{10}-1_{01}$)        & 37                  & 1.6\phantom{11}   &  16.6  &  \phantom{0}12 \\
$p$-H$_2^{34}$S($2_{20}-2_{11}$)        & \phantom{0}3        & 1.2\phantom{11}   &  15.0  &  \phantom{00}6.7 \\
$p$-H$_2^{34}$S($2_{02}-1_{11}$)        & \phantom{0}9        & 5:\phantom{111}   &   6: &  \phantom{0}12 \\
\hline
\end{tabular}
\label{t:obs_ede}
\tablefoot{(1) See Sect.~\ref{s:obs_desc} for a discussion of the flux uncertainties. (2) Centred on the continuum peak. (3) FWZP of the line within the given aperture at 1.5\,km\,s$^{-1}$ resolution. (4) Data obtained within a velocity range of 20\,km\,s$^{-1}$ centred on the systemic velocity. (5) This is based on APEX data and the split into emission from the CCS and EDE components obtained using Gaussian decomposition is uncertain. (6) The integrated intensity of the sum of the hyperfine components.}
\end{table*}

\begin{figure*}
   \centering
   \includegraphics[width=14cm]{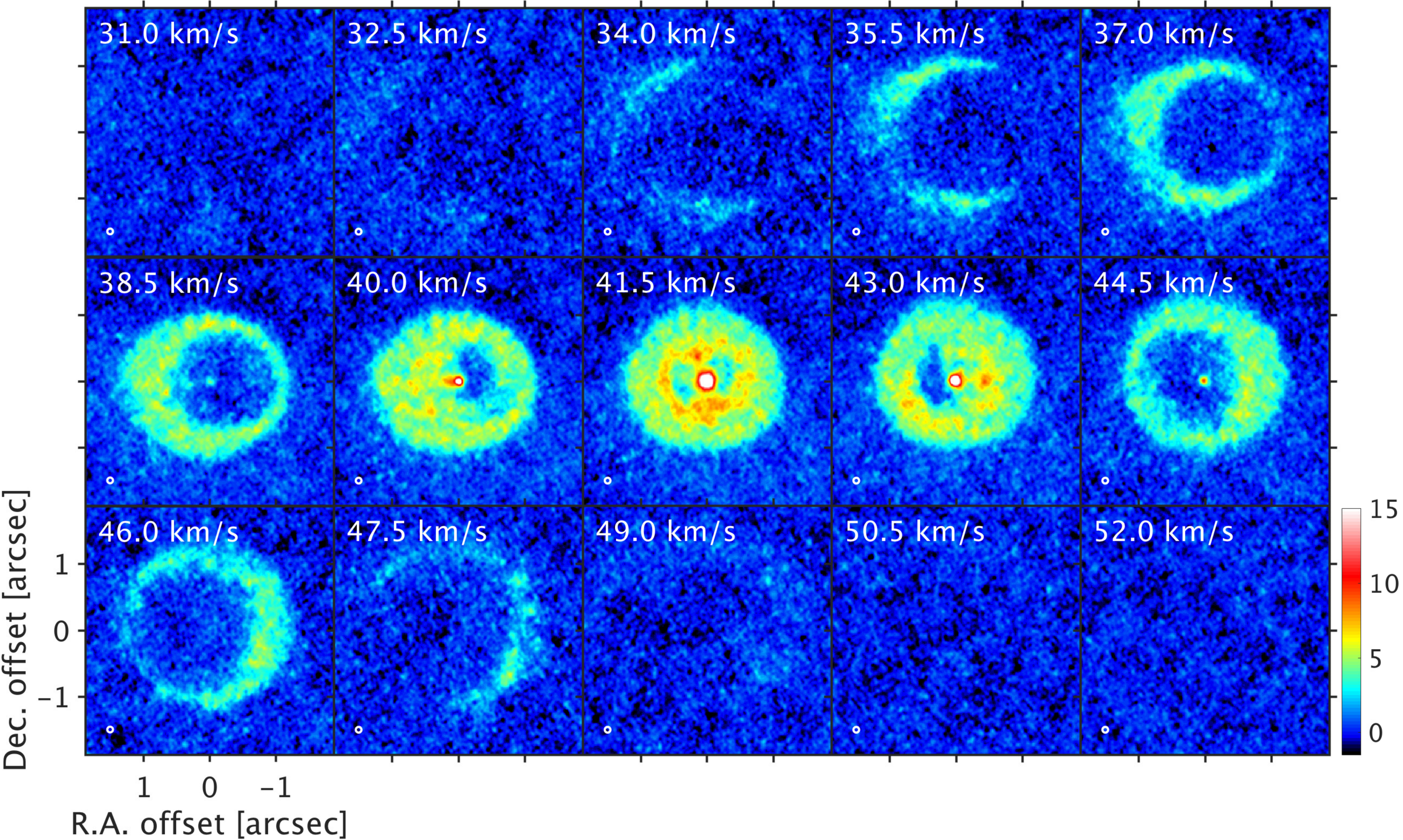}
      \caption{$p$-H$_2$S(\mbox{2$_{20}$--2$_{11}$}) channel maps with a width and spacing of 1.5\,km\,s$^{-1}$ at a resolution of 0\farcs085 (he beam is shown in the lower left corner of each panel). The flux scale is in mJy\,beam$^{-1}$. Emission from the EDE dominates for this line, although emission from the CCS is present at the centre.}
   \label{f:h2s_channels}
   \end{figure*}

\begin{figure}
   \centering
   \includegraphics[width=8cm]{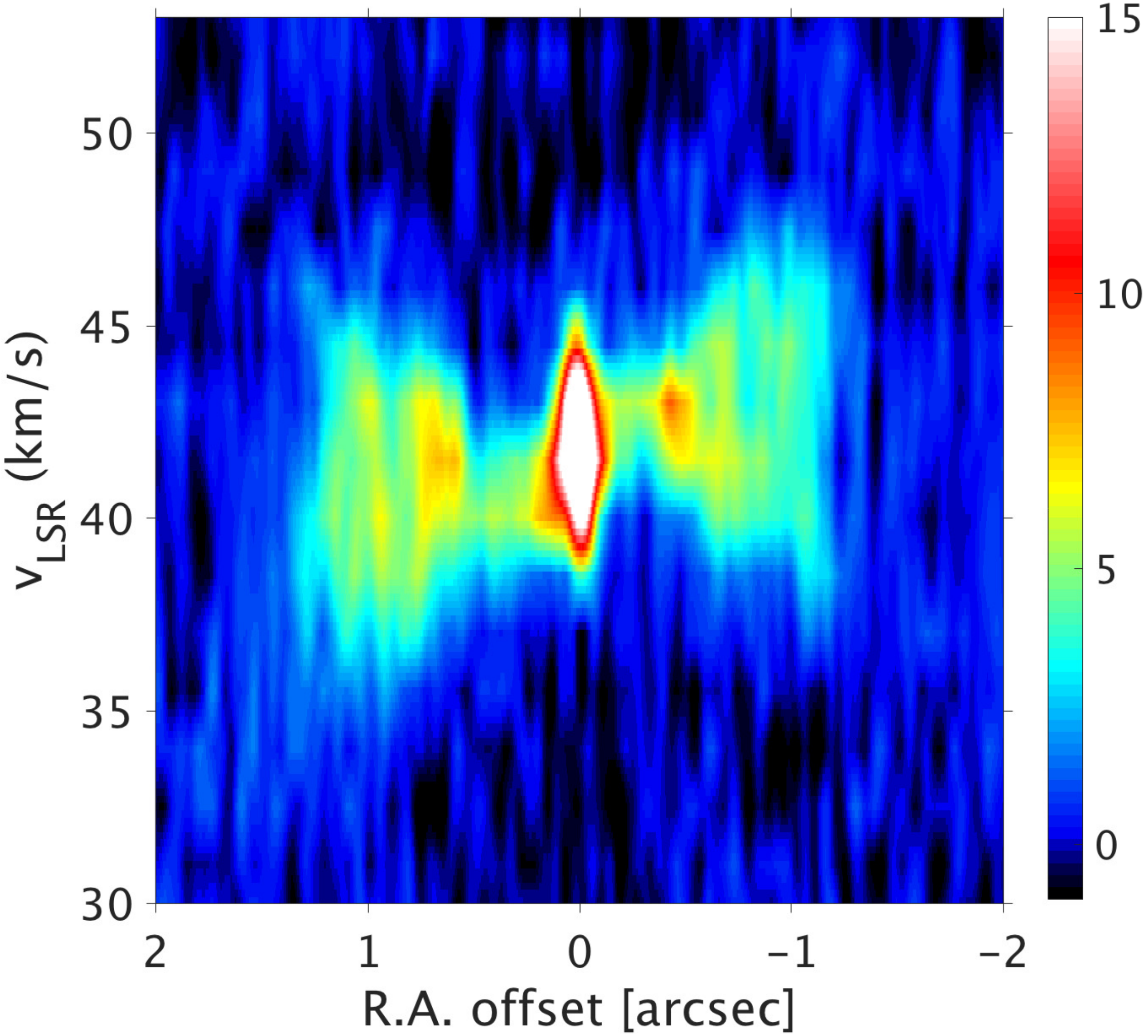}
      \caption{$p$-H$_2$S(\mbox{2$_{20}$--2$_{11}$}) PV-diagram along PA\,=\,90$^\circ$ at resolutions of 0\farcs085 and 1.5\,km\,s$^{-1}$. The flux scale is in mJy\,beam$^{-1}$. }
   \label{f:h2s_posvel}
   \end{figure}

The $p$-H$_2$S(\mbox{2$_{20}$--2$_{11}$}) brightness distribution is dominated by emission from an essentially circular structure of size 2\farcs3$\times$2\farcs1 (PA\,$\approx$\,90$^\circ$) centred on the CCS, Fig.~\ref{f:h2s_channels}.  The emission is very sharply truncated at the edge and limb-brightened. Notably, the size of the emitting region is independent of the line-of-sight velocity, and both blue- and redshifted emission is seen on either side of the centre channel map. As can be seen both in the channel maps and the PV-diagram, Fig.~\ref{f:h2s_posvel}, the velocity gradient over the EDE is opposite to that of the HVO (compare right panel of Fig.~\ref{f:sketch_co21}, and further discussed in Sect.~\ref{s:obs_hvo}). The most straightforward interpretation is that the EDE has an expanding, flattened (possibly flared) density distribution that is oriented orthogonal to the HVO, that is, it is seen almost face-on. To estimate the expansion velocity and its dependence on the distance to the centre is difficult due to the essentially face-on orientation and unknown density distribution. It could be a disk or a torus. The dust modelling as described in Sect~\ref{s:sed_results} suggests a disk, while the molecular line data appear more consistent with a torus.

\begin{figure}
   \centering
   \includegraphics[width=8.5cm]{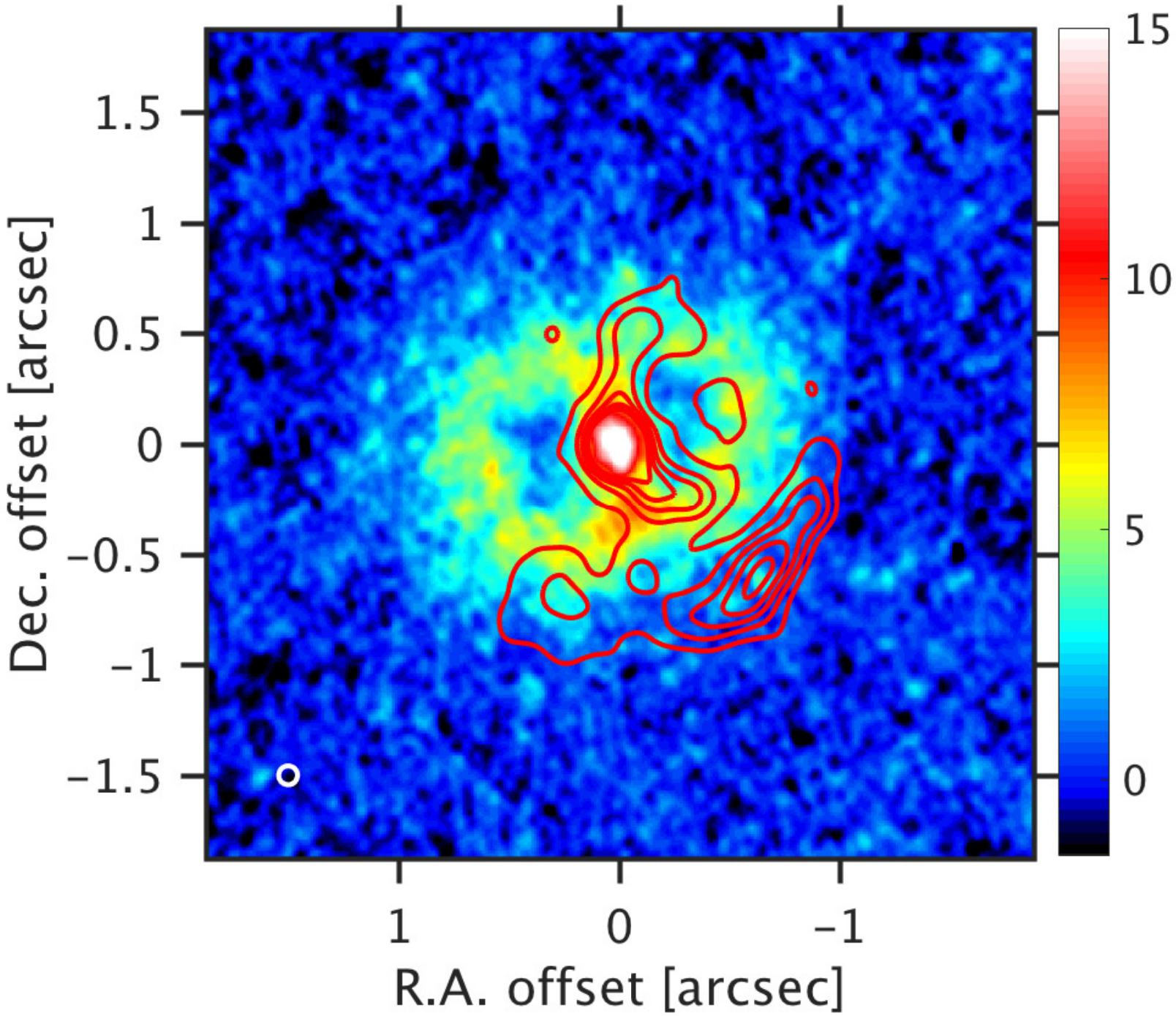}
      \caption{$p$-H$_2^{34}$S(\mbox{2$_{20}$--2$_{11}$}) image in a 1.5\,km\,s$^{-1}$ channel at 41.5\,km\,s$^{-1}$ at a resolution of 0\farcs085 (the beam is shown in the lower left corner). 1.3\,mm continuum, at 0\farcs15 resolution, is shown in red contours. The flux scale is in mJy\,beam$^{-1}$, and the contours start at 0.6\,mJy\,beam$^{-1}$ with a spacing of 0.3\,mJy\,beam$^{-1}$ (the maximum value is 9.5\,mJy\,beam$^{-1}$).}
   \label{f:ede_h2(34)s_continuum}
   \end{figure}

When looked at in detail the EDE morphology exhibits some complications. There is an inner structure in the form of ``ears'' attached to the CCS (not necessarily physically though), that is particularly prominent in the rarer isotopologue $p$-H$_2^{34}$S(\mbox{2$_{20}$--2$_{11}$}) line image at the systemic velocity, Fig.~\ref{f:ede_h2(34)s_continuum}. This feature is also seen (weakly) in the CO(\mbox{2--1}) and SiO(\mbox{5--4}) data. The PA of the minor axis of this inner structure is $\approx$\,20$^\circ$. Its velocity coverage is much lower than that of the outer circular structure. Both inner and outer structures can be partly traced in the form of ``arcs'' in the 1.3\,mm continuum emission as seen in Fig.~\ref{f:ede_h2(34)s_continuum}.

\begin{figure*}
   \centering
   \includegraphics[width=14cm]{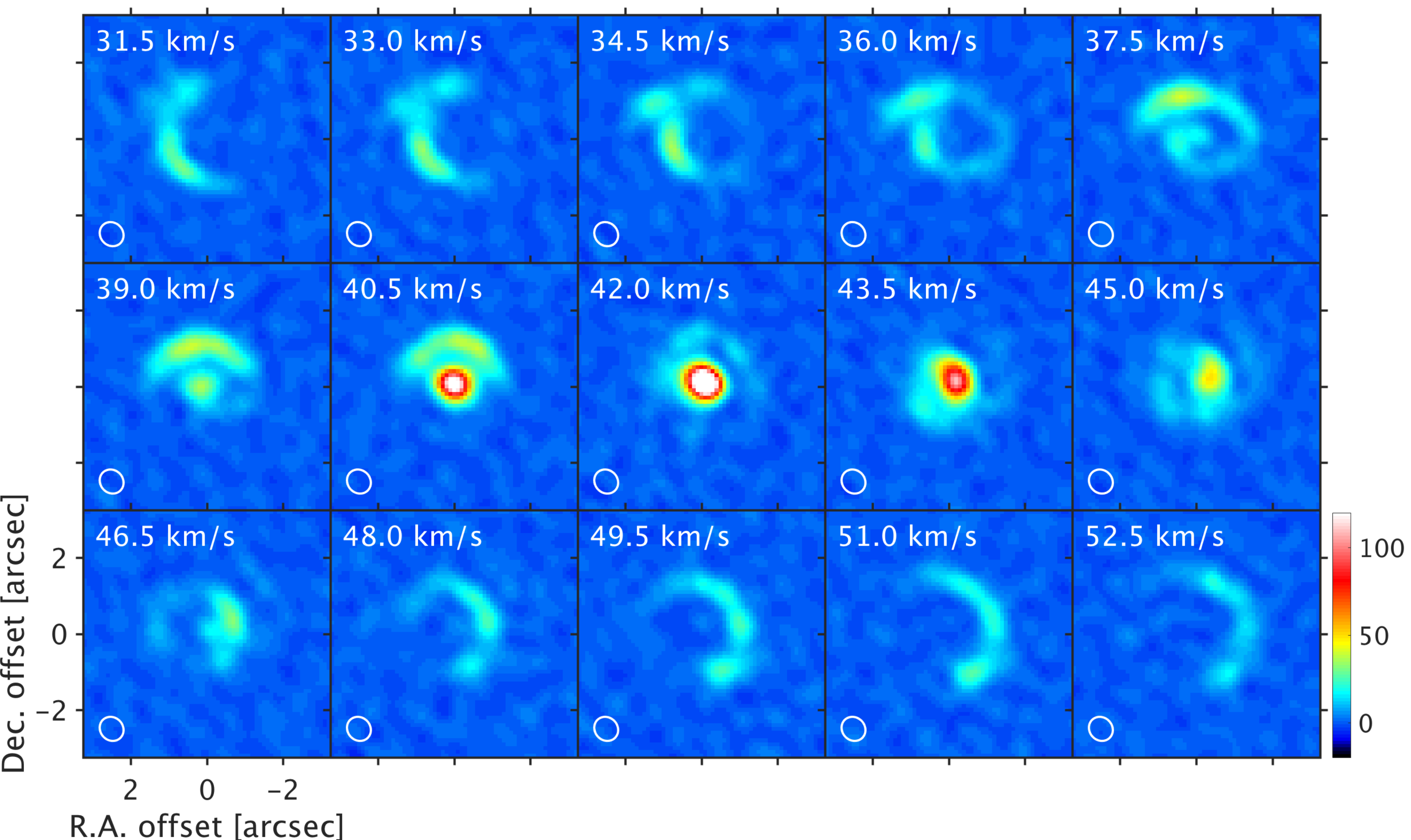}  \includegraphics[width=14cm]{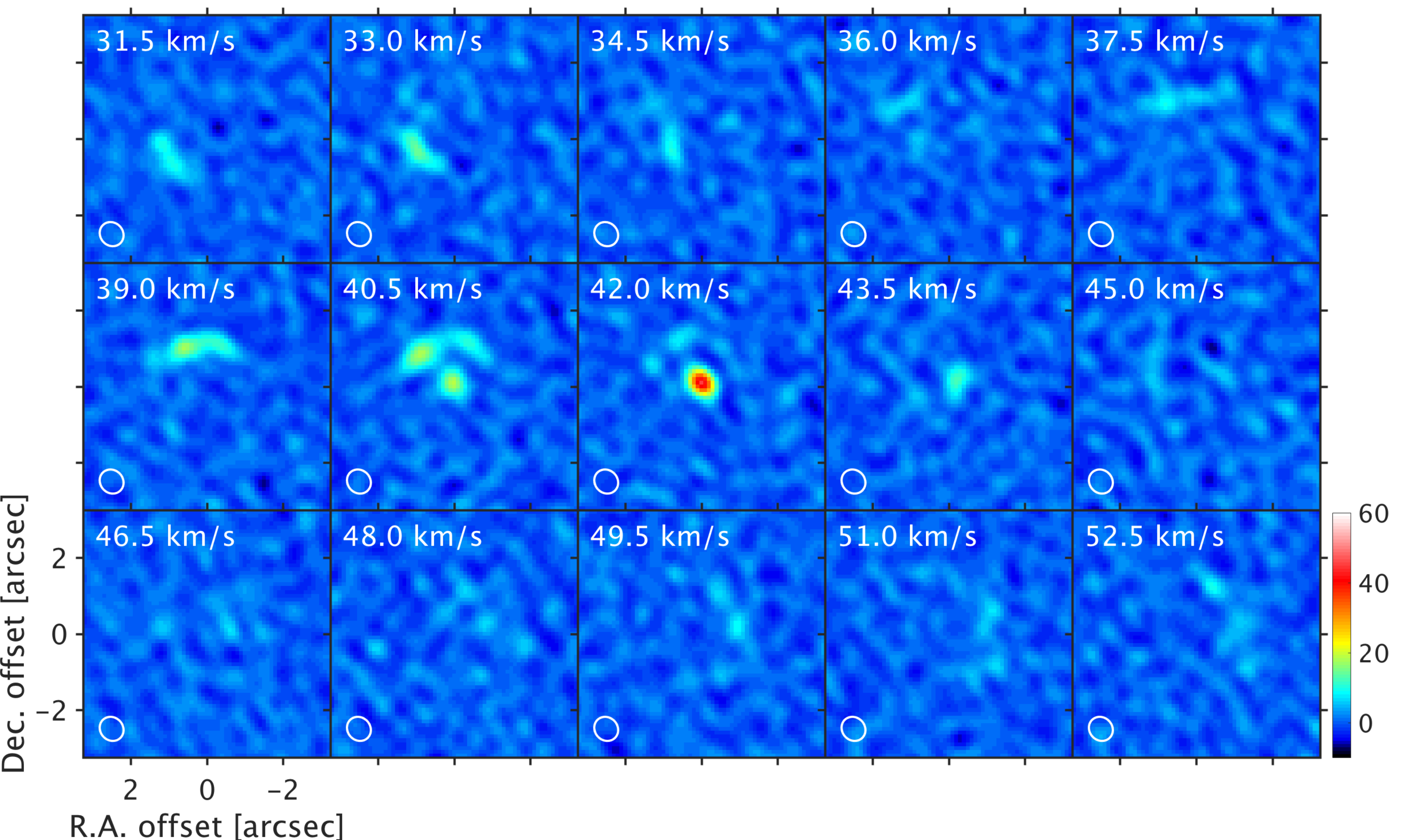}
      \caption{{\bf Upper:} SO(\mbox{5$_6$--4$_5$}) line channel maps. {\bf Lower:} SO$_2$(\mbox{$4_{22}-3_{13}$}) line channel maps. The channel width and spacing is 1.5\,km\,s$^{-1}$ at a resolution of 0\farcs6 (he beam is shown in the lower left corner of each panel). The flux scale is in mJy\,beam$^{-1}$. Emission from the EDE is clearly visible for these lines, but also emission from the CCS is present at the centre. }
   \label{f:ede_so_so2}
   \end{figure*}

\begin{figure}
   \centering
    \includegraphics[width=8cm]{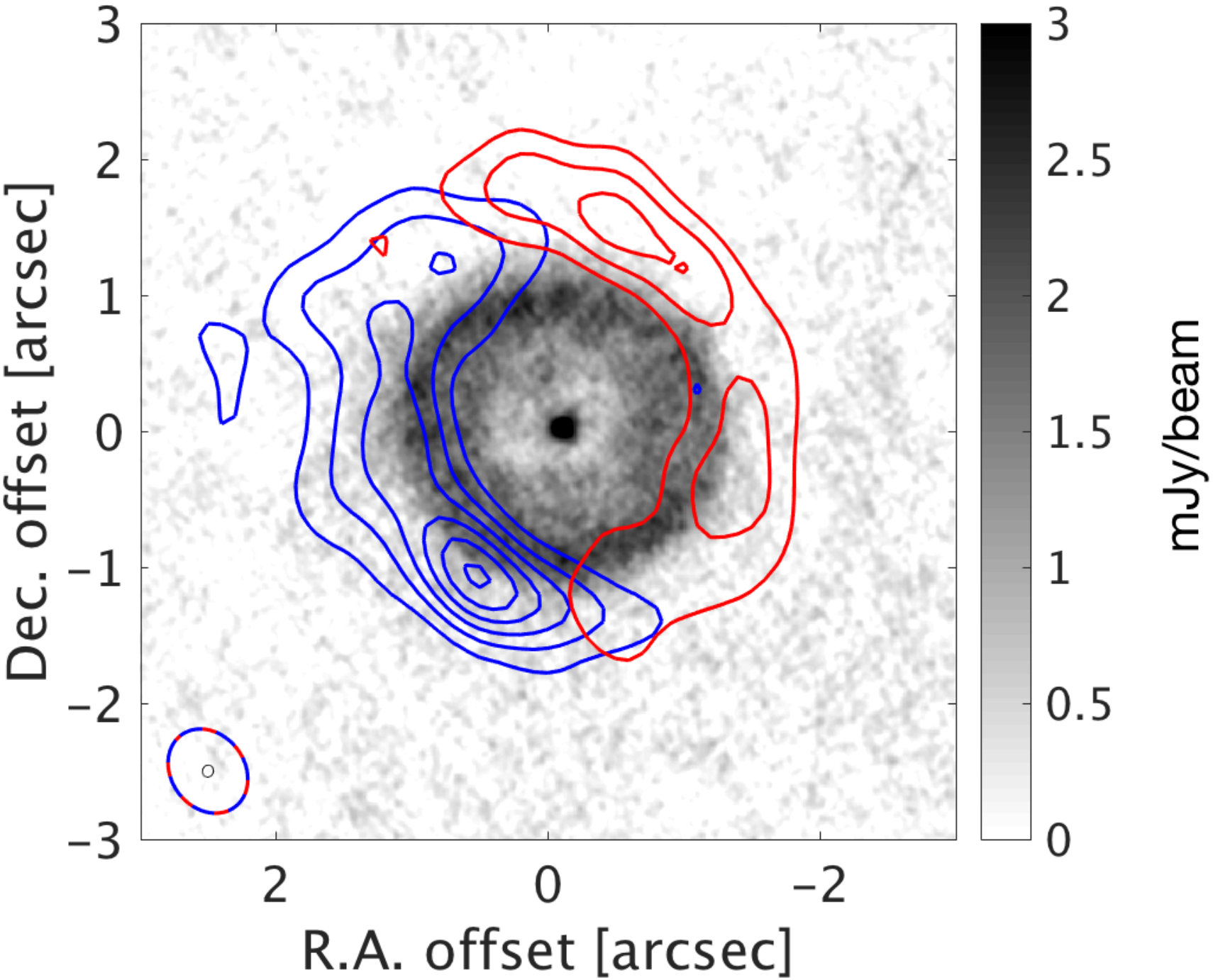}
      \caption{EDE component as seen in the $p$-H$_2$S(\mbox{2$_{20}$--2$_{11}$}) (grey scale: velocity range \mbox{32--52}\,km\,s$^{-1}$; 0\farcs085 resolution) and SO(\mbox{5$_6$--4$_5$}) lines (blue contours: velocity range \mbox{22--32}\,km\,s$^{-1}$, red contours: velocity range \mbox{52--62}\,km\,s$^{-1}$; 0\farcs6 resolution). The contours start at 2\,mJy\,beam$^{-1}$ with a spacing of 3\,mJy\,beam$^{-1}$.}
   \label{f:ede_h2s_so}
   \end{figure}

Apart from H$_2$S, the EDE component is safely identified only in the ALMA SO and SO$_2$ images. Figure~\ref{f:ede_so_so2} shows the channel maps of the SO(\mbox{5$_6$--4$_5$}) and SO$_2$(\mbox{$4_{22}-3_{13}$}) lines (note that the $4_{22}-3_{13}$ line is the one lowest in energy of our observed SO$_2$ lines, and hence less dominated by the CCS component). The EDE component is clearly visible. However, when looked at in more detail, a significant difference compared to the H$_2$S line emission from the EDE can be seen. In particular, the SO(\mbox{5$_6$--4$_5$}) line extends over a velocity range larger than that of the H$_2$S line emission. It covers the range $\approx$\,$\pm$20\,km\,s$^{-1}$ around the systemic velocity, Fig.~\ref{f:ede_spectra}, and its emission in the velocity range outside that of the $p$-H$_2$S(\mbox{2$_{20}$--2$_{11}$}) line comes from a ring-like region that lies just outside that of the latter emission, Fig.~\ref{f:ede_h2s_so}. The symmetry axis of this emission has a PA\,$\approx$\,20$^\circ$, that is, the same as the PA of the inner structure seen in the $p$-H$_2^{34}$S(\mbox{2$_{20}$--2$_{11}$}) line data and discussed above, and the diameter of the ring-like region is $\approx$\,3\arcsec . In addition, there is also an arc-like structure to the N at slightly blueshifted velocities, and a feature that stretches to the NE at slightly larger velocity offsets, in both the SO(\mbox{5$_6$--4$_5$}) and SO$_2$(\mbox{$4_{22}-3_{13}$}) lines. These have no apparent counterparts in the H$_2$S data. At present, we have no interpretation of these features. 

It is difficult to reconcile the $p$-H$_2$S(\mbox{2$_{20}$--2$_{11}$}) and SO(\mbox{5$_6$--4$_5$}) line brightness distributions. The sharp truncation of the H$_2$S line emission is most reasonably explained by a sharp density drop, and this is supported by the dust emission that is also truncated at roughly the same radius, see Fig.~\ref{f:ede_h2(34)s_continuum}. Less likely explanations are excitation and/or chemistry. However, the apparent continuation in space and velocity of the SO line emission with respect to that of H$_2$S rather suggests a smooth density distribution, and a chemistry where H$_2$S is destroyed at the expense of forming SO (for example, from S\,+\,OH). It is also possible that gas further away from the centre has been more accelerated through interaction with the HVO, and that this favours formation of SO. In fact, one may speculate that the SO line emission is coming from the HGS rather than the EDE. Against this interpretation, it may be argued that the SO line brightness distribution is circular and centred on the CCS emission. Further, it is not clear why, as opposed to the behaviour of the H$_2$S line emission that shows both blue- and redshifted emission over the area of the EDE, blue- and redshifted SO line emission is only seen towards the E and the W, respectively.

In summary, it is not clear whether the EDE component is defined by the spatial and kinematical characteristics of the H$_2$S line emission or whether it extends further both in space (reaching a diameter of $\approx$\,3\arcsec) and velocity (reaching a maximum velocity of $\approx$\,20\,km\,s$^{-1}$). Higher angular resolution data for also the SO line emission may shed light on this issue. Furthermore, it may be that the CCS component gradually tapers into the EDE component, and that they are part of the same phenomenon.
 
\begin{figure}
   \centering
   \includegraphics[width=7cm]{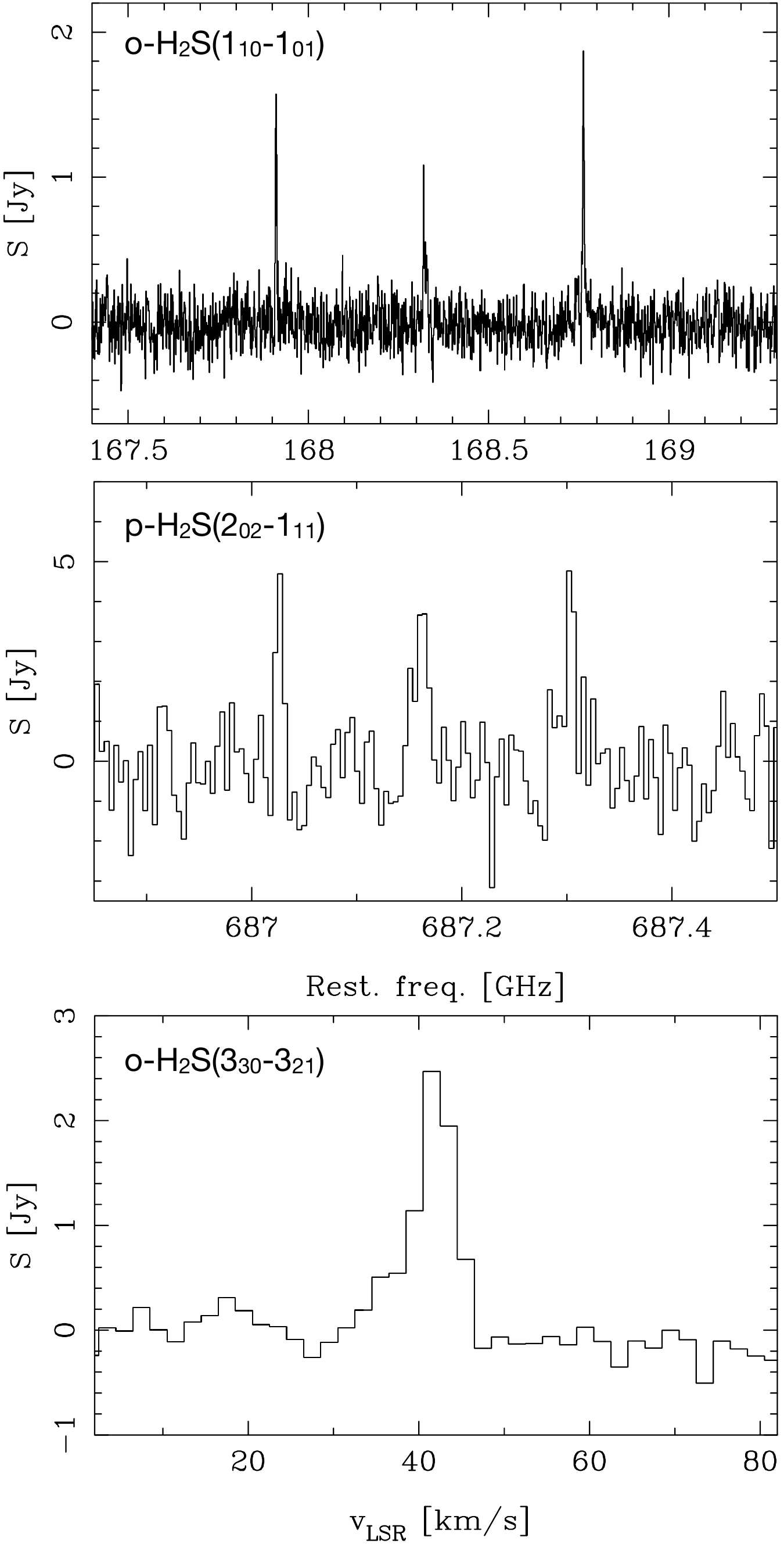}
      \caption{H$_2$S line spectra observed with APEX (the velocity resolution is 2\,km\,s$^{-1}$). {\bf Top:} \mbox{1$_{10}$--1$_{01}$} lines of $o$-H$_2$S (right), $o$-H$_2^{33}$S (middle), and $o$-H$_2^{34}$S (left). 
      {\bf Middle:} \mbox{2$_{02}$--1$_{11}$} lines of $p$-H$_2$S (right), $p$-H$_2^{33}$S (middle), and $p$-H$_2^{34}$S (left). 
      {\bf Bottom:} \mbox{3$_{30}$--3$_{21}$} line of $o$-H$_2$S.
      }
   \label{f:h2s_ede_spectra}
   \end{figure}

We have complemented the ALMA H$_2$S data with APEX observations of the \mbox{$1_{10}-1_{01}$}, \mbox{$2_{02}-1_{11}$}, and \mbox{$3_{30}-3_{21}$} lines (including isotopologues for the first two) and they emphasise the abundance of H$_2$S in this component, Fig.~\ref{f:h2s_ede_spectra}. The relative contributions by emission from the CCS and the EDE in these lines are unknown, but the fact that the ALMA $p$-H$_2$S(\mbox{2$_{20}$--2$_{11}$}) line is $\approx$\,75 times stronger in the latter is a strong argument in favour of also the APEX lines coming predominantly from the EDE component. The line widths of the \mbox{$1_{10}-1_{01}$} and \mbox{$3_{30}-3_{21}$} lines are consistent with this. On the other hand, the narrow widths of the \mbox{2$_{02}$--1$_{11}$} lines (these are the lowest-energy lines of the observed H$_2$S lines) are more characteristic of the CCS component. Some guidance to the interpretation of these lines is obtained from the relative isotopologue line strengths. The three isotopologue \mbox{2$_{02}$--1$_{11}$} lines are about equally strong indicating very high optical depths in the main isotopologue (the solar sulphur isotope ratios are $^{32}$S:$^{33}$S:$^{34}$S\,=\,127:1:23, and there is no reason to expect a low-mass star to alter this in any significant way during its evolution). Further, the \mbox{2$_{02}$--1$_{11}$} line is $\approx$\,80 times stronger than the CCS emission in the \mbox{2$_{20}$--2$_{11}$} line, while the expected ratio is about ten for optically thick emission at the same temperature, that is, following the black-body law of radiation. Thus, we conclude that also the \mbox{2$_{02}$--1$_{11}$} lines originate mainly in the EDE, but we have no explanation for why the lines are so narrow.

%
%
%
%
\subsection{The bipolar high-velocity outflow, HVO}
\label{s:obs_hvo}

\begin{figure}
   \centering
   \includegraphics[width=8cm]{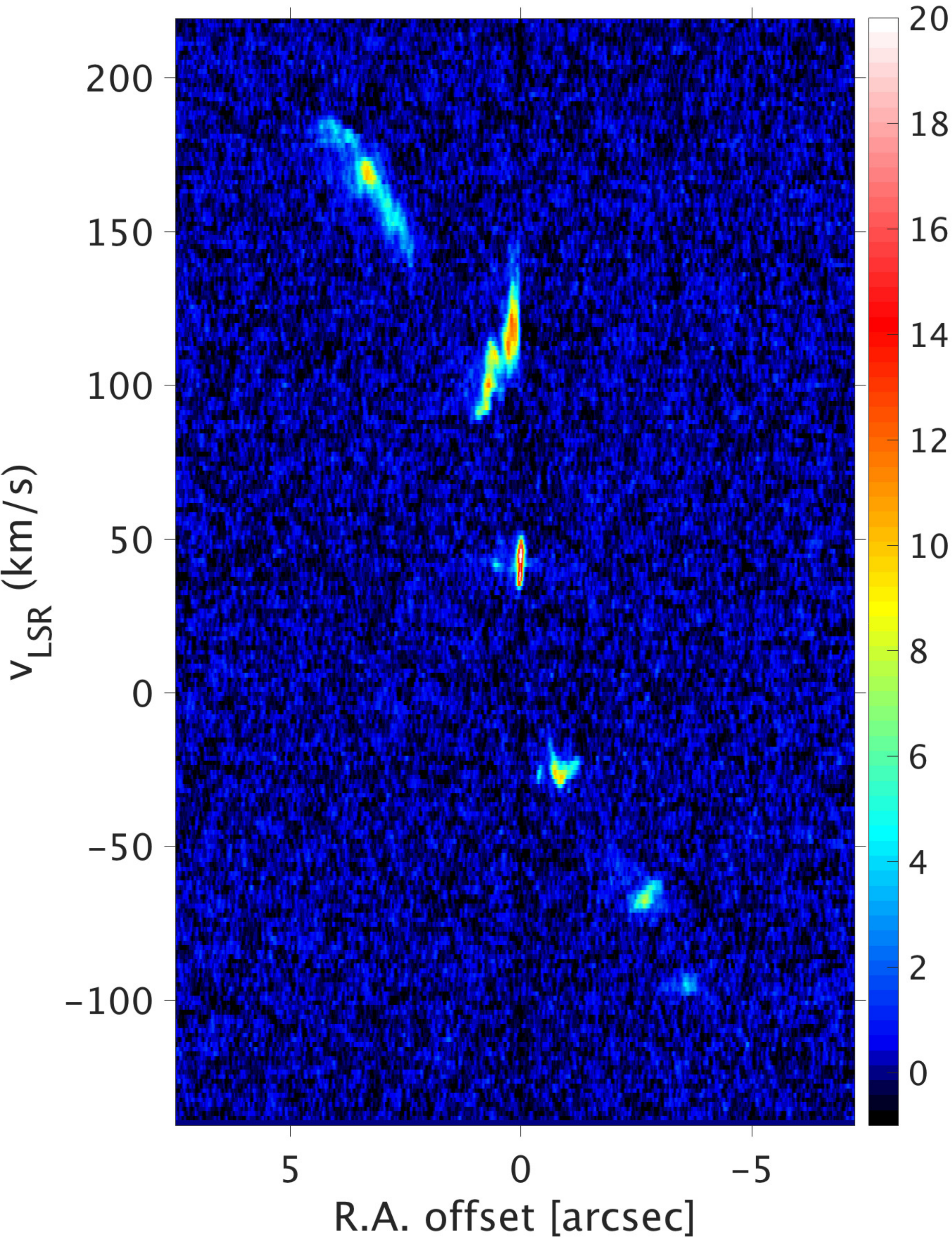}
      \caption{SiO(\mbox{5--4}) PV-diagram along PA\,=\,90$^\circ$ at resolutions of 0\farcs085 and 1.5\,km\,s$^{-1}$.  The flux scale is in mJy\,beam$^{-1}$. }
   \label{f:sio_posvel}
   \end{figure}

The HVO component is clearly seen in the CO(\mbox{2--1}) and SiO(\mbox{5--4}) line emissions, but it is also present in the line emissions of SO, CS, and OCS, and it completely dominates the line emissions from HCN, HCO$^+$, H$_2$CO, and CH$_3$OH, where the EVSs are particularly prominent at $\pm$\,140\,km\,s$^{-1}$ and offset by $\approx$\,4\arcsec\ on either side of the centre, as discussed in Sect.~\ref{s:obs_evs}. On the contrary, H$_2$S line emission, which is very strong in the EDE component, is markedly absent here. We exemplify the characteristics of the HVO through the CO(\mbox{2--1}) and SiO(\mbox{5--4}) channel maps (Figs \ref{f:co_channels} and \ref{f:sio_channels}) and an SiO(\mbox{5--4}) PV diagram along PA\,=\,90$^\circ$ (Fig.~\ref{f:sio_posvel}).

The characteristics of the HVO as shown in the SiO(\mbox{5--4}) data suggest that the driving outflow is highly collimated. The HVO is very symmetric with respect to the centre with a PA close to 90$^\circ$ initially, turning gradually beyond an offset of $\approx$\,3\arcsec\ to reach $\approx$\,100$^\circ$ at the EVSs. The HVO gas reaches a maximum velocity (not corrected for inclination angle) of $\approx$\,150\,km\,s$^{-1}$ at $\pm$\,4\farcs2 from the centre.  Particularly noticeable in the PV-diagram is the close to Hubble-like velocity dependence on distance to the centre, a phenomenon common in proto-PNe \citep[for example,][]{alcoetal01}. 

There are spots of enhanced line emission, symmetrically placed at $\approx$\,$\pm$\,0\farcs8, $\pm$\,3\farcs0, and $\pm$\,4\farcs2 with respect to the centre. They outline a slightly S-shaped figure in the SiO(\mbox{5--4}) PV-diagram, presumably an effect of a precessing driving jet. The EVSs are particularly prominent, both in the maps and as distinct features at the extreme velocities in the CO and $^{13}$CO single-dish spectra, Fig.~\ref{f:co_spec}. The most reasonable explanation is that at these spots there are also piled-up material from an interaction between the HVO and a remnant wind [\citet{olofetal17} provided evidence that the $^{12}$C/$^{13}$C ratio is the same, $\approx$\,13, in the EVSs as in the CCS]. Another indication of this is that the bubbles of the HGS close at the EVSs.

%
%
%
%
\subsection{The extreme-velocity spots, EVSs}
\label{s:obs_evs}

\begin{figure}
   \centering
   \includegraphics[width=7cm]{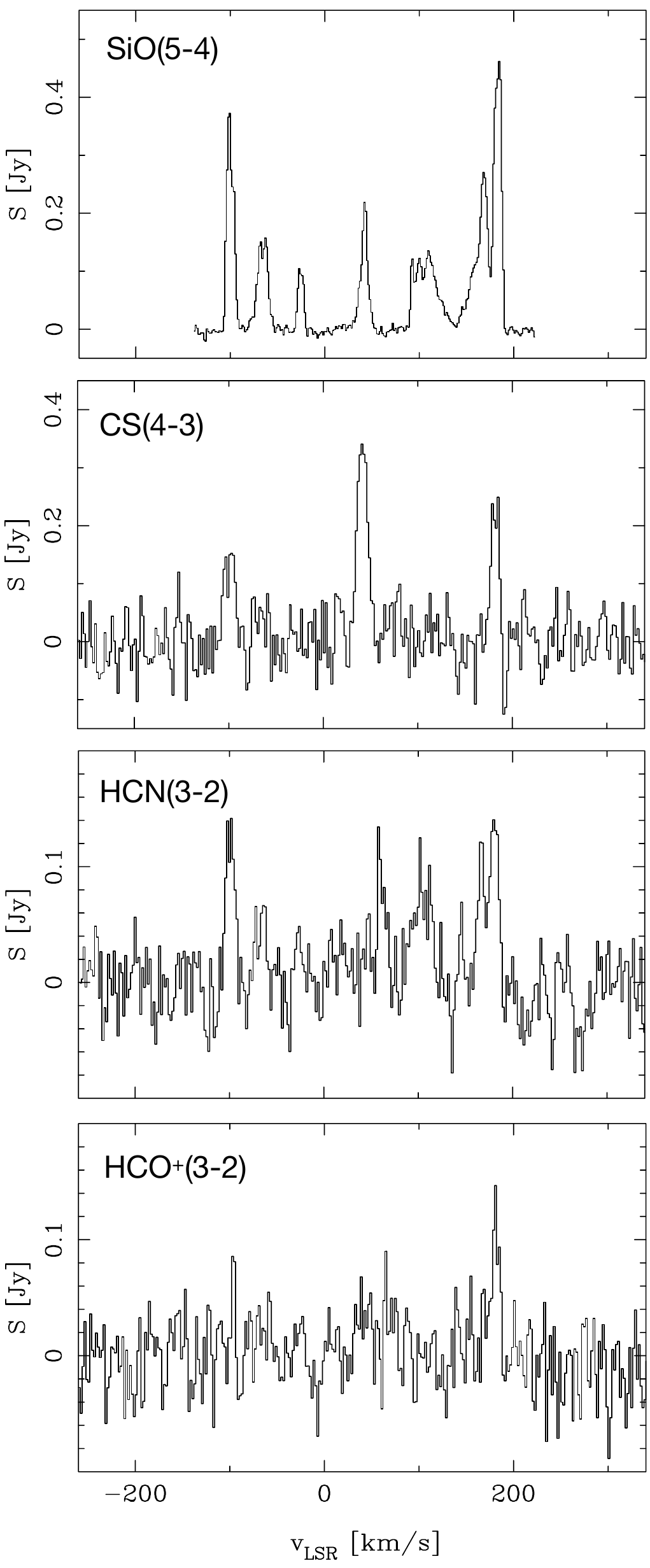}
      \caption{Spectra obtained with APEX, and for comparison the SiO(\mbox{5--4}) line extracted from the ALMA data integrated over the source (velocity resolution 2\,km\,s$^{-1}$). {\bf From top to bottom:} Global ALMA SiO(\mbox{5--4}) line, CS(\mbox{4--3}) emission from the EDE and EVS components, HCN(\mbox{3--2}) emission from the EVS components (and other features possibly related to the HVO component), and HCO$^+$(\mbox{3--2}) emission from the EVS components.}
   \label{f:cs_hcn_hco+_spec}
   \end{figure}

\citet{olofetal17} showed that the EVSs at the terminations of the HVO are particularly chemically rich (in a relative sense). They detected line emission from CO, $^{13}$CO, C$^{18}$O, $^{13}$CS, SO, SiO, $^{29}$SiO, H$_2$CO, H$_2^{13}$CO, and CH$_3$OH at these spots. Here we report the detections of also OCS using ALMA, as well as detections of CS, HCN, and HCO$^+$ using APEX where emissions from the two EVSs are clearly seen, Fig.~\ref{f:cs_hcn_hco+_spec}. We searched unsuccessfully for the $p$-H$_2$O(\mbox{3$_{13}$--2$_{20}$}) line with APEX and report an upper limit (which is high compared to the detection levels of the ALMA data).

\begin{table*}
\caption{Molecular line emission from the e-EVS\,$^1$}
\centering
\begin{tabular}{l l c c c}
\hline \hline
Line                         & Instr.  & $S$                & $\Delta \upsilon$\,$^2$ & $I$ \\
                             &         & [Jy]               & [km s$^{-1}$]           & [Jy km s$^{-1}$]   \\
\hline 
CO(1--0)                     & SEST    & 0.38\phantom{0}    & \ldots              & \phantom{0}6.2\phantom{0}             \\
CO(2--1)                     & ALMA    & 1.38\phantom{0}    & 10.8              & 16\phantom{.00}                \\
                             & SEST    & 1.48\phantom{0}    & \ldots              & 21\phantom{.00}              \\
CO(3--2)                     & APEX    & 1.64\phantom{0}    & \ldots            & 35\phantom{.00}                  \\
CO(4--3)                     & APEX    & 0.86\phantom{0}    & \ldots            & 54\phantom{.00}                  \\
CO(6--5)                     & APEX    & 0.86\phantom{0}    & \ldots            & 68\phantom{.00}                  \\
$^{13}$CO(2--1)              & ALMA    & 0.54\phantom{0}    & \phantom{1}7.5    & \phantom{0}4.3\phantom{0}   \\
                             & SEST    & 0.49\phantom{0}    & \ldots            & \phantom{0}8.6\phantom{0}        \\
$^{13}$CO(3--2)              & APEX    & 0.37\phantom{0}    & \ldots            & \phantom{0}7.0\phantom{0}        \\
C$^{18}$O(2--1)              & ALMA    & 0.06\phantom{0}    & \phantom{1}4.8    & \phantom{0}0.31              \\
SiO(5--4)                    & ALMA    & 0.45\phantom{0}    & \phantom{1}9.5    & \phantom{0}4.5\phantom{0}    \\
$^{29}$SiO(5--4)             & ALMA    & 0.17\phantom{0}    & \phantom{1}7.2    & \phantom{0}1.3\phantom{0}   \\
CS(4--3)                     & APEX    & 0.2\phantom{00}    & \ldots             & \phantom{0}1.9\phantom{0} \\
$^{13}$CS(5--4)              & ALMA    & 0.065              & \phantom{1}9.6    & \phantom{0}0.66            \\
SO($5_5-4_4$)                & ALMA    & 0.078              & \phantom{1}9.2    & \phantom{0}0.76                \\
SO($5_6-4_5$)                & ALMA    & 0.10\phantom{0}    & \phantom{1}5.8    & \phantom{0}0.61              \\
OCS(18--17)                  & ALMA    & 0.016              & \phantom{1}8.6    & \phantom{0}0.15            \\
OCS(19--18)                  & ALMA    & 0.016              & \phantom{1}7.9    & \phantom{0}0.13            \\
HCN(3--2)                    & APEX    & 0.08\phantom{0}    & \ldots             & \phantom{0}1.4\phantom{0} \\ 
HCO$^+$(3--2)                & APEX    & 0.06\phantom{0}    & \ldots            & \phantom{0}1.0\phantom{0} \\
$p$-H$_2$O($3_{13}-2_{20}$)  & APEX    & \ldots             & \ldots            & $<$\,3\\
$p$-H$_2$CO($3_{03}-2_{02}$) & ALMA    & 0.10\phantom{0}    & \phantom{1}9.3    & \phantom{0}1.0\phantom{0}  \\
$p$-H$_2$CO($3_{22}-2_{21}$) & ALMA    & 0.034              & \phantom{1}9.4    & \phantom{0}0.34            \\
$p$-H$_2$CO($3_{21}-2_{20}$) & ALMA    & 0.039              & \phantom{1}7.4    & \phantom{0}0.31            \\
$o$-H$_2^{13}$CO($3_{12}-2_{11}$) & ALMA & 0.023            & \phantom{1}8.7    & \phantom{0}0.21             \\
$E$-CH$_3$OH($4_2-3_1$)      & ALMA    & 0.051              & \phantom{1}5.2    & \phantom{0}0.28                \\
$E$-CH$_3$OH($8_{-1}-7_0$)   & ALMA    & 0.065              & \phantom{1}6.2    & \phantom{0}0.43              \\
  \hline
\end{tabular}
\label{t:obs_evs}
\tablefoot{(1) The line intensities apply to the velocity interval and region specified in the text for the ALMA data, and within the velocity interval specified in the text for the APEX and SEST data. The single-dish data are observed with the beam positioned at the centre of the source, and the reported intensities are not corrected for the beam response. Flux uncertainties are discussed in Sect.~\ref{s:obs_evs}. (2) FWHM of a Gaussian fit to the line within the given aperture at 1.5\,km\,s$^{-1}$ resolution.}
\end{table*}

In Table~\ref{t:obs_evs} we summarise the observational results for the eastern EVS (e-EVS; the western EVS shows very much the same line brightness pattern, but the emission is somewhat weaker). The data extraction is based on the appearance of the SiO(\mbox{5--4}) line brightness distribution. Its emission at the most extreme redshifted velocities produces a line profile extending from 175 to 193\,km\,s$^{-1}$ at zero power (this is also the velocity range that contains for example all of the redshifted CH$_3$OH and essentially all of the redshifted H$_2$CO line emissions), and a brightness distribution that is essentially circular with a diameter of about 1\arcsec\ (determined from a 2D Gaussian fit) and centred 4\farcs14\,E and 0\farcs37\,S of the continuum peak. We have determined the data for all molecules observed with ALMA at this position, within an aperture of 1\arcsec\ and within the given velocity range (within this velocity range, all the brightness distributions are close to circular and have deconvolved sizes, FWHMs of 2D Gaussian fits, of 1\arcsec\ to within 0\farcs3). The APEX and SEST data are estimated within the observing beams and in the above velocity range. The uncertainties in the observational results are dominated by the complexity of the brightness distributions for the ALMA data and by the uncertainty in which velocity range to use for the APEX and SEST data. They are difficult to estimate in a formal way, but may reach 50\,\% for some lines (especially for the single-dish data). 

%
%
%
%
\subsection{The hourglass structure, HGS}
\label{s:obs_hgs}

The CO(\mbox{2--1}) brightness distribution (and also those of its observed isotopologues, for example, Fig.~5 in \citet{olofetal15}), Fig.~\ref{f:co_channels}, shows an ellipse-like distribution whose size increases with velocity offset from the systemic velocity in the range $\approx$\,$\pm$\,30\,km\,s$^{-1}$. Assuming that the distance from the centre scales with the velocity offset  (see below for a discussion on this), this gives an hourglass-like structure whose cross section increases with the distance from the centre. Also, the centre of the ``ellipse'' as a function of velocity is shifted consistently with the same sign of the velocity gradient as that of the HVO. Thus, an interpretation in the form of an hourglass structure (HGS), having the HVO along its symmetry axis, produced by pressure towards the sides from the outflow appears the most likely explanation for this component. The HGS is not seen in any of the other molecular line emissions, for example, it is absent in the SiO data, suggesting that the conditions in the walls are less extreme than in the HVO. 

The velocity range in which we lose flux in the ALMA data is also the velocity range of the HGS, Sect.~\ref{s:alma_flux}. This, most likely, means that we are not detecting material that has been accelerated to the same extent as the gas of the HGS component, because it is distributed in a more diffuse way.

A slight distortion of the ellipse form starts at velocity offsets of $\approx$\,$\pm$\,30\,km\,s$^{-1}$ from the systemic velocity. This turns into a major complex structure in the velocity-offset ranges $\approx$\,40--80\,km\,s$^{-1}$ on either side of the systemic velocity, whose major components are two bright spots at velocity offsets of $\pm$\,60\,km\,s$^{-1}$. The morphology of the distortion, despite its complexity, is very symmetric with respect to the centre. We will come back to a possible explanation of this phenomenon in Sect.~\ref{s:multi_polar}. Beyond these velocities the HGS becomes much fainter, but it can be traced as a bubble, on either side of the centre, that closes at the EVS. This provides further evidence for a connection between the HGS and the HVO. We will come back to this in Sects~\ref{s:inc_angle} and \ref{s:3D}.

%
%
%
%
\subsection{Inclination angle of the HVO}
\label{s:inc_angle}

   \begin{figure*}
   \centering
   \includegraphics[width=16cm]{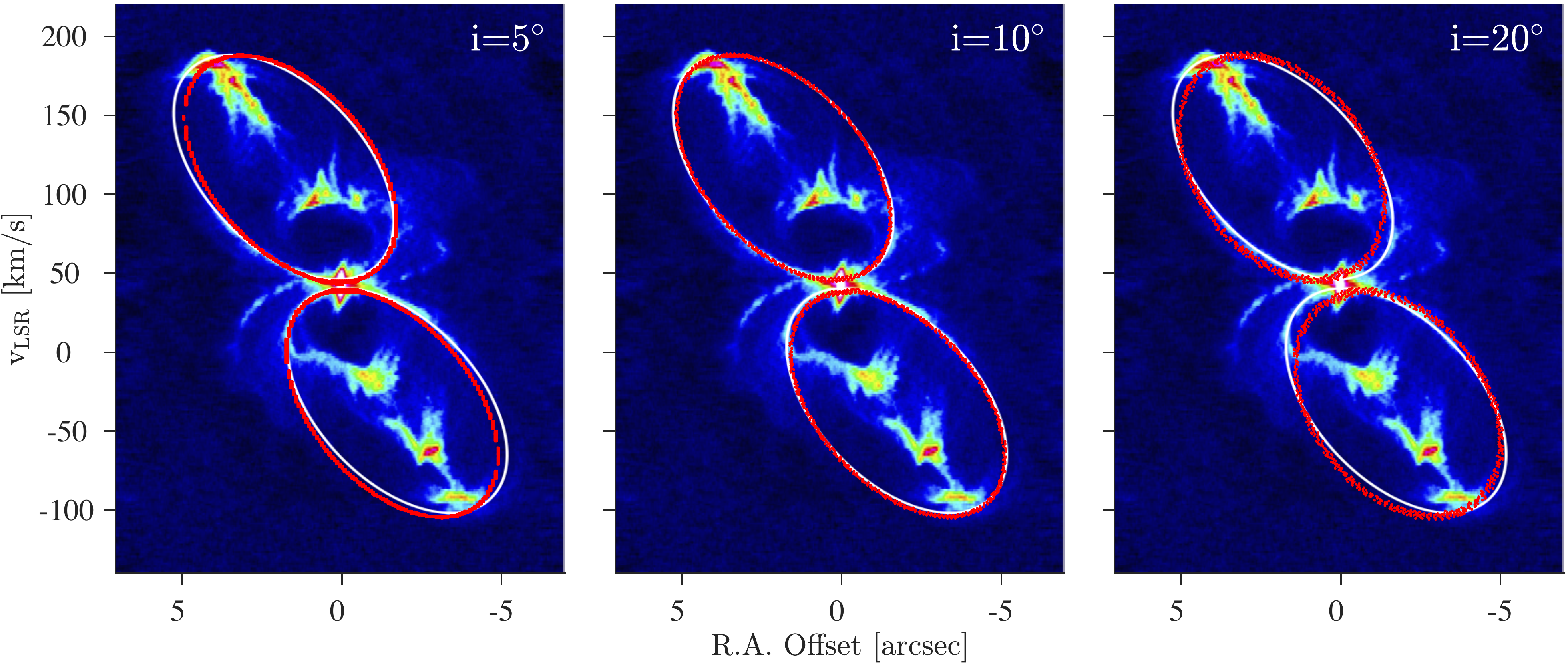} 
         \caption{CO(\mbox{2--1}) PV-diagram along PA\,=\,90$^\circ$ with an ellipse fitted to the observational HGS/bubble structure shown in white. Results of the best-fit model (see text for details) for $i$\,=\,5$^\circ$, 10$^\circ$, and 20$^\circ$ are shown as a red solid line.}
   \label{f:inclination_angle}
   \end{figure*}   

The inclination angle, $i$, of the HVO cannot be estimated from the HVO emission itself since the maximum outflow velocity is not known. However, emission from the HGS/bubble structure, as seen in the PV diagram (Fig.~\ref{f:sketch_co21}, right panel) and channel maps (Fig.~\ref{f:co_channels}) of the CO(\mbox{2--1}) line, can be used to constrain it under certain assumptions. The latter are a source symmetric with respect to its centre, and an HVO axis direction constant with time (its projection on the sky is at PA\,=\,90$^\circ$ in this case). We further assume that the expansion velocity of the HGS/bubble structure is at each point linearly proportional to its distance from the centre. The latter is suggested by the close to Hubble-like appearance of the relation between line-of-sight velocity and apparent offset from the centre as shown by the HVO line emission, for example, Fig.~\ref{f:sio_posvel}, that is, $\upsilon_{\rm z}$\,$\propto$\,$p$ suggests $\upsilon_{\rm r}$\,$\propto$\,$r$ for at least the emission involved in the HVO, that is, including the HGS.

We have used the publicly available code \texttt{SHAPE} \citep{stefwetal11} to determine the geometrical properties of the HGS/bubble structure. The model is described by two, diametrically oriented, expanding, and thin-walled ellipsoids of homogeneous density (cigar-like and of the same geometry) that reach the maximum velocity at their tips. The results of this are compared to the observed PV-diagram and channel maps. The best fit to the data is found for $i$\,$\approx$\,10$^\circ$. This is examplified in Fig.~\ref{f:inclination_angle} where the results of this model for three inclination angles, 5$^\circ$, 10$^\circ$, and 20$^\circ$, are shown. A change of the cross section of the ellipsoids in order to make the results for the inclination angles of 5 and 20 degrees resemble better the observational data, leads to model channel maps that are inconsistent with the observed channel maps (primarily the cross section width of the ellipsoids is determined by the vertical sizes of the ellipses in the channel maps, and these are independent of the inclination angle for the chosen geometry). The estimated inclination angle of the HVO is therefore 10$^\circ$\,($-5^\circ, +10^\circ$). We conservatively estimate, using inspection by the eye, that the inclination angle lies in the range 5 to 20 degrees, that is, the quoted errors can be seen as 2$\sigma$ limits. As argued above, the most reasonable conclusion is that the CCS and EDE components have flattened density distributions that are orthogonal to this direction.
%
%
%
%
\subsection{A 3D-reconstruction}
\label{s:3D}

The assumption that the line-of-sight velocity can be used as a measure of the spatial coordinate along the line of sight can be used to make a 3D-reconstruction of the source structure. The estimated inclination angle allows a determination of the correct scaling between the $z$ and $\upsilon_{\rm z}$ axes. Since the emission lines become gradually narrower towards the centre, the same relation can in principle be used also here to properly locate the emission, while the emergent morphology is more doubtful in this case. 

\begin{figure*}
   \centering
   \includegraphics[width=17cm]{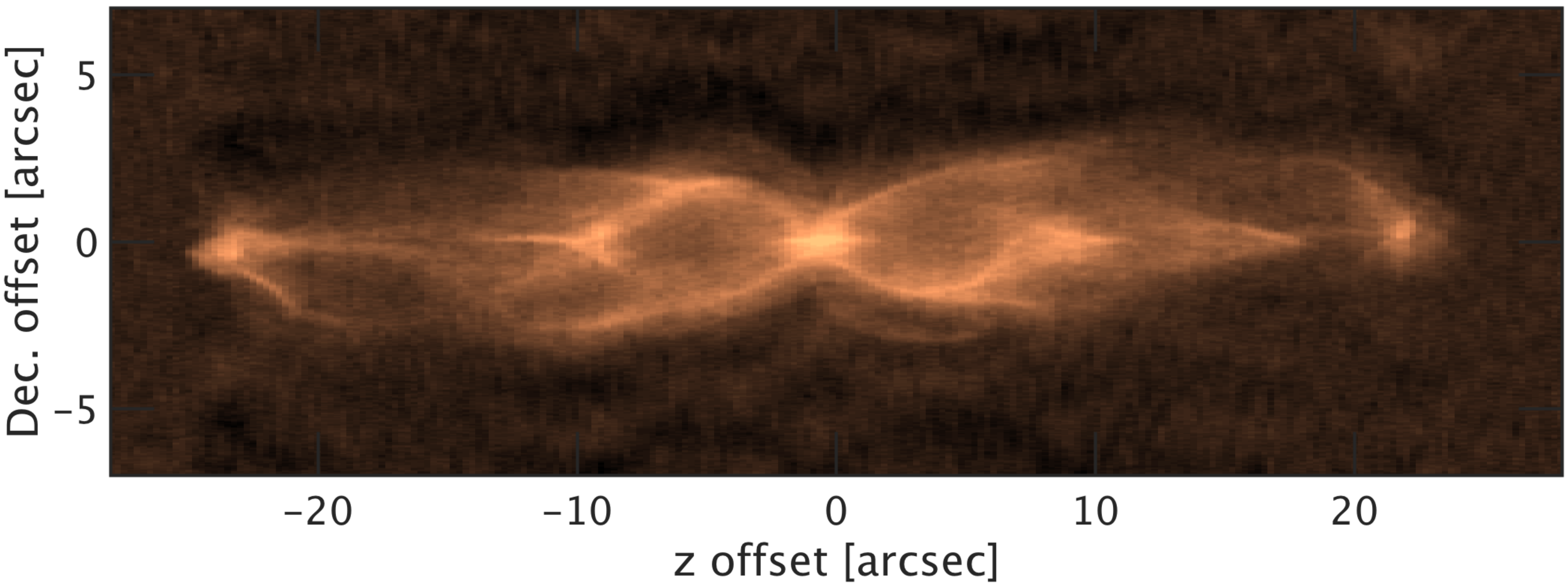}  
         \caption{Circumstellar environment of HD\,101584 as seen from the side (the righthand side is facing towards us). The image is obtained by assuming radial expansion with a velocity that scales linearly with the distance from the centre, and using the estimated inclination angle. It gives the average intensity of the CO(\mbox{2--1}) line in the R.A. direction at each pixel.}
   \label{f:3d}
   \end{figure*}

The final 3D-reconstruction of the circumstellar medium of HD\,101584 is shown in Fig.~\ref{f:3d} in the form of an image of the CO(\mbox{2--1}) line average intensity in the R.A. direction as seen from the side. There are a number of limitations to this 3D reconstruction. Among them, no corrections for radiative transfer effects, and no correction for the fact that within the same observed velocity channel, that is, gas moving with the same line-of-sight velocity, there will be gas moving at different absolute velocities and hence different distances to the centre (the small inclination angle and the high collimation of the outflow strongly limit the problem in our case). The movie in Fig.~\ref{f:movie} shows the relation between the channel maps of the CO(\mbox{2--1}) line emission and the 3D structure. 

%
%
%
%
\subsection{A second bipolar outflow?}
\label{s:multi_polar}

There is evidence of a second bipolar, bubble-like structure in the CO(\mbox{2--1}) line data, see Sect.~\ref{s:decomp}. It covers $\approx$\,80\,km\,s$^{-1}$ in velocity, and has a velocity gradient opposite to that of the HVO in the direction of PA\,$\approx$\,$-50^\circ$. The most reasonable explanation to this structure is a second bipolar outflow, but there are no indications of bright spots in this case, for example, SiO line emission is not present. The inclination angle is unknown, but a difference in direction between this and the HVO may not be particularly large if we see both of them almost pole-on. Likewise, the opposite directions of the velocity gradients may become a natural consequence of only a smaller change in the direction of the outflow axis. The velocity of the second outflow is difficult to estimate since in the CO data we see only the bubbles, presumably of the same character as the HGS surrounding the HVO, and the inclination angle is unknown. The full velocity width of the CO line data indicates a maximum line-of-sight velocity of $\approx$\,40\,km\,s$^{-1}$, but the outflow velocity can be substantially higher.

The PA, the velocity gradient, and the velocity coverage of this bipolar outflow are interestingly close to the results for the OH\,1667\,MHz maser line data presented by \citet{telietal92}. It is tempting to believe that this is more than a coincidence. The OH 1667\,MHz maser peaks are distributed in two clusters, one with blueshifted spots and one with redshifted spots, separated by $\approx$\,3\farcs5 along PA\,$\approx$\,$-55^\circ$. Based on these data \citet{zijletal01} proposed the existence of a second bipolar outflow. In fact, the innermost OH maser spots border exactly outside the edge of the EDE emission as traced in the $p$-H$_2$S(\mbox{$2_{20}-2_{11}$}) line, and coincides well with the blue- and red-shifted SO(\mbox{5$_6$--4$_5$}) line emissions shown in Fig.~\ref{f:ede_h2s_so}, except that the OH maser emission avoids the regions of peak emission in the SO line. The full velocity extent of the OH emission is $\approx$\,80\,km\,s$^{-1}$, and the dominant maser peaks are separated by $\approx$\,50\,km\,s$^{-1}$. Furthermore, the two clusters of red- and blue-shifted OH maser spots, and hence the SO emission peaks, are nearly coincident, in space but less so in velocity, with the areas where the ellipse-like pattern of the HGS in the CO(\mbox{2--1}) channel maps is disturbed (on either side of the centre) and from where it develops eventually into the innermost bright spots as discussed in Sect.~\ref{s:obs_hgs}. 

We speculate here that the disturbance of the HGS structure and the existence of the OH 1667\,MHz masers are connected to an interaction between the two outflows. It is noteworthy in this context that the HD\,101584 OH maser has some rather peculiar characteristics: there is no detectable emission from the three other 18\,cm lines (at 1612, 1665, and 1720\,MHz) at very low levels (in a relative sense), the 1667\,MHz emission lacks any time variability over a time scale of 25 years (time variability is a well-known characteristic of both circumstellar and interstellar cosmic masers), and there is no detectable polarisation (hence no indication of a magnetic field) (Vlemmings et al., in prep.). 
%
%
%
%
\section{Quantitative estimates: gas}
\label{s:quant_est_gas}

We will here derive some quantitative results for the circumstellar molecular medium. However, our data base consists of only 13\,GHz of ALMA data and some complementary APEX and SEST data. This means that the observational constraints on densities and temperatures are limited. Furthermore, the object is seen almost pole-on which restricts the information on its structure along the line of sight and its kinematics orthogonal to it. We will therefore perform rather simple analyses at this stage.
%
%
%
%
%
\subsection{Kinematical ages}
\label{s:age}

A simple estimate of the age of an outflow is obtained by using its maximum apparent outflow velocity, its apparent length, and its inclination angle, that is, its kinematical age. This results in an age of $\approx$\,770\,[$D$/1\,kpc]\,yr for the HVO. An estimate like this suffers from a lack of knowledge of some key parameters required for a proper estimate, notably the velocity of the driving agent, presumably a collimated jet, and its evolution through interactions with the surrounding medium. Therefore, it should be regarded as a reasonable upper limit estimate of the time scale of the phenomenon \citep{bujaetal01}. 

A comparison can be made with the kinematical age of the EDE using a line-of-sight expansion velocity of 10\,km\,s$^{-1}$ (from the H$_2$S line width, but the possible flaring of the EDE makes this an uncertain estimate), the estimated inclination angle, and the measured size. The result is an estimated age of $\approx$\,110\,[$D$/1\,kpc]\,yr, that is, considerably shorter than that of the HVO (though the HVO estimate is an upper limit). This goes against the conclusion by \citet{hugg07} that in proto-PNe jets and tori develop nearly simultaneously, with the torus appearing first and the jet typically a few hundred years later. A possible explanation to our finding is that the EDE is more of a ``pattern'' structure, that is, a region where matter flows through, becomes excited, and hence observable. This would, on the other hand, mean that there must be an inner reservoir of gas and for this we find no evidence. Currently, we have no explanation for the relative ages of the EDE and HVO components.

The age of the second bipolar outflow is more difficult to estimate since there are no bright spots at the end of this outflow in the CO(\mbox{2--1}) data, and the inclination angle is unknown. Using the separation of $\approx$2\arcsec\ of the strongest OH masers,  separated by $\approx$\,50\,km\,s$^{-1}$ in velocity, and the same inclination angle as for the HVO, we derive an age of $\approx$\,2100\,[$D$/1\,kpc]\,yr, that is, considerably older than the HVO. An inclination angle of $\approx$\,70$^\circ$ for the second bipolar outflow is required to make the outflows of similar age. Alternatively, the outflow velocity is much higher than estimated from the bubble structure. Irrespective of this, the most important conclusion is that there is a recurrence aspect in the phenomenon responsible for the circumstellar structure of HD\,101584.

%
%
%
%
\subsection{Mass, density, and temperature estimates}

As argued already above, we restrict ourselves here to simple calculations, which we think, nevertheless, provide us with good order-of-magnitude estimates. Only for the e-EVS do we attempt a radiative transfer analysis, since here we can use single-dish data on a number of CO isotopologue lines from different transitions as constraints. 
%
%
%
%
\subsubsection{Molecular gas estimates}

We will here extensively use the CO(\mbox{2--1}) line brightness temperatures ($T_{\rm b}$) estimated in the various components. If the CO(\mbox{2--1}) line emission is optically thick, which appears to be the case in most parts of the circumstellar medium of HD\,101584, this will give us the excitation temperature of the CO \mbox{2--1} transition ($T_{\rm ex}$\,=\,$T_{\rm b}$). If the population distribution of the rotational levels is thermalised by collisions ($T_{\rm rot}$\,=\,$T_{\rm ex}$\,=\,$T_{\rm b}$), which considering the high densities estimated below also appears to be the case throughout the circumstellar medium of HD\,101584, this will also give a good (but averaged over the beam) estimate of the gas kinetic temperature ($T_{\rm k}$\,=\,$T_{\rm b}$). 

We will use the C$^{18}$O(\mbox{2--1}) line data to estimate column densities assuming that this line is optically thin (which appears to be the case in most regions). These estimates are converted to H$_2$ column densities assuming that the CO/C$^{18}$O ratio reflects the solar O/$^{18}$O ratio of 480 \citep{scotetal06}, and that the fractional CO abundance with respect to H$_2$, $f_{\rm CO}$, is the one expected for O-rich AGB CSEs, 4$\times$10$^{-4}$ (close to full association of carbon into CO assuming solar abundances, and close to the value that fits the e-EVS data, see Sect.~\ref{s:physics_evs}), that is, $f_{\rm C^{18}O}$\,=\,8$\times$10$^{-7}$. The justification for using a solar value for the O/$^{18}$O ratio is presented in Sect.~\ref{s:evol_status}.

In the same way, a simple estimate of the gas mass is obtained using the equation,
\begin{equation}
\label{e:gasmass}
M_{\rm g} = \frac{16\pi m_{\rm H}}{hc g_{\rm u} A_{\rm ul} f_{\rm C^{18}O}}\,
     I_{\rm C^{18}O}\, D^2\, Q(T_{\rm rot})\, e^{E_{\rm u}/kT_{\rm rot}}\, ,
\end{equation}
where the usual symbols are used for the constants, and $I_{\rm C^{18}O}$ is the C$^{18}$O(\mbox{2--1}) flux density integrated over a velocity range and an area, $Q$ the partition function, and $E_{\rm u}$ the energy of the upper level. 
%
%
%
%
\subsubsection{The CCS}
\label{s:physics_ccs}

The molecular line emissions come from a region $\approx$\,0\farcs15 in diameter (corresponding to $\approx$\,150\,[$D$/1\,kpc]\,au), Table~\ref{t:obs_ccs}. An estimate of the gas temperature can be obtained from the brightness temperature in the CO(\mbox{2--1}) line, $\approx$\,160\,K. As argued above, this is likely close to the kinetic temperature. This gas temperature estimate is very comparable to the equilibrium temperature of low-albedo dust at a distance of 75\,[$D$/1\,kpc]\,au from a star with the adopted HD\,101584 characteristics, $\approx$\,200\,K.

An estimate of the H$_2$ column density is obtained using the C$^{18}$O(\mbox{2--1}) data. The strength of the ALMA line and assuming a source size of 0\farcs15 and an excitation temperature of 160\,K lead to a source-averaged C$^{18}$O column density of $\approx$\,10$^{18}$\,cm$^{-2}$ (corresponds to a, source-averaged, optical depth in this line of $\approx$\,1, that is, there is some uncertainty in this estimate due to opacity). Assuming $f_{\rm C^{18}O}$\,=\,8$\times$10$^{-7}$ we find a source-averaged H$_2$ column density of $\approx$\,10$^{24}$\,cm$^{-2}$. We use also the C$^{18}$O(\mbox{2--1}) data and Eq.~(\ref{e:gasmass}) to estimate the mass. The observed C$^{18}$O(\mbox{2--1}) line intensity, its fractional abundance, and the estimated gas temperature result in a gas mass of $\approx$\,0.029\,[$D$/1\,kpc]$^2$\,$M_\odot$. Both the H$_2$ column density and the mass are likely lower limits considering the opacity of the C$^{18}$O(\mbox{2--1}) line. A crude lower limit to the density can be obtained by making the reasonable assumption that the CCS has an extent along the line of sight that is not larger than its extent in the plane of the sky. With the estimated H$_2$ column density this points to an H$_2$ density in excess of 10$^9$\,[1\,kpc/$D$]\,cm$^{-3}$. This is a very high density meaning that for all molecules observed the excitation is collisionally dominated and the lines are thermalized.

The most likely interpretation of the CCS component is that of a circumbinary disk, presumably in slow rotation. Unfortunately, the spatial and velocity resolutions of our data are not high enough to allow a determination of the detailed kinematics of this component. 
%
%
%
%
\subsubsection{The EDE}
\label{s:physics_ede}

An estimate of the gas column density and mass can be obtained in the same way as for the CCS using the C$^{18}$O(\mbox{2--1}) data. The obtained CO(\mbox{2--1}) line brightness temperature in the EDE area is $\approx$\,50\,K. The strength of the ALMA C$^{18}$O(\mbox{2--1}) line and assuming a source size of 3\arcsec\ combined with an excitation temperature of 50\,K leads to a source-averaged C$^{18}$O column density of $\approx$\,2$\times$10$^{17}$\,cm$^{-2}$ (corresponds to a, source-averaged, optical depth in this line of $\approx$\,0.3, that is, there is some uncertainty in this estimate due to opacity). With the same assumptions on C$^{18}$O fractional abundance ratio as for the CCS, we estimate a source-averaged H$_2$ column density of $\approx$\,2$\times$10$^{23}$\,cm$^{-2}$. The gas mass is estimated using Eq.~(\ref{e:gasmass}), the C$^{18}$O(\mbox{2--1}) line intensity, its fractional abundance, and the estimated gas temperature. The result is 0.24\,[$D$/1\,kpc]$^2$\,$M_\odot$, that is, a significant fraction of the mass of the circumstellar medium around HD\,101584 lies in this component, since the masses of the CCS (above) and HGS and HVO (below) components are lower. Making the same assumption on the geometry of the EDE as for the CCS, we get a lower limit to the H$_2$ density of $\approx$\,10$^7$\,[1\,kpc/$D$]\,cm$^{-3}$, that is, much lower than the lower limit for the CCS, but still a high-density region. Also here, all the observed molecular species are expected to be effectively excited by collisions.

The EDE component, probably a disk or a torus, contains most of the circumstellar mass, and it is in expansion. The connection between the CCS and EDE components, if any, is not clear.  
%
%
%
%
\subsubsection{The HGS and HVO}

The HGS and HVO components are only clearly seen in the CO line data (except for SiO in the bright spots of the HVO). As for the CCS and EDE, the gas temperature is estimated using the CO brightness temperatures in these regions. The results are brightness temperatures of $\approx$\,25\,K and $\approx$\,50\,K in the HGS and HVO (for the latter this is estimated in the e-EVS). The observed CO/$^{13}$CO \mbox{2--1} line intensity ratios are about 1.5 and 3 in the HGS and HVO (as estimated for the e-EVS). Taking the estimated CO/$^{13}$CO abundance ratio of $\approx$13 \citep{olofetal17} into account, we conclude that the CO optical depths are high in both regions, and higher in the HGS than in the HVO, and that the brightness temperatures are good estimates of the CO excitation temperature, and presumably the kinetic temperature. The C$^{18}$O(\mbox{2--1}) integrated intensities are 8.3\,Jy\,km\,s$^{-1}$ for the HGS (integrated over the velocity range 10\,$\le$\,$\mid$\,$\upsilon-\upsilon_{\rm sys}$\,$\mid$\,$\le$\,45\,km\,s$^{-1}$) and 1.5\,Jy\,km\,s$^{-1}$ for the HVO ($\mid$\,$\upsilon-\upsilon_{\rm sys}$\,$\mid$\,$\ge$\,45\,km\,s$^{-1}$). Using this, the adopted C$^{18}$O fractional abundance, and the estimated gas temperatures, the resulting masses are 0.12\,[$D$/1\,kpc]$^2$ and 0.030\,[$D$/1\,kpc]$^2$\,$M_\odot$ for the HGS and HVO, respectively. A separate mass estimate for the e-EVS is given below.  
%
%
%
%
\subsubsection{The EVSs}
\label{s:physics_evs}

In \citet{olofetal17} we introduced a simple physical model for an EVS in order to derive molecular abundances through a radiative transfer analysis, noting that this is the only component for which we can identify emission from many CO transitions, hence providing observational constraints on the excitation. The observational data give limited information on the geometry except that the emission is largely confined to a region of diameter $\la$\,1\arcsec . We will use the same method, although simplified to only one component, and assumptions here.

Here we present the results of the CO line modelling, the result for the other molecules will be presented in a paper focussing on the circumstellar chemistry of HD\,101584. The radiative transfer is solved using an Accelerated-Lambda-Iteration code, taking into account excitation through collisions with H$_2$. Collisional coefficients for CO is taken from \citet{yangetal10}. Radiative excitation due to central star light (too distant) and dust emission inside the clump (too low optical depth) can be ignored. The CO line intensities for the e-EVS were used since they are slightly stronger than those of the western EVS, Table~\ref{t:obs_evs}. We assume a spherical, homogenous, iso-thermal clump of radius 0\farcs5. 

The density and kinetic temperature are reasonably well-constrained by the observed CO line intensities. The best fit is obtained for $n_{\rm H_2}$\,=\,(5$\pm$2)$\times$10$^5$\,cm$^{-3}$ and $T_{\rm k}$\,=\,60$\pm$10\,K, that is, consistent with thermal excitation, high optical depths, and the CO(\mbox{2--1}) brightness temperature estimate above. Hence, the gas mass of the e-EVS becomes $\approx$\,10$^{-3}$\,[$D$/1\,kpc]$^2$\,M$_\odot$. 

The analysis also gives a best-fit CO abundance of (7$\pm$2)$\times$10$^{-4}$, that is, close to the level expected in an O-rich circumstellar gas (full association of CO and solar values for O and C results in a fractional CO abundance of 5$\times$10$^{-4}$). The resulting CO/$^{13}$CO ratio is 14$\pm$6, that is, in very good agreement with the value 13$\pm$6 estimated from C$^{17}$O and $^{13}$C$^{17}$O line emission from the CCS by \citet{olofetal17}. An important conclusion from this is that the e-EVS material is dominated by circumstellar gas (possibly swept-up from a previous wind), and not by swept-up interstellar material. The C$^{16}$O/C$^{18}$O ratio of 225$\pm$75 is somewhat low compared to the solar value of 480 \citep{scotetal06}. Low- and intermediate-mass stars are expected to destroy rather than produce $^{18}$O, but considering our simple model, the uncertainty in identifying emission from the EVS, and the use of a single C$^{18}$O line, we draw no conclusions based on this result.
%
%
%
%
\subsubsection{Summary of gas properties}
\label{s:mass_summary}

We summarize our findings on the densities, temperatures, and masses of the identified gas components in Table~\ref{t:phys_char}. We conclude that the estimated total gas mass of the circumstellar material around HD\,101584 is $\approx$\,0.42\,[$D$/1\,kpc]$^2$\,$M_\odot$. Finally, we note that, due to missing flux in the velocity range of the HGS component, there exists gas whose contribution to the total gas mass is not known, and the estimate of the latter should therefore be seen as a lower limit. If we assume that the characteristics of this gas is similar to that of the HGS component, we estimate the mass of this gas to be $\approx$\,0.12\,[$D$/1\,kpc]$^2$\,$M_\odot$, that is, a total circumstellar gas mass of $\approx$\,0.5\,[$D$/1\,kpc]$^2$\,$M_\odot$.

\begin{table}
\caption{Physical characteristics of identified components}
\centering
\begin{tabular}{l c c c c c}
\hline \hline
Component       & Size           & $n_{\rm H_2}$    &  $M_{\rm g}$    &    $T_{\rm k}$    &  $\Delta \upsilon$\,$^1$  \\
                & [au]      &  [cm$^{-3}$]     &  [$M_\odot$]    &    [K]            &  [km\,s$^{-1}$] \\
\hline \hline
\multicolumn{6}{c}{Low-$L_\ast$ case (500\,$L_\odot$, 0.56\,kpc):}\\
\hline
CCS             & \phantom{000}80  & $>$10$^9$           &  0.01\phantom{00} &    160            &  \phantom{00}3 \\
EDE             & \phantom{0}1700  & $>$10$^7$           &  0.08\phantom{00} &    \phantom{0}50  &  \phantom{0}20 \\
HGS             & \ldots           & \ldots              &  0.04\phantom{00} &    \phantom{0}25  &  \phantom{0}80\\
HVO             & 27000            & \ldots              &  0.01\phantom{00} &    \phantom{0}50  &  150 \\
EVS\,$^2$       & \phantom{00}560  & \phantom{0}10$^6$   &  0.0002             &   \phantom{0}60             &   \phantom{00}8 \\
\hline \hline
\multicolumn{6}{c}{High-$L_\ast$ case (5000\,$L_\odot$, 1.8\,kpc):}\\
\hline
CCS             & \phantom{00}270  & $>$10$^{9}$         &  0.1\phantom{00} &    160            &  \phantom{00}3 \\
EDE             & \phantom{0}5400  & $>$10$^7$           &  0.8\phantom{00} &    \phantom{0}50  &  \phantom{0}20 \\
HGS             & \ldots           & \ldots               &  0.4\phantom{00} &    \phantom{0}25      &  \phantom{0}80\\
HVO             & 87000            & \ldots              &  0.1\phantom{00} &    \phantom{0}50      &  150 \\
EVS\,$^2$       & \phantom{0}1800  & \phantom{0}10$^6$       &  0.005      &   \phantom{0}60             &   \phantom{00}8 \\
\hline
\end{tabular}
\label{t:phys_char}
\tablefoot{(1) FWHMs of Gaussian fits for CCS and EVS, FWZPs for EDE and HGS, and maximum expansion velocity for the HVO. (2) Using estimates for the e-EVS, see Sect.~\ref{s:physics_evs}.}
\end{table}
%
%
%
%
%
\subsection{The energetics of the outflowing material}
\label{s:energy}

The energetics, that is, the energy, scalar momentum, and scalar momentum rate, of the outflowing material is of great interest in connection with discussing the driving mechanism and the energy source. We repeat here essentially the analysis in \citep{olofetal15} since it is based on the same C$^{18}$O(\mbox{2--1}) data. The differences are that we now apply this analysis to the EDE and HGS components in addition to the HVO component, since it is likely that the accelerations of these gas components have the same origin, and using the excitation temperatures estimated for these regions separately. We still use the line-of-sight velocities, so in this sense the estimates should be regarded as lower limits. The correction for the inclination angle will be largest for the EDE component, but its contributions to the energy and momentum are limited.

The results are that the kinetic energy of the accelerated gas is 7.2$\times$10$^{45}$\,[$D$/1\,kpc]$^2$\,erg (360\,[$D$/1\,kpc]$^2$\,$M_\odot$\,km$^2$\,s$^{-2}$) and its scalar momentum is 1.8$\times$10$^{39}$\,[$D$/1\,kpc]$^2$\,g\,cm\,s$^{-1}$ (9.0\,[$D$/1\,kpc]$^2$\,$M_\odot$\,km\,s$^{-1}$). Should this momentum be supplied by radiation, the corresponding time scale [momentum/($L_\ast/c$)] is close to 3$\times$10$^5$\,yr which is of course unreasonably long in the case of HD\,101584. Another acceleration mechanism must be at work. The corresponding values for the HVO component only is 6.4$\times$10$^{45}$\,[$D$/1\,kpc]$^2$\,erg for the energy and 1.0$\times$10$^{39}$\,[$D$/1\,kpc]$^2$\,g\,cm\,s$^{-1}$ for the scalar momentum.
%
%
%
%
\section{Quantitative estimates: dust}
\label{s:quant_est_dust}

For the analysis of the dust emission the situation is better in the sense that we have the full spectral energy distribution (SED) at our disposal. However, the analysis is still plagued by the uncertainty imposed by the geometry and the additional lack of knowledge of the characteristics and distribution of different dust types. We start by looking at the dust distribution, as traced by the 1.3\,mm continuum, and its relation to the different components identified in the molecular line data.
%
%
%
%
\subsection{The CCS}

The CCS is estimated to have a 1.3\,mm continuum flux density of 7\,mJy coming from a source of size $\la$\,0\farcs01 (corresponding to $\la$\,10\,[$D$/1\,kpc]\,au). Since this component is not (or just barely) resolved, it is, in principle, possible that this flux is coming from the central star. However, using the adopted values for HD\,101584 (8500\,K and $L_\ast/D^2$\,=\,1600\,$L_\odot$/kpc$^2$) its flux density at 1.3\,mm is estimated to be only about 0.01\,mJy (the flux scales as $L_\ast/D^2$, hence the same result for the low- and high-$L_\ast$ cases). It is in principle possible that also free-free emission from a region surrounding the warm star contributes to the flux, but it remains for the future to determine the properties of the immediate circumstellar surroundings of HD\,101584. 

Another possible explanation is that the emission is coming from heated dust in a disk, presumably the innermost warmest part of the region responsible for the narrow-line-width molecular line emission. The equilibrium temperature of low-albedo dust at a distance of 5\,[$D$/1\,kpc]\,au from a star with the adopted HD\,101584 characteristics is $\approx$\,800\,K. The blackbody emission at 1.3\,mm from an optically thick dust disk of these characteristics is $\approx$\,3\,mJy. Thus, this provides a possible explanation, but it requires a dust optical depth close to one (or higher) at 1.3\,mm. Adopting a dust opacity of 0.5\,cm$^2$\,g$^{-1}$ [\citet{liseetal15}; the average of the five values listed in their Table~6] such a disk become optically thick at a dust mass of about 2$\times$10$^{-5}$\,[$D$/1\,kpc]$^2$\,$M_\odot$ (corresponding to a gas mass of about 0.004\,[$D$/1\,kpc]$^2$\,$M_\odot$, assuming a ``canonical'' gas-to-dust-mass ratio of 200). This value is not unreasonable. It can be further noted that the 0\farcs028 disk with a brightness temperature of 650\,K at 10.7\,$\mu$m observed by \citet{hilletal17}  will have a 1.3\,mm continuum flux of 7\,mJy (if optically thick at 1.3\,mm and assuming that the brightness temperature equals the dust temperature). That is, it is very likely that this disk and the central 1.3\,mm continuum source are two aspects of the same object. As noted in Sect.~\ref{s:obs_ccs} increasing the aperture to 0\farcs3 increases the flux by only $\approx$\,70\,\%, so the continuum flux from the central region is dominated by the very central part.
%
%
%
%
\subsection{The EDE}

The 1.3\,mm continuum emission from HD\,101584 is dominated by emission from the EDE, $\approx$\,120\,mJy within an aperture of 3\arcsec , Fig.~\ref{f:morphology} and  Table~\ref{t:obs_ede}. This is likely also the emission that dominates at the far-IR wavelengths of the SED. An order of magnitude estimate of the dust mass can be obtained assuming optically thin dust emission,
\begin{equation}
\label{e:dustmass}
M_{\rm d} = \frac{S_\nu D^2}{\kappa_\nu B_\nu}\,\, ,
\end{equation}
where $S_{\nu}$ is the flux density at the frequency $\nu$ and within a given aperture, and $B_{\nu}$ and $\kappa_{\nu}$ the blackbody brightness and the dust opacity at this frequency, respectively. Assuming a dust temperature of 50\,K (the estimated gas temperature of the EDE component), and an opacity of 0.5\,cm$^2$\,g$^{-1}$ result in $M_{\rm d}$\,$\approx$\,0.014\,[$D$/1\,kpc]$^2$\,$M_\odot$ (the dust optical depth at 1.3\,mm of a uniform disk of this size is well below one). This is not unreasonable, and consequently the EDE will be an important component of the SED modelling below. 

%
%
%
%
\subsection{The SED}
\label{s:sed}

\begin{figure}
\centering
    \includegraphics[width=9cm]{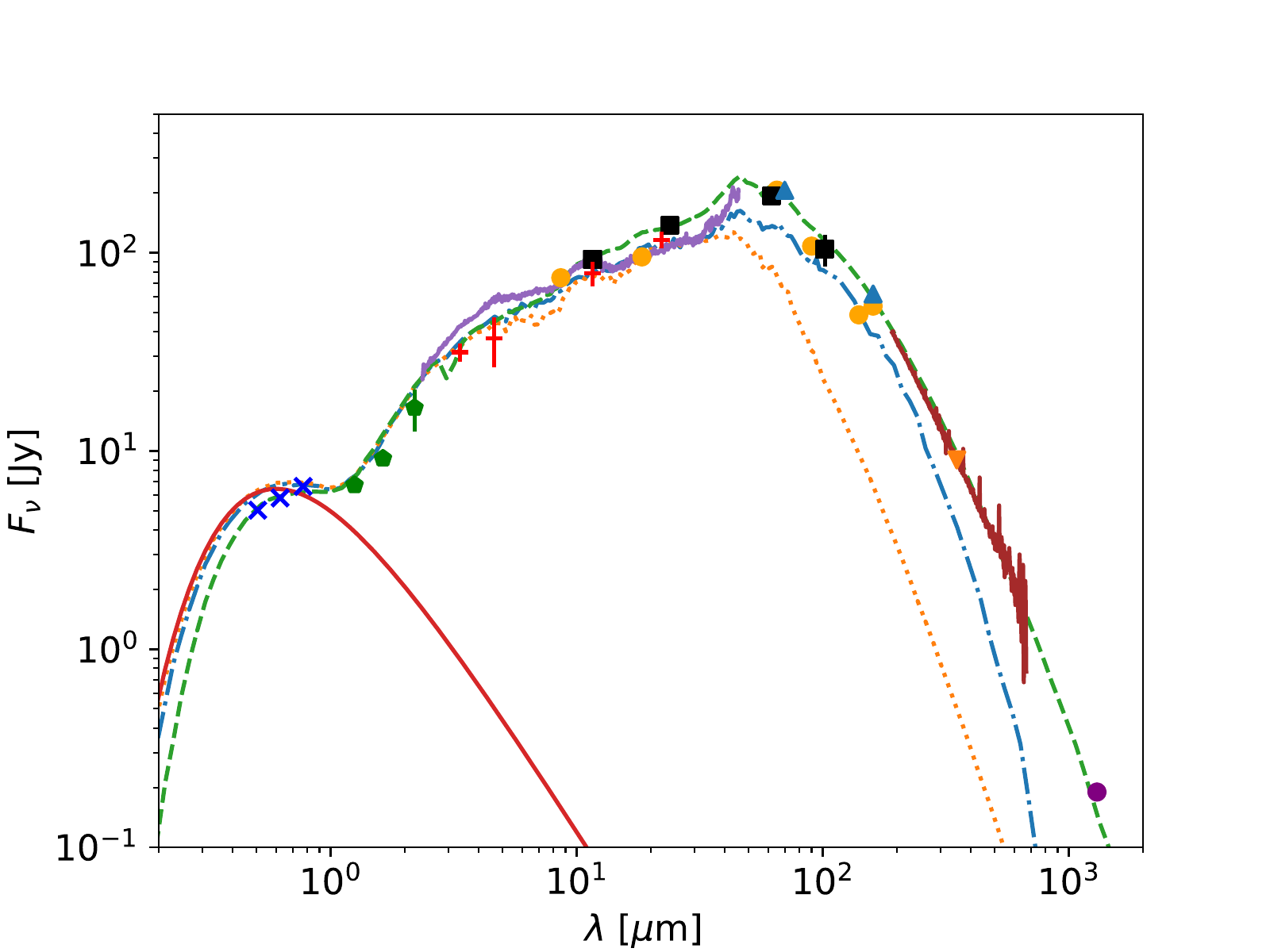}
    \caption{Observed SED of HD\,101584, and a comparison with the model results. ISO SWS and Herschel/SPIRE spectra are shown by the full violet and brown lines, respectively (note the presence of high-$J$ CO lines in the SPIRE spectrum). Photometric measurements are from Gaia (blue x:s), 2MASS (green pentagons), AKARI (orange circles), IRAS (black squares), WISE (red crosses), ALMA (purple circle), PACS (blue triangles), and ArTeMiS (orange inverted triangle). The red solid line represents a black-body at a temperature of 8500\,K, and for a distance of 1\,kpc the luminosity is 1600\,$L_\odot$. We show models with $a_{\rm max}$\,=\,1\,mm ($M_{\rm d}$\,=\,10$^{-2}$\,$M_\odot$, green dashed line), $a_{\rm max}$\,=\,0.1\,mm ($M_{\rm d}$\,=\,4$\times$10$^{-3}$\,$M_\odot$, blue dash-dotted line), and $a_{\rm max}$\,=\,0.01\,mm ($M_{\rm d}$\,=\,10$^{-3}$\,$M_\odot$, orange dotted line).}
    \label{f:sed}
\end{figure}

The SED of HD\,101584 was constructed using photometric and spectroscopic measurements available from astronomical catalogues, Fig.~\ref{f:sed}. The ISO short-wavelength spectrometer \citep{degretal96} data were obtained from the NASA/IPAC Infrared science archive, and the Herschel/SPIRE \citep{grifetal10} data from the level 2 product in the Herschel Space Observatory archive. Photometric measurements were obtained from the point-source catalogue of the IRAS satellite \citep{neugetal84}, the WISE All-Sky Data Release \citep{cutretal12}, the 2MASS All-Sky Catalog of Point Sources \citep{cutretal03}, the MESS program \citep{groeetal11} for Herschel/PACS \citep{pogletal10}, the AKARI Infrared Camera Mid-IR All-sky Survey \citep{ishietal10}, the AKARI Far-infrared Surveyor \citep[from catalogue;][]{kawadetal07}, and the Gaia second data release \citep{gaiaetal18}. Finally, we have the results of our ArTeMiS and ALMA observations. There are several notable features: the very strong far-IR excess, the presence of a 10\,$\mu$m feature, strong features around 45\,$\mu$m [identified as due to crystalline water, \citet{hoogetal02}], and the presence of high-$J$ CO lines in the SPIRE data.

%
%
%
\subsection{SED modelling}
\label{s:sed_fit}

\begin{figure}
\centering
    \includegraphics[width=7cm]{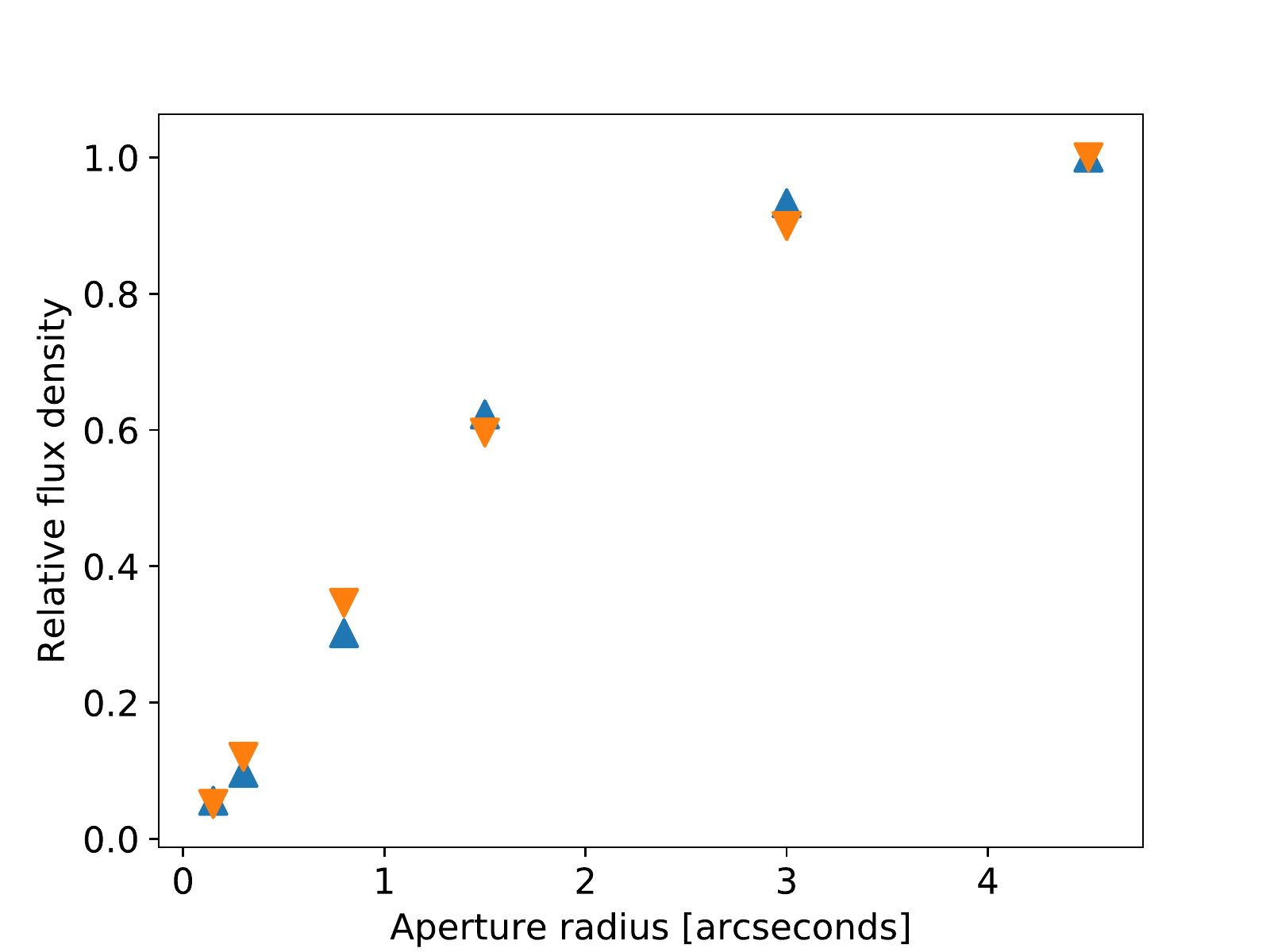}
    \caption{Radial profile of the 1.3\,mm continuum emission observed using ALMA (blue triangles) and the results of the best-fit model to the SED (orange inverted triangles).}
    \label{f:cont_radial}
\end{figure}

The ALMA 1.3\,mm continuum emission is dominated by the EDE component, and there is also a contribution from extended, diffuse emission. It is reasonable to assume that the optically thin emission at 1.3\,mm is a good measure of the dust distribution, that is, we will model the SED assuming that the dust is located in a disk embedded in a spherically symmetric envelope. The dust density distribution is given by a simplified version of the analytical expression provided by \cite{meixetal02}, with their $C$, $D$, and $E$ parameters set to 0, 0, and 1, respectively,
\begin{equation}
\rho(r,\theta) = \rho_{\rm in} \left( \frac{r}{R_{\rm in}}\right)^{-B} \times \left\{ 1+A(1-cos\theta)^F\left[\frac{e^{(-r/R_{\rm disk})}}{e^{(-R_{\rm in}/R_{\rm disk})}}\right]\right\}.
\end{equation}
The dust density, both in the equatorial and the polar directions (defined by the radius, $r$, and the polar angle, $\theta$), decreases radially following a power law. At the disks outer edge, the density in the equator decreases exponentially until it reaches the lower values of the density in the polar direction. The free parameters that define the density distribution are the inner radius of the envelope, $R_{\rm in}$, the exponent of the radial profile, $B$, the equatorial density enhancement in the disk, $1+A$, the parameter that controls the shape (mainly opening angle) of the disk, $F$, and the disk radius, $R_{\rm disk}$. The total dust mass, $M_{\rm d}$, is used to calculate the overall density and define the density at the inner radius in the equatorial direction, $\rho_{\rm in}$. The outer radius of the envelope, $R_{\rm out}$, is an input parameter that we set to 3000\,[$D$/1\,kpc]\,au. We consider an inclination angle of 10$^\circ$ for the system, as determined from the analysis of the molecular lines, and a blackbody stellar spectrum with an effective temperature of 8500\,K. For the continuum radiative transfer we used the state-of-the-art code \texttt{MCMax} \citep{minetal09} in its axisymmetric mode.

The dust opacity was calculated using the Mie approximation and optical constants for astronomical silicate dust \citep{osseetal92} and water ice \citep{warr84}. Astronomical silicates were used to simplify the fitting procedure, since a detailed study of the dust composition using optical constants for different species measured in the laboratory is out of the scope of this study. The water ice abundance by mass relative to the total amount of dust is $f_{\rm ice}$. We considered a size distribution of the dust grains between a minimum size, $a_{\rm min}$ (0.01\,$\mu$m), and a maximum size, $a_{\rm max}$, which we vary to fit the observations. A value of $-3.5$ was used for the exponent of the size distribution, $p$, but also the effects of making the distribution steeper or shallower were explored as discussed below. The different grain sizes and the two species are all in thermal contact, and, hence, have a single temperature at a given radius and polar angle.

We apply no correction for interstellar extinction for two reasons. First, the total Galactic reddening, $E(B-V)$, in the direction of HD\,101584 (Galactic coordinates: $\ell$\,=\,293$^\circ$ and $b$\,=\,6$^\circ$) is 0.25$^{\rm m}$ (using the estimator in the NASA/IPAC Infrared Science Archive), while the distance to the source is only of the order 1\,kpc. Second, most of the radiated energy emerges at near- to far-IR wavelengths where the extinction is negligible.
%
%
%
%
\subsection{SED model results}
\label{s:sed_results}

The comparison between the model results and the data was carried out by eye. We obtain good fits using models with disks that are optically thick at visual and near-infrared wavelengths in the equatorial direction. This causes the source to appear much more luminous when viewed pole-on than edge-on. For HD\,101584, with an inferred low inclination, this implies a lower luminosity for a given distance than previously presented in the literature. For instance, for a distance of 0.7\,kpc, we find a luminosity of $\approx$\,750\,$L_\odot$, which is significantly smaller than the value of 5000\,$L_\odot$ considered by \citet{bakketal96a}. The value of $A$ affects the circumstellar extinction along the line of sight and the relative flux density between visible and far-IR wavelengths. By decreasing $A$, and increasing the opacity along the light of sight, the stellar luminosity can be made larger, but we were unable to obtain values significantly larger than $\approx$\,1000\,$L_\odot$ for a distance of 0.7\,kpc in the context of our models. The circumstellar dust optical depth in the visual and in the direction of HD\,101584 is 0.3 for the best-fit model.

We are able to reproduce the compact, strong continuum emission at 1.3\,mm observed using ALMA, and also the Herschel/SPIRE data, only if we consider a grain-size distribution including relatively large grains, $\sim$\,1\,mm in size, Fig.~\ref{f:sed}). Interestingly, we find that the water ice grains must be kept smaller in order to reproduce the strength of the 45\,$\mu$m feature. Therefore, we have used a water ice opacity calculated for small grains, sizes $\lesssim$\,10\,$\mu$m. The value of $B$ is well constrained by fitting the radial profile of the ALMA 1.3\,mm continuum image, Fig.~\ref{f:cont_radial}, while the other parameters do not have a strong effect on it. The values of $A$ and $F$ are constrained from fitting the SED and we estimate that they are uncertain by a factor of a few. The estimated dust mass is $\approx$\,0.01\,[$D$/1\,kpc]$^2$\,$M_\odot$. A summary of the best-fit model is given in Table~\ref{t:sed_fit}. A comparison between the radial brightness distribution of the $p$-H$_2$S(\mbox{$2_{20}-2_{11}$}) line and the dust density distribution suggests that the inner region, except for the CCS component, is dust-dominated.

\begin{table}
\caption[]{Derived parameters for the dust envelope of HD\,101584\,$^1$.}
\begin{tabular}{lcl}
\hline \hline
Parameter                  & Preferred value     & Constraint\\
\hline
$R_\star$ [$R_\odot$]      & 18.6                & SED ($L_\star$\,=\,1600\,$L_\odot$) \\
$R_{\rm in}$ [au]          & 2.15                & Near-IR interferometry\\ 
$A$                        & 750                 & SED\\ 
$B$                        & 1.2                 & ALMA B6 image\\ 
$F$                        & 5.0                 & SED \\ 
$R_{\rm disk}$ [au]        & 1500                & ALMA B6 image \\ 
$R_{\rm out}$ [au]         & 3000                & ALMA B6 image\\ 
$M_{\rm d}$ [$M_\odot$]    & 0.010               & SED \\ 
$a_{\rm min}$ [$\mu$m]     & $\sim$\,10$^{-2}$   & Not well constrained \\ 
$a_{\rm max}$ [$\mu$m]     & $\sim$\,10$^3$      & SED + ALMA B6 image \\ 
$p$                        & -3.5                & SED \\
$f_{\rm ice}$              & 7\,\%               & Water ice feature in the far-IR \\
\hline
\end{tabular}
\label{t:sed_fit}
\tablefoot{(1) Values for a distance of 1\,kpc.}
\end{table}

The dust mass we derive is relatively large compared to the gas mass obtained from the observation of molecular lines, $M_{\rm d}$\,$\approx$\,0.01\,$M_\odot$ and $M_{\rm g}$\,$\approx$\,0.5\,$M_\odot$, respectively, at a distance of 1\,kpc, that is, a gas-to-dust-mass ratio of 50 (distance-independent estimate). This is mainly caused by the large amount of mass in large grains. We attempted to decrease the dust mass required by our model by changing the exponent of the size distribution of the dust grains and by adding metallic iron grains \citep[using optical constants from][]{ordaetal88} to the dust mix but without success. Hollow or fluffy grains may decrease the dust mass somewhat, but not substantially. We also managed to get a good fit to the SED using small, pure iron grains in combination with water ice. The total dust mass remains the same, but it requires an excessive amount of iron. Nevertheless, the uncertainties in the gas and dust mass estimates, due to many simplifying assumptions, are such that a more physically reasonable gas-to-dust mass ratio lies within their limits.

A simple, but independent of any assumptions on the dust density distribution, estimate of the dust mass can be obtained using Eq.~(\ref{e:dustmass}). A flux density of 202\,mJy at 1.3\,mm, a dust temperature of 50\,K (the estimated gas temperature of the EDE component), and an opacity of 0.5\,cm$^2$\,g$^{-1}$ result in a dust mass of 0.02\,$M_\odot$ at a distance of 1\,kpc, only a factor of two different from the more sophisticated estimate above.
%
%
%
%

\section{Discussion}

As discussed already in Sect.~\ref{s:hd_distance} the evolutionary status of HD\,101584 remains unclear. We will here provide arguments that favour an interpretation in the form of a post-RGB object. We will also compare the characteristics of HD\,101584 with those of seemingly similar objects, that is, objects with recent, energetic, bipolar outflows and early-spectral-type central stars, as well as discuss to what extent premature termination of the red giant evolution is a common phenomenon or not. 
%
%
%
%
\subsection{Evolutionary status}
\label{s:evol_status}

The estimated circumstellar CO isotopologue ratios \citep{olofetal17}, and the reasonable assumption that in this case  they directly reflect the C and O isotope ratios, give crucial information on the characteristics and evolutionary status of HD\,101584. The low $^{12}$C/$^{13}$C ratio, $\approx$\,13, provides strong evidence of CNO-processing and hence a location on or beyond the RGB \citep{tsuj07}. A classification as a RSG or a YSO is highly unlikely, as it is only during red giant evolution that such nucleo-processed material is brought to the surface of the star. The high effective temperature means that it must be beyond the RGB or the AGB. Furthermore, the essentially solar $^{17}$O/$^{18}$O ratio, $\approx$\,0.2, is a strong indication of a low-mass star, $\la$\,1\,$M_\odot$ \citep{denuetal17}. For more massive stars the ratio will go up, and for stars going though hot-bottom-burning on the AGB the ratio is expected to be very high [for example, $>$\,10 for a sample of OH/IR stars, \citet{justetal15}]. The fact that HD\,101584 shows a (close to) solar abundance of N is not at odds with this, since the abundance of a lower-abundance element (like $^{13}$C) changes (in a relative sense) more easily than that of a higher-abundance element (like N). Likewise, the abundances of $^{17}$O and $^{18}$O are not affected by CNO-processing for a low-mass star \citep{denuetal17}. This is also the justification for adopting a solar value for the O/$^{18}$O ratio, and hence a solar CO/C$^{18}$O ratio, in the circumstellar mass estimates.

Beyond this, it is, however, difficult to make a firm conclusion. At the time that \citet{partpott86} and \citet{bakketal96a} studied this object, a post-AGB identification seemed the most reasonable. However, the detections of post-RGB objects in the Magellanic Clouds with many characteristics similar to those of post-AGB objects \citep{kamaetal16} have opened up the possibility of an alternative identification. We will discuss these two possibilities in the light of our findings.

In the absence of a reliable distance estimate, in Sect.~\ref{s:hd_distance} we introduced two characteristics of the primary star that are examples of a post-RGB ($L_\ast$\,=\,500\,$L_\odot$, $D$\,=\,0.56\,kpc) and a post-AGB ($L_\ast$\,=\,5000\,$L_\odot$, $D$\,=\,1.8\,kpc) object, both having an effective temperature of 8500\,K. The distance dependence of the circumstellar mass estimate means that the circumstellar mass, and hence the ejected mass from the primary in our scenario, will be very different in the two cases. This will make one identification more likely than the other, although it is too early to make a firm statement on this.

In the low-$L_\ast$ case we use a $M_{\rm c}$--$L_\ast$ ($M_{\rm c}$ is the core mass) relation for RGB stars \citep{bootsack88} to estimate a present stellar mass of 0.36\,$M_\odot$ (the high surface temperature makes it likely that only the core of the star remains). The circumstellar mass is estimated to be 0.18\,$M_\odot$ (where we have made a correction also for the missing flux as discussed in Sect.~\ref{s:mass_summary}), that is, an initial stellar mass of $\approx$\,0.54\,$M_\odot$. The estimated dust mass and a canonical gas-to-dust mass ratio of 200 would raise the circumstellar mass to $\approx$\,0.6\,$M_\odot$ and the stellar mass to $\approx$\,1\,$M_\odot$. Thus, in the low-$L_\ast$ case we have a low-mass star in the mass range \mbox{0.5--1}\,$M_\odot$. This is fully compatible with the estimated low $^{17}$O/$^{18}$O ratio. 

In the high-$L_\ast$ case we obtain a present stellar mass of 0.55\,$M_\odot$ using instead an $M_{\rm c}$--$L_\ast$ relation for AGB stars \citep{bootsack88}. The circumstellar mass is estimated to be 1.8\,$M_\odot$, hence an initial stellar mass of $\approx$\,2.4\,$M_\odot$. The estimated dust mass and a canonical gas-to-dust mass ratio of 200 would raise the circumstellar mass to $\approx$\,6.5\,$M_\odot$ and the stellar mass to $\approx$\,7\,$M_\odot$, that is, close to, or maybe even above, the upper mass limit for an AGB star. Thus, in the high-$L_\ast$ case we have a intermediate-mass star in the mass range \mbox{2--7}\,$M_\odot$. This is not compatible with the estimated low $^{17}$O/$^{18}$O ratio, which suggests a low-mass star.

Taken together, these results favour a post-RGB scenario over a post-AGB scenario. As far as we understand, it is observationally very difficult, if not impossible, to distinguish between a post-RGB and a post-AGB star if the distance is not known. Circumstellar-wise there may be a difference. In the former case we expect no, or very little, remnant circumstellar material, while this is not necessarily the case for a post-AGB star where the AGB mass loss can be substantial. Unfortunately, it is difficult to estimate from our data whether or not a remnant CSE exists in the case of HD\,101584. It may be that complementary observations with the Atacama Compact Array (ACA) of ALMA can shed light on this.
%
%
%
%
\subsection{The energetics and common-envelope evolution}

The kinetic energy of the outflowing gas is estimated to be 2$\times$10$^{45}$\,erg and 2$\times$10$^{46}$\,erg in the low-$L_\ast$ and high-$L_\ast$ cases. Both values are very high, and it is remarked already in Sect.~\ref{s:energy} that the scalar momentum rate cannot, by a large margin, be supplied by the stellar radiation. It is difficult to reconcile such high kinetic energies with anything but a CE evolution scenario where gravitational binding energy is released when the companion is captured by, and falls towards, the primary. Even more energy may be released if material falls towards the companion, possibly forming an accretion disk.

Following the same procedure as in \citet{olofetal15} we estimate the orbital characteristics from the mass function (assuming circular orbit; ps = primary star, cs = companion star),
\begin{equation}
\frac{(M_{\rm cs} \sin i)^3}{(M_{\rm ps} + M_{\rm cs})^2} = \frac{4\pi^2 (a_{\rm ps} \sin i)^3}{G P^2} \,,
\end{equation}
and the expression for the semi-major axis of the primary star's orbit
\begin{equation}
a_{\rm ps} \sin i = \frac{K_{\rm ps}P}{2\pi} \,.
\end{equation}
where $K_{\rm ps}$ is the semi-amplitude of the velocity curve of the primary, $P$ the orbital period, and $G$ the gravitational constant. The semi-major axis of the companion's orbit is obtained from $M_{\rm ps}a_{\rm ps}$\,=\,$M_{\rm cs}a_{\rm cs}$. The released gravitational energy is obtained using
\begin{equation}
E_{\rm rel} = -\frac{G\,M_{\rm ps,i}\,M_{\rm cs}}{2a_{\rm i}} + \frac{G\,M_{\rm ps}\,M_{\rm cs}}{2a}\,, 
\end{equation}
where $a$\,=\,$a_{\rm ps}$\,+\,$a_{\rm cs}$, and $a_{\rm i}$ is the initial separation, and $M_{\rm ps,i}$ the initial mass of the primary star. 

We adopt here the orbit period estimated from the radial velocity data \citep{diazetal07}, $P$\,=\,144$^{\rm d}$, $K_{\rm ps}$\,=\,3\,km\,s$^{-1}$ \citep{bakketal96a}, and the inclination angle of 10$^\circ$ estimated in this paper. For the two estimated $M_{\rm ps}$ of 0.36 and 0.55\,$M_\odot$ we find companion masses of 0.27 and 0.41\,$M_\odot$, respectively, and a separations of 0.53\,au in both cases. The released gravitational energy depends on the initial separation, $a_{\rm i}$. We assume 1 and 4\,au in the post-RGB and post-AGB cases, respectively [the latter is larger because an AGB star is larger, especially during a thermal pulse; for a detailed study of the capture process, see \citet{madaetal16}]. The resulting released gravitational energies are 3$\times$10$^{44}$ and 2$\times$10$^{45}$\,{\rm erg} in the post-RGB and post-AGB cases, respectively. Consequently, the ratio between kinetic and released gravitational energy is about 10 in both cases. There is also energy released due to hydrogen recombination, but according to \citet{sokeetal18} it contributes little to removing the stellar envelope.

It must be emphasized that there are considerable uncertainties in both the estimated energy released when the companion spirals inwards and the estimated kinetic energy of the outflowing gas. Furthermore, the CE evolution scenario is complex with an uncertain energy transfer efficiency \citep{ivanetal13}. Nevertheless, the discrepancy between the released energy and kinetic energy estimates is so large that it must be concluded that energy released by the inward motion of the companion alone is not enough to explain the characteristics of HD\,101584. A further possibility is that additional gravitational energy is released as circumstellar material falls towards the companion. The effect of such an infall may be the formation of a circum-companion accretion disk.

In turn, this may provide an efficient mechanism for driving an outflow via a jet \citep{gorletal12, gorletal15}. \citet{blaclucc14} have estimated the required accretion rate to drive an outflow, of given characteristics, under such circumstances. Assuming that the accretion disk surrounds an 0.27\,$M_\odot$ MS-star in the RGB case, we derive a required accretion rate of 2\,$\times$\,10$^{-5}$\,($Q$/2)\,$M_\odot$\,yr$^{-1}$ for the outflowing gas of HD\,101584 ($Q$ is a numerical factor typically in the range 1 to 5 in jet models). The corresponding values for an 0.55\,$M_\odot$ MS-star or WD in the post-AGB case are 7\,$\times$\,10$^{-5}$\,($Q$/2) and 10$^{-5}$\,($Q$/2)\,$M_\odot$\,yr$^{-1}$, respectively. This is below the Eddington accretion rates in all cases, but well above what can be obtained from Bondi-Hoyle-Lyttleton accretion and wind-Roche-lobe overflow. Only accretion in connection with CE evolution, or possibly what is termed ``Red Rectangle Roche-lobe overflow'' in the MS-star case, will fit the requirements.
%
%
%
%
\subsection{Termination of red giant evolution}

There is good evidence that the red-giant evolution of HD\,101584 was prematurely ended by a CE evolution. Whether it happened on the RGB or the AGB remains open. If sufficiently common, premature termination of red giant evolution may have a significant effect on for example the elemental synthesis of AGB stars and their contribution to the integrated light of galaxies, since such estimates are based on results from single-star evolution models. 

Based on the probability of having a companion and the period distribution of main sequence stars \citep{raghetal10}, one can estimate how common binary systems, with the required characteristics to achieve CE evolution, are (it should be noted that the frequency of post-RGB binaries of this type will strongly affect the frequency of post-AGB binaries of this type). The estimate has been done for AGB stars. By considering only binaries with initial separations between that of a couple of maximum radii of an RGB star and the same for an AGB star, \citet{madaetal16} estimated that only a few per cent of the AGB stars can interact strongly, that is, go through CE evolution, with a companion. 

The observational value is substantially higher, but far from 100\,\%. Observationally-based constraints can be obtained from studies of PNe. Looking at purely statistical information, that is, disregarding whether or not close binaries are required to shape PNe, \citet{miszetal09} used PNe in the Galactic bulge to estimate a close binary fraction of 12--21\,\%. \citet{nieetal12} used a Monte Carlo simulation and the observed frequency of sequence E binaries in period-luminosity diagrams in the Large Magellanic Cloud as observational constraint. They found that the fraction of PNe with close binary central stars is 7--9\,\%, and the fraction having separations capable of influencing the nebula morphology (set as orbital periods less than $<$\,500\,yr) is 23--27\,\%. In summary, only about 1 in 5 PNe seem to have their origin in CE evolution. However, this may be a (significant) underestimate as argued by for example \citet{demaetal17}. Therefore the question whether or not a close companion is a pre-requisite for the formation of PNe appears to remain open. A related issue is that post-CE-evolution objects may for some reason dominate in observed samples, hence giving a distorted view of the characteristics of AGB stars and PNe.

Looking at the problem from the side of the PN morphology and kinematics, the evidence is strong that in those cases where the central star of a PNe has gone through CE evolution, there is also a link to the spatio-kinematic evolution of the nebula \citep{hillwetal16}, and some understanding how this works \citep{garcetal18}. This adds strength to the question whether the estimated low percentage of close-binary systems among PNe is wrong. Currently, there exist no viable way in which also single stars produce PNe with complex morphologies and kinematics.

If low-luminosity PNe have their origin in RGB stars, the CE-evolution process must be crucial for a successful result, since normal stellar mass loss on the RGB occurs at substantially lower rates than on the AGB and the evolution must be terminated before reaching the tip of the RGB and the start of He-burning. It remains to determine the fraction of this type of PNe that have their origin in RGB stars.
%
%
%
%
\subsection{Comparison with objects of similar characteristics}

\begin{table*}
\caption{Characteristics of selected evolved objects with extreme high-velocity outflows.}
\centering
\begin{tabular}{l c c c c c c c}
\hline \hline
Name                     & Spectral type & $^{12}$C/$^{13}$C    & $\upsilon_{\rm max}$  & Kinematical age & $M_{\rm HVO}$  &   $E_{\rm HVO}$             &  $P_{\rm HVO}$ \\
                         &               &                      & [km\,s$^{-1}$]        & [yr]            &[$M_\odot$]     &   [erg]                     &  [g\,cm\,s$^{-1}$] \\
\hline 
HD\,101584\,$^1$               & A6Ie          & 13\,\phantom{$^0$}   &  150\,\phantom{$^0$}  & 770             &  0.04\phantom{0}  & \phantom{$>$}\,6$\times$10$^{45}$ &  \phantom{$>$}\,1$\times$10$^{39}$ \\
IRAS\,08005--2356        & F5Ie          & 10\,\phantom{$^0$}   &  200\,$^2$            & 190             &  0.08\phantom{0}  & \phantom{$>$}\,3$\times$10$^{45}$ &  \phantom{$>$}\,3$\times$10$^{39}$\\
IRAS\,16342--3814\,$^3$  & obscured      & \phantom{0}\,3\,$^4$   &  250\,\phantom{$^0$}  & 110             &  0.006            & $>$\,3$\times$10$^{45}$           &  $>$\,2$\times$10$^{38}$\\
IRAS\,22036+5306         & F5\,$^5$      &\phantom{0}\,6\,\phantom{$^0$} &  250\,$^2$     & \phantom{0}25   &  0.03\phantom{0}  & \ldots                              &  \ldots \\
\hline
\end{tabular}
\label{t:hvos}
\tablefoot{(1) Values for a distance of 1\,kpc. (2) Inclination angles of 30$^\circ$ assumed. (3) This source also has an extreme-high-velocity outflow with a maximum velocity of $\approx$\,430\,km\,s$^{-1}$. (4) The ratio is estimated to be very low, but the value of 3 is assumed. (5) Uncertain, but not later than F5.}
\end{table*}

There exists in the literature a number of sources with characteristics similar to that of HD\,101584, in particular the presence of a bipolar outflow with a very high maximum velocity, $>$\,100\,km\,s$^{-1}$. These are particularly interesting since the estimated ages of the outflows are of the order only a few\,$\times$\,10$^2$ years, i.e, they are objects for which the observed phenomenon is very energetic and recent. A number of them have been imaged in CO line emission, for example, IRAS\,08005--2356 \citep{sahapate15}, IRAS\,16342--3814 \citep{sahaetal17a}, and IRAS\,22026+5306 \citep{sahaetal06}. Some of their relevant characteristics (and those of HD\,101584) are summarised in Table~\ref{t:hvos}, and we will here give a brief comparison between HD\,101584 and these sources. The central stars, when identified, are all of relatively early spectral type, F5 or earlier, and they all have low $^{12}$C/$^{13}$C ratios, $\la$\,10. The maximum expansion velocities of the high-velocity outflows lie in the range 150 to 250\,km\,s$^{-1}$, and their masses are of the order 10$^{-(1-2)}$\,$M_\odot$. The energies and scalar momenta lie within a factor of a few, and within the ranges found by \citet{bujaetal01} for a larger sample of proto-PNe. The momentum rates of their outflows are much higher than can be supplied by radiation. 

IRAS\,16342--3814 has been studied in most detail using ALMA \citep{sahaetal17a}, and some further comparison can therefore be made with this object. It has an additional extremely-high-velocity bipolar outflow with a maximum expansion velocity of $\approx$\,430\,km\,s$^{-1}$ (adopting an inclination angle of 43$^\circ$) and a PA slightly different than that of the high-velocity outflow. In terms of mass, energy, and scalar momentum, this outflow has 7\,\%, 51\,\%, and 21\,\% of that of its high-velocity outflow, and it is estimated to be older suggesting a recurrent phenomenon. IRAS\,16342--3814 further has an expanding equatorial density enhancement in the form of a torus \citep[that obscures the central star even at 12\,$\mu$m;][]{verhetal09}. Its estimated size, expansion velocity, and particle density are 1300\,au, 20\,km\,s$^{-1}$, and 10$^{(6-8)}$\,cm$^{-3}$, respectively (the expansion age is 160\,years). Thus, its characteristics are comparable to those of the EDE component of HD\,101584. The molecular species detected towards this source is still limited, but include CO, SiO, SO, SO$_2$, and HCN (in some cases through their rare isotopologues). 

Another type of possibly related objects are the red novae, most likely the stellar-merger end products of a CE evolution. These are objects that also show spectacular circumstellar characteristics in molecular line emission \citep{kamietal18b}. An interesting example is CK~Vul, an object where the radioactive molecule $^{26}$AlF was recently detected \citep{kamietal18a}. The Boomerang nebula could be of this type \citep{sahaetal17b}. Its circumstellar characteristics are remarkable, in particular the spherical outflow of high velocity where the gas has cooled down to temperatures below that of the cosmic microwave background. In addition, there is an equatorial density enhancement and a bipolar high-velocity outflow. The circumstellar mass is high, $\ga$\,3\,$M_\odot$, and consequently the mass of the star, $\ga$\,4\,$M_\odot$. \citet{sahaetal17b} argued that this object is in a post-CE-evolution phase after the companion merged into the core of an RGB star.

In terms of chemistry, OH231.8+4.2, an object with a morphology similar to that of HD\,101584 \citep{alcoetal01, bujaetal02}, is a suitable object to compare with \citep{sacoetal15, velietal15}, but the chemical aspects will be discussed in a forthcoming paper. Also the red novae are interesting  comparison sources in terms of circumstellar chemistry \citep{kamietal18b}.

%
%
%
%
%
\section{Conclusions}

We have used ALMA and single-dish data to determine the physical and chemical characteristics of the circumstellar environment of the star HD\,101584, a binary system with a period in the range 150--200 days consisting of a luminous, evolved star and a low-mass companion. The circumstellar medium is rich in molecules of different types, 12 (not counting isotopologues) have been detected, and a significant fraction of the line emissions have been mapped with angular resolutions in the range 0\farcs1 to 0\farcs6. The different chemistry and excitation conditions required by different molecules have been utilised in the analysis presented in this paper, where we focus on the physical characteristics of the circumstellar medium, and its consequences for the interpretation of the evolutionary status of HD\,101584. An SED has been constructed, using also our ArTeMiS and ALMA data, to provide complementary information on the dusty circumstellar medium.

The circumstellar chemistry will be discussed in a forthcoming paper, but we note that the detected lines are typical for an oxygen-rich chemistry with a significant presence of sulphur species, but also weak lines from carbon-species, other than CO, are present. In addition, the existence of more complex species (in the circumstellar context) like H$_2$CO and CH$_3$OH points towards a chemistry that is most likely affected by also shocks and/or dust grains. 

We have identified four distinct components of the circumstellar medium, and the most likely interpretations are: {\it i)} a slowly rotating circumbinary disk of diameter $\approx$\,150\,[$D$/1\,kpc]\,au, {\it ii)} an expanding disk or torus of diameter $\approx$\,3000\,[$D$/1\,kpc]\,au, {\it iii)} a bipolar high-velocity outflow reaching 24\,000\,[$D$/1\,kpc]\,au from the centre, surrounded by {\it iv)} an hourglass structure forming the inner part of two bubbles that enclose the outflow. The outflow is oriented essentially along the line of sight, an inclination angle of only 10$^\circ$\,(-5$^\circ$,+10$^\circ$, {\bf 2$\sigma$ errors}), and the central star is seen through a region that has been (at least partly) cleared from material. The circumbinary disk and the expanding equatorial disk/torus are seen close to face-on. There is strong evidence of a second bipolar outflow in a direction different from that of the major outflow. In addition, there is structure in the hourglass component that is not understood, but it can possibly be related to an interaction between the two outflows. A 3D-reconstruction of the source has been attempted based on the assumption that the radial outflow velocity scales linearly with the distance to the centre. The mass of the circumstellar gas is $\approx$\,0.5\,[$D$/1\,kpc]$^2$\,$M_\odot$, about half of it lies in the expanding equatorial density enhancement. 

The 1.3\,mm continuum is dominated by emission from the equatorial disk/torus, but 30--50\,\% of the flux comes from extended low-brightness emission whose morphology is difficult to determine. The innermost region of the circumbinary disk is particularly prominent, and the estimated size is $\approx$\,10\,[$D$/1\,kpc]\,au. The position of the continuum peak coincides with the Gaia position of HD\,101584 within the uncertainties. Modelling the full SED in terms of a flared disk seen close to face-on (the equatorial density enhancement), surrounded by a much thinner spherical envelope, leads to a dust mass estimate of $\approx$\,0.01\,[$D$/1\,kpc]$^2$\,$M_\odot$, and a substantial fraction of this mass must be in the form of large-sized, up to 1\,mm, grains. About 7\,\% of the grains are in the form of crystalline water.

The absence of a reliable distance estimate makes the identification of the evolutionary status of HD\,101584 difficult. The low estimated $^{12}$C/$^{13}$C ratio and the high effective temperature are strong arguments for a phase beyond the RGB, that is, either post-RGB or post-AGB. We have therefore looked at two separate cases, one of lower luminosity (a post-RGB star) and one of higher luminosity (a post-AGB star). Relations between core mass and luminosity provide estimates of the present day masses of the stars. Combined with the circumstellar masses, this lead to estimated initial masses that lie in the ranges 0.5--1\,$M_\odot$ and 2--7\,$M_\odot$ in the post-RGB and post-AGB cases, respectively. The low estimated $^{17}$O/$^{18}$O is consistent with a low-mass post-RGB star, while it is inconsistent with an intermediate-mass post-AGB star. Thus, based on these data we advocate a post-RGB identification.

Irrespective of the evolutionary status, the results presented in this paper favour an interpretation of HD\,101584 as an object where the red-giant evolution was terminated prematurely due to a CE evolution that ended avoiding a stellar merger. The remaining hydrogen envelope of HD\,101584 was ejected during the interaction and it now forms a circumstellar medium of considerable complexity. The size and kinematics of the bipolar high-velocity outflow provide a time scale of $\approx$\,770\,[$D$/1\,kpc]\,yr for the circumstellar evolution. This is a time scale in line with those estimated for apparently similar objects. The considerably shorter kinematical age of the expanding equatorial disk/torus is at odds with the conclusion by \citet{hugg07} that the torus develops first and the jet follows shortly thereafter. However, the almost face-on orientation of the equatorial disk/torus around HD\,101584 makes it difficult to estimate reliably its kinematics and hence its age. Most likely, the ejection of the stellar envelope was powered by released gravitational energy. We estimate that the kinetic energy of the accelerated gas is 7$\times$10$^{45}$\,[$D$/1\,kpc]$^2$\,erg, and that the kinetic to released gravitational energy is about 10, irrespective of the evolutionary status. This indicates that substantial additional energy must have been released, for example, due to material falling towards the companion. As a consequence a circum-companion accretion disk may have formed that now drives a highly collimated jet that, in turn, drives the expanding high-velocity molecular gas. The existence of the second bipolar outflow points to a recurrent phenomenon rather than a single explosive event, although its age is uncertain. Its interpretation in the light of a CE-evolution scenario is not clear.

%
%
%
%
\begin{acknowledgements}
HO and WV acknowledge support from the Swedish Research Council. WV acknowledges support from the ERC through consolidator grant 614264.
This paper makes use of the following ALMA data: ADS/JAO.ALMA\#2012.1.00248.S and \#2015.1.00078.S. ALMA is a partnership of ESO (representing its member states), NSF (USA) and NINS (Japan), together with NRC (Canada) and NSC and ASIAA (Taiwan) and KASI (Republic of Korea), in cooperation with the Republic of Chile. The Joint ALMA Observatory is operated by ESO, AUI/NRAO and NAOJ.
This paper makes use of the following APEX data: O-093.F-9307, O-096.F-9303, O-098.F-9314, O-099.F-9310, O-0100.F-9302, O-0101.F-9302, and O-0102.F-9402A (SEPIA Band 9 Science Verification data). The Atacama Pathfinder EXperiment (APEX) is a collaboration between the Max-Planck-Institut f{\"u}r Radioastronomie, the European Southern Observatory, and the Onsala Space Observatory. This paper has also made use of the NASA/ IPAC Infrared Science Archive, which is operated by the Jet Propulsion Laboratory, California Institute of Technology, under contract with the National Aeronautics and Space Administration, and data products from the Wide-field Infrared Survey Explorer, which is a joint project of the University of California, Los Angeles, and the Jet Propulsion Laboratory/California Institute of Technology, funded by the National Aeronautics and Space Administration.  Frederic Sch{\"u}ller is gratefully acknowledged for help with the reduction of the ArTeMiS data. Finally, we are grateful to the anonymous referee who provided a very constructive report.
\end{acknowledgements}


\bibpunct{(}{)}{;}{a}{}{,}

\newpage

\begin{appendix}

\section{Channel maps}

   \begin{figure*}
   \centering
   \includegraphics[width=18cm]{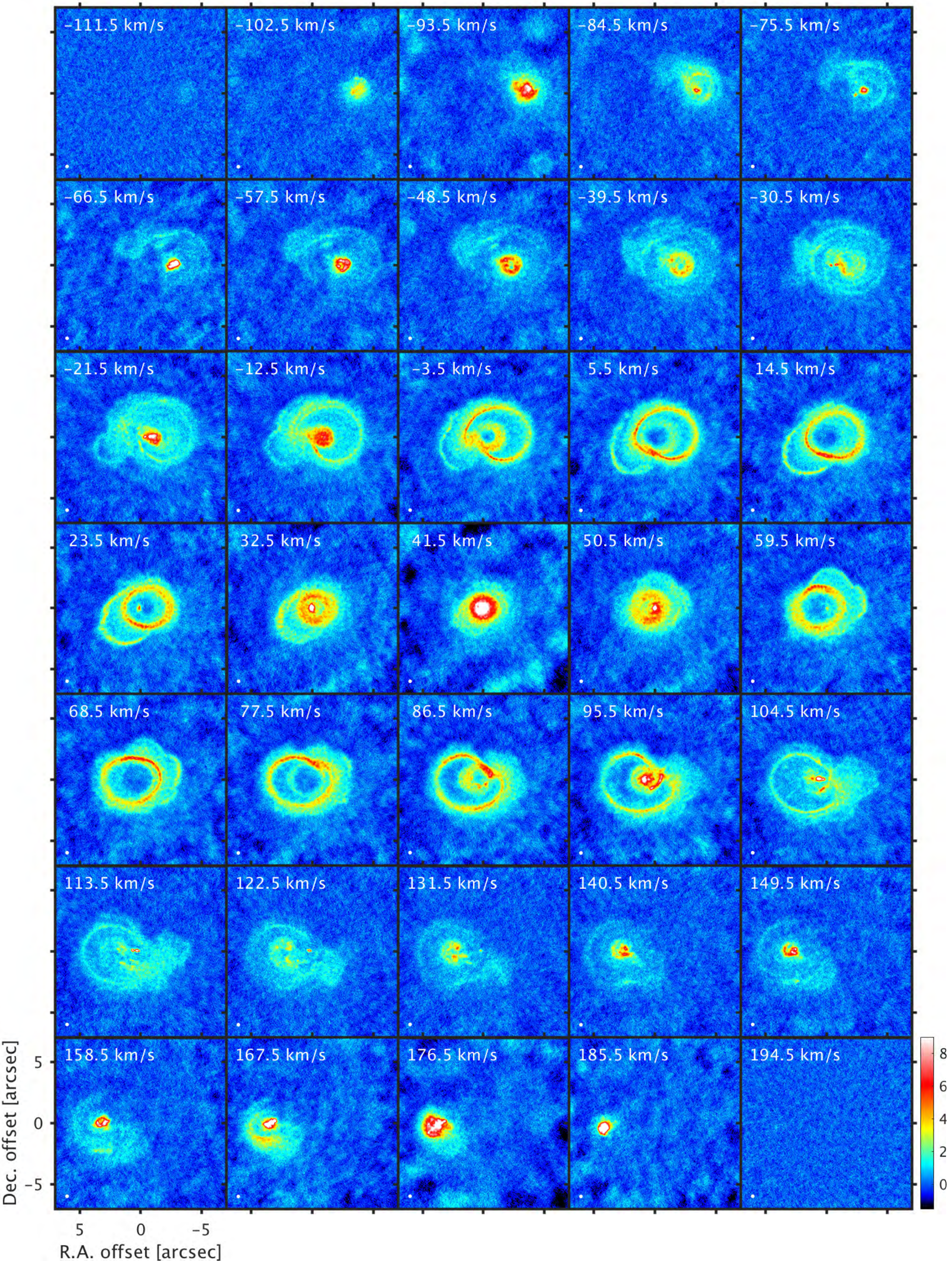}
      \caption{CO(\mbox{2--1}) channel maps with a width and spacing of 9\,km\,s$^{-1}$ at a resolution of 0\farcs085 (the beam is shown in the lower left corner of each panel). The flux scale is in mJy\,beam$^{-1}$. Emission from all the identified components of the circumstellar medium of HD\,101584 are present in this line. }
   \label{f:co_channels}
   \end{figure*}   

   \begin{figure*}
   \centering
   \includegraphics[width=18cm]{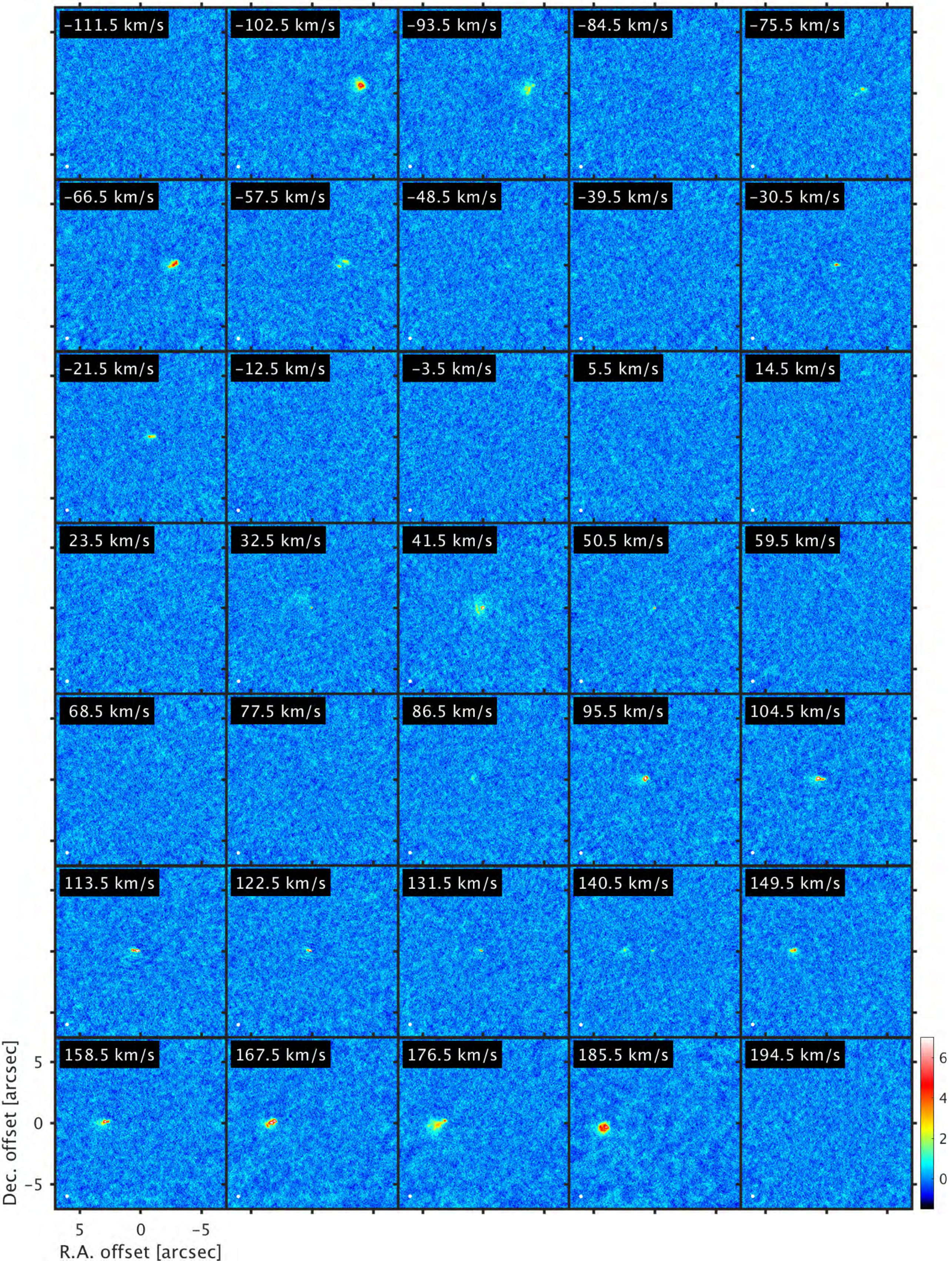}
      \caption{SiO(\mbox{5--4}) channel maps with a width and spacing of 9\,km\,s$^{-1}$ at a resolution of 0\farcs085 (the beam is shown in the lower left corner of each panel). The flux scale is in mJy\,beam$^{-1}$. Emission from the HVO (especially the bright spots) dominates for this line, but also emission from the CCS is present at the centre. }
   \label{f:sio_channels}
   \end{figure*}   
%
%
%
%
\section{The 3D reconstruction}

   \begin{figure*}
   \centering
   \includegraphics[width=14cm]{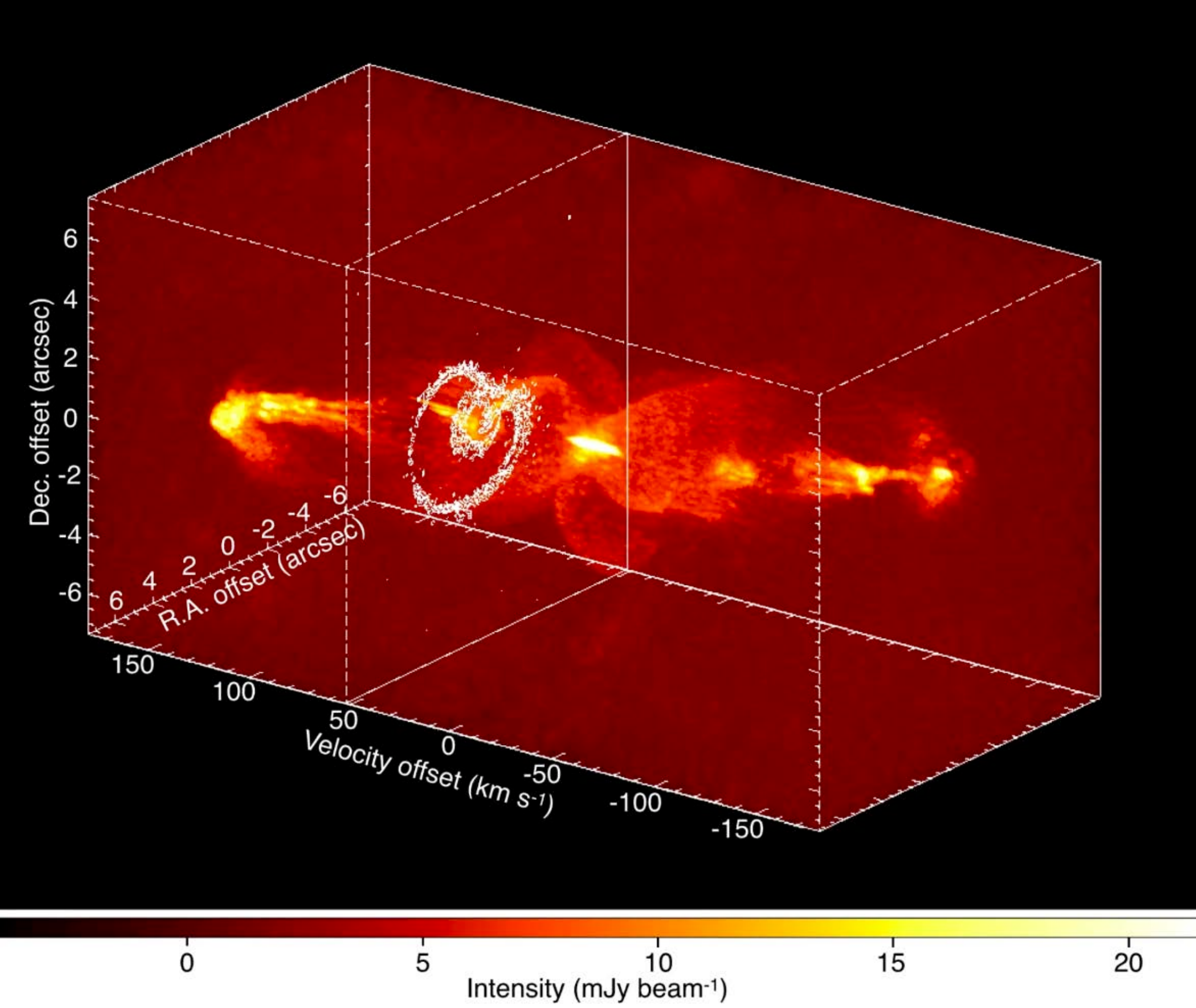}
      \caption{Single velocity step of a movie showing the relation between the CO(\mbox{2--1}) channel maps and the 3D structure (online movie). The velocity is given with respect to the systemic velocity (41.7\,km\,s$^{-1}$). The conversion to spatial scale along the line of sight is 0\farcs165 per km\,s$^{-1}$.}
   \label{f:movie}
   \end{figure*}   

\end{appendix}

\end{document}